%% file: cexp.tex
\documentclass[ppnp]{article}
\usepackage{amssymb,amsmath} 
\usepackage[pdftex]{graphicx}
\usepackage{subfig}
\usepackage{verbatim} 
\newfont{\thiplo}{msbm10 scaled\magstep 2}
\newfont{\gothic}{eufb10 scaled\magstep 2}
\newfont{\unc}{eurb10} 
\newskip\humongous \humongous=0pt plus 1000pt minus 1000pt
\def\caja{\mathsurround=0pt}
\def\eqalign#1{\,\vcenter{\openup1\jot \caja
        \ialign{\strut \hfil$\displaystyle{##}$&$
        \displaystyle{{}##}$\hfil\crcr#1\crcr}}\,}
\newif\ifdtup

\def\eqright #1\cr{\noalign{\hfill$\displaystyle{{}#1}$}}
\def\eqleft #1\cr{\noalign{\noindent$\displaystyle{{}#1}$\hfill}}

\def\oldreffmt#1{\rlap{[#1]} \hbox to 2\parindent{}}

\def\figfmt#1{\rlap{Figure {#1}} \hbox to 1in{}}

\def\sectioneq{\def\theequation{\thesection.\arabic{equation}}{\let
\holdsection=\section\def\section{\setcounter{equation}{0}\holdsection}}}%

\newcounter{holdequation}

\def\begineq #1\endeq{$$ \refstepcounter{equation}\eqalign{#1}\eqno
	(\theequation) $$}
\def\contlimit{\,{\hbox{$\longrightarrow$}\kern-1.8em\lower1ex
\hbox{${\scriptstyle (a\rightarrow0)}$}}\,}
\def\centeron#1#2{{\setbox0=\hbox{#1}\setbox1=\hbox{#2}\ifdim
\wd1>\wd0\kern.5\wd1\kern-.5\wd0\fi
\copy0\kern-.5\wd0\kern-.5\wd1\copy1\ifdim\wd0>\wd1
\kern.5\wd0\kern-.5\wd1\fi}}
\def\centerover#1#2{\centeron{#1}{\setbox0=\hbox{#1}\setbox
1=\hbox{#2}\raise\ht0\hbox{\raise\dp1\hbox{\copy1}}}}
\def\centerunder#1#2{\centeron{#1}{\setbox0=\hbox{#1}\setbox
1=\hbox{#2}\lower\dp0\hbox{\lower\ht1\hbox{\copy1}}}}
\def\lsim{\;\centeron{\raise.35ex\hbox{$<$}}{\lower.65ex\hbox
{$\sim$}}\;}
\def\gsim{\;\centeron{\raise.35ex\hbox{$>$}}{\lower.65ex\hbox
{$\sim$}}\;}

\def\super#1{\ifmmode \hbox{\textsuper{#1}}\else\textsuper{#1}\fi}
\def\textsuper#1{\newcount\holdspacefactor\holdspacefactor=\spacefactor
$^{#1}$\spacefactor=\holdspacefactor}

\def\getcite#1,{\advance\citenumber by1
\def\getcitearg{#1}\def\lastarg{@}
\ifnum\citenumber=1
\ref{#1}\let\next=\getcite\else\ifx\getcitearg\lastarg\let\next=\relax
\else ,\ref{#1}\let\next=\getcite\fi\fi\next}

\def\pom{I\!\!P}
\def\spom{{\rm P\kern -0.36em\llap \small I\,}}
\def\sspom{{\rm P\kern -0.33em\llap \footnotesize I\,}}

\relax
\def\contlimit{\,{\hbox{$\longrightarrow$}\kern-1.8em\lower1ex
\hbox{${\scriptstyle (a\rightarrow0)}$}}\,}
\def\upon #1/#2 {{\textstyle{#1\over #2}}}
\relax
\renewcommand{\thefootnote}{\fnsymbol{footnote}}

\sectioneq

\def\til#1{\centeron{\hbox{$#1$}}{\lower 2ex\hbox{$\char'176$}}}
\def\tild#1{\centeron{\hbox{$\,#1$}}{\lower 2.5ex\hbox{$\char'176$}}}
\def\sumtil{\centeron{\hbox{$\displaystyle\sum$}}{\lower
-1.5ex\hbox{$\widetilde{\phantom{xx}}$}}}

\def\d{\mathrm{d}}

\newcommand{\Mhmax}{M_{h,{\mathrm{max}}}}
\newcommand{\B}{\ensuremath{\mathbf}}
\def\pom{I\!\!P}
\def\GeVc{GeV/c}
\def\GeVcc{GeV/c$^{2}$}

\let\oldmarginpar\marginpar
\renewcommand\marginpar[1]{\-\oldmarginpar[\raggedleft\footnotesize #1]%
{\raggedright\footnotesize #1}}
\reversemarginpar

\begin{document} 
\renewcommand{\thefootnote}{\arabic{footnote}}
\begin{titlepage} 
\rightline{\vbox{\halign{&#\hfil\cr
&MAN/HEP/2010/1\cr}}}
\rightline{\vbox{\halign{&#\hfil\cr
&\today\cr}}} 
\vspace{0.25in} 

\begin{center} 
  
{\large\bf Central Exclusive Particle Production at High Energy Hadron
  Colliders}

\medskip

M.G. Albrow\footnote{albrow@fnal.gov; corresponding author}
\vspace{0.5cm}
\centerline{Fermi National Accelerator Laboratory, USA.}
T.D. Coughlin\footnote{coughlin@hep.ucl.ac.uk} 
\vspace{0.5cm}
\centerline{University College London, Gower Street, London, WC1E 6BT, UK.}
J.R. Forshaw\footnote{jeff.forshaw@manchester.ac.uk} 
\centerline{The University of Manchester, Oxford Road, Manchester M13
  9PL, UK.}
\end{center}
\setcounter{footnote}{0}

\begin{abstract} 
 We review the subject of central exclusive particle production at
 high energy hadron colliders. In particular we consider reactions of the
 type $A+B\rightarrow A + X + B$, where $X$ is a fully specified
 system of particles that is well separated in rapidity from the
 outgoing beam particles. We focus on the case where the colliding
 particles are strongly interacting and mainly they will be protons
 (or antiprotons) as at the ISR, S$p\bar{p}$S, Tevatron and
 LHC. The data are surveyed and placed
 within the context of theoretical developments.    
\end{abstract} 

\end{titlepage} 
\tableofcontents

\newpage
\input{cxpintro}
\input{cxpomeron}

\input{cexptheory}
\input{cxpisr}

\input{cxpft}

\input{cxspps}

\input{cxptev}

\input{cexplhc}

\bibliographystyle{elsarticle-num}
\bibliography{cexp}

\end{document}

%% file: cxpintro.tex
\section{Introduction}
\label{sec:intro}
Central  exclusive particle production (CEP) is defined as the class of reactions $A+B\rightarrow A+X+B$, where the
colliding particles $A$ and $B$ emerge intact and a produced state, $X$, is fully measured. We will review the particularly interesting case where the centre-of-mass energy
$\sqrt{s}$ is large, such that one can have rapidity regions $\Delta y \gsim 3$ completely devoid of particles
between $X$ and the outgoing particles $A$ and $B$ (rapidity gaps).
With this definition, only high energy colliders are relevant, and we shall focus on hadron colliders.
The field of central production in nucleus-nucleus collisions at the Relativistic Heavy Ion Collider (RHIC) 
and at the Large Hadron Collider (LHC) merits a review in itself and we will not attempt to cover it.
The first hadron collider was the CERN Intersecting Storage Rings (ISR), which carried out $pp$ and $p\bar{p}$ 
collisions (also $dd$ and $\alpha\alpha$). The $Sp\bar{p}S$ collider at CERN followed and, most recently, 
the Tevatron at Fermilab has examined $p\bar{p}$ collisions. In the near future, the LHC will furnish us 
with $pp$ (and $AA$) collisions with much higher energy and luminosity than before. It
opens up the exciting new possibility of producing heavy systems $X$, such as Higgs bosons, 
weak vector bosons, supersymmetric (SUSY) particles and other ``Beyond the Standard Model'' (BSM) particles. 

Much of the renewed interest in central exclusive production comes from the higher mass reach that will be opened
up at the LHC. Added to that are strong constraints that are unique to exclusive processes. In some
cases these allow measurements of particle properties that can be obtained no other way. The ``strong constraints'' are both kinematic and dynamic. In an exclusive
process such as $p+p\rightarrow p+X+p$, four-momentum conservation means that knowing the incoming proton momenta,
and measuring the outgoing proton momenta (in high precision forward spectrometers), 
the mass $M_X$ is determined~\cite{Albrow:2000na}. This is expected to be possible with a resolution of $\sigma(M_X) \approx 2$ GeV$/c^2$
\emph{per event}~\cite{Albrow:2008pn}, independent of the decay of $X$, even if $X$ decays with a large amount of missing energy, e.g. $X\rightarrow W^+W^- \rightarrow e^+\mu^- \nu \bar{\nu}$. With a precise
calibration of the forward spectrometers (using the QED process $p+p\to p+\mu^+\mu^-+p$), the central mass can be found with a resolution
improving as $\sigma(M_X)/\sqrt{N}$, and if (in the case of a resonance) the width $\Gamma_X$ exceeds a few GeV, that too can be measured. In addition, the transverse momenta, $p_T$, of the scattered protons are typically $p_T \lsim$ 1~GeV/c and so $p_T(X)$
is small. In this case, dynamic constraints, discussed later, allow one to
also determine the spin $J$, parity P and charge-parity C of any exclusively produced particle, such as a Higgs boson.

The possibility to instrument the LHC in order to measure the small angle protons produced in CEP 
has been explored through the FP420 R\&D project (so named because the forward protons would be 
detected 420~m from the collision region)\footnote{There are also plans to install detectors at 220/240~m.} involving a consortium of ATLAS and CMS physicists 
and theorists~\cite{Albrow:2008pn}. Suitable sub-detectors have now been proposed: AFP (ATLAS 
Forward Protons) in ATLAS and HPS (High Precision Spectrometers) in CMS. Precision tracking and 
timing detectors at $220/240$~m and 420~m will measure the momenta of both protons.

In this review our attention will focus on CEP mainly through $I\!\!PI\!\!P \rightarrow X$ and $\gamma\gamma\rightarrow X$,  although photoproduction, i.e. $\gamma\pom \rightarrow X$, is briefly mentioned. We use the symbol ``$\pom$'' to signify pomeron exchange (see Section 2). We will neglect contributions from 
odderon exchange and consider data that are not much contaminated by sub-leading Regge exchanges. In
classifying the production this way we are making explicit the theoretical property that in both QED and in Regge theory
the scattering amplitudes can often be factorized so as to separate the
dependence on the beam particles from the dependence on the production of the central system. However, it is
not always the case that the scattering amplitudes factorize this way. In particular, the QCD production of
high-mass, short-distance ($Q^2 \gg \Lambda_{QCD}^2$) systems can be computed in perturbation theory (at least up to soft
rescattering corrections) and the amplitudes do not factorize in quite this manner. Generally, they do not
factorize wherever the dominance of a single pomeron pole does not hold. Nevertheless, we shall still
follow popular nomenclature and refer to the production of a central system via strong dynamics as
$I\!\!PI\!\!P \rightarrow X$, but it is to be understood that this is more of a placeholder than a description
of the underlying strong dynamics. It is our aim in the following sections to clarify these matters and to
present the theory underpinning central exclusive particle production.

In the next section we introduce the relevant ideas in Regge theory and show how one may
describe CEP in terms of pomeron exchange in the $t$-channel. References~\cite{forshaw} 
and \cite{donnachie} are treatises on diffraction and the pomeron and our treatment here will 
be brief. In Section 3 we explain how one can use QCD perturbation theory to compute the CEP of 
high mass systems. While perturbative Quantum Chromodynamics (QCD) is the well established theory 
of high-$Q^2$ (short distance) strong
interaction processes, diffractive processes almost always involve some low-$Q^2$ physics, 
where perturbation theory cannot be used and factorization theorems do not hold. We are thus 
still far from a complete understanding, and that makes for an interesting field, both for 
theory and for experiment.

We shall then move on to a survey of the experimental data, covering the published results from hadron
colliders up to the end of 2009. We do not attempt to review CEP in $e^+e^-$ or $ep$ machines, and only briefly
mention fixed target experiments. We end our review with a forward look to the LHC era.

Some conventions need stating. We will frequently use the term proton to mean $p$ or $\bar{p}$ whenever the distinction is
unimportant or obvious. In a  reported cross section, such as $2.09\pm0.90\pm0.19$ pb, the first uncertainty is statistical and the 
second is systematic. Feynman-$x$, $x_F = p_z/p_{\mathrm{beam}}$, represents the ratio of longitudinal momentum to the
beam momentum and $\xi = 1 - x_F$ denotes the fractional momentum loss of a beam particle (typically a $p$ or $\bar{p}$). The rapidity of a particle is defined as $y = \frac{1}{2} \mathrm{ln} \left( \frac{E + p_z}{E - p_z} \right)$, and the 
pseudorapidity $\eta = - \mathrm{ln(tan} \frac{\theta}{2})$ with $\theta$ the polar angle; $\eta = y$ for massless particles.

%% file: cxpomeron.tex
\section{The Pomeron}
\label{pomeron}

A detailed treatment of Regge theory is beyond the scope of this
review and we refer instead to Refs.~\cite{forshaw,donnachie,Collins:1977jy,Kaidalov:1979jz,Ganguli:1980mp}.
Here we provide a very basic introduction. Pomeron phenomenology was developed originally to describe the
behaviour of total hadronic cross sections and diffractive/elastic
scattering at high energies.  Most recently, it has been taken to a new
level by experiments conducted mainly at the HERA $ep$ collider (see
Refs.~\cite{Wolf:2009jm,Newman:2009bq} and references therein) and the
Tevatron $p\bar{p}$ collider (see Section~\ref{tev}).

When the CERN ISR came into operation in 1971, the quark model had been formulated but not yet QCD~\cite{isrrep}; 
asymptotic freedom was uncovered in 1973. At that time, strong
interaction physics tended to focus on total cross sections, elastic
scattering and few-body reactions (e.g. $\pi^- + p \rightarrow \pi^0 +
n$). General methods, based on
     the unitarity, analyticity and crossing symmetry of the
     scattering matrix provided the foundations, with Regge theory
     providing the principal tool. With the advent of  QCD the
     emphasis shifted to the investigation of scattering processes at
     short distances for which the strong coupling is small and
     perturbative methods can be exploited.
However, soft diffraction and elastic scattering processes cannot be
described by perturbative QCD, and Regge theory remains an
important tool. In Regge theory these process are described as
     the $t$-channel exchange of ``reggeons'' ($I\!\!R$), which correspond to a sum of mesons ($\rho^0, \omega^0$, etc.) with the same
     quantum numbers. The contribution of the reggeons to the elastic scattering
     cross section falls with increasing centre-of-mass energy as
     $s^{\alpha_{I\!\!R}(0)-1} \sim 1/\sqrt{s}$, where
     $\alpha_{I\!\!R}(t)$ is the reggeon trajectory which is a
     function of the Mandelstam four-momentum transfer squared,
     $t$. By the Optical Theorem, the reggeon contribution to total cross sections likewise falls
     as the centre-of-mass energy increases. The observed rise of total hadronic cross sections therefore
     mandated the emergence of a new reggeon,  with intercept 
     $\alpha_{I\!\!P}(0) > 1.0$. 
     To generate a non-falling total cross section, the exchange must have isospin zero and even charge parity, C $= +1$,
     i.e. it has the quantum numbers of the vacuum. The new reggeon
     was dubbed the pomeron
     ($I\!\!P$) after Pomeranchuk, who had
     previously studied the behaviour of vacuum exchange in Regge theory.

\begin{figure}[htbp]
\begin{center}
\includegraphics[width=0.4\textwidth,angle=0]{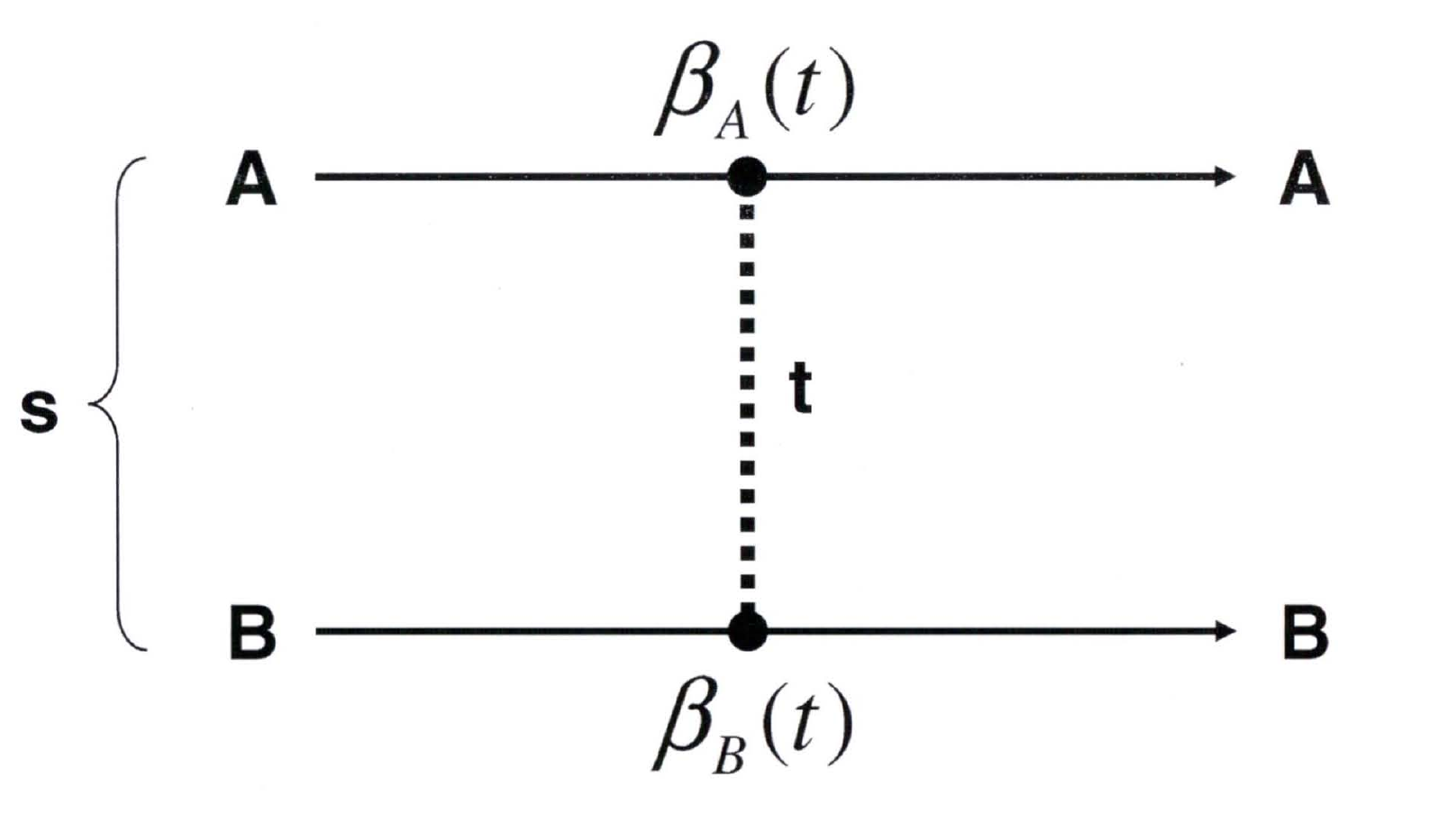}
\caption{Elastic scattering between two hadrons $A$ and $B$, at centre-of-mass energy $\sqrt{s}$.
The four-momentum transfer squared is $t$.
\label{elastic2}} 
\end{center}
\end{figure}
     
At high enough centre-of-mass energy, if one assumes the dominance of a single Regge pole, the elastic scattering of strongly interacting particles may be described by pomeron exchange, see Fig.~\ref{elastic2}. The elastic scattering amplitude for $AB\to AB$ is thus approximated by
\begin{equation}
\frac{A(s,t)}{s} = \beta_{A}(t)
  \beta_B(t) \, \eta(t)\left(\frac{s}{s_0}\right)^{\alpha_{\pom}(t)-1}
\end{equation} 
where
\begin{equation}
\eta(t) = i - \cot\left(\frac{\pi \alpha_{\pom}(t)}{2}\right)
\end{equation}
is the signature factor, $\alpha_{\pom}(t)$ is the pomeron trajectory,
$\beta_{A,B}(t)$ fixes the coupling of the
pomeron to the external particles and $s_0$ is a constant.
The Optical Theorem then relates the total cross section for $AB \to
X$, $\sigma_T$, to the imaginary part of the forward ($t=0$) scattering amplitude via
\begin{equation}
{\mathrm{Im}}A(s,0) = s \; \sigma_T(s)
\end{equation}
and so
\begin{equation}
\sigma_T(s) = \beta_A(0) \beta_B(0)
\left(\frac{s}{s_0}\right)^{\alpha_{\pom}(0)-1}~.
\label{eq:sigmat}
\end{equation}
Thus we see that if the pomeron intercept $\alpha_{\pom}(0) >1$ the
total cross section rises with energy, in accord with the
data. Conversely, the contributions of any Regge
poles with $\alpha_{I\!\!R}(0) <1$ (such as those containing the
$\rho$ and $\pi$) become negligible at sufficiently high energy.

One might hope that the properties of
Regge poles, i.e. their intercepts and couplings, would emerge from
calculations based on QCD. To an extent that is what happens. For
example, the gluon is
known to ``reggeize'' to a simple Regge pole in perturbative
QCD after re-summing to all orders in $\alpha_s \ln(s)$ (see for example Ref.~\cite{forshaw} and references therein). 
However,
calculations are generally plagued by the need to focus on processes
at short distances, where perturbation theory is valid, and total
hadronic cross sections certainly do not fall into that class. It is
also far from clear that amplitudes are dominated by a single Regge
pole at high energies, although there is some indication that this is
so in the case of hadron-hadron elastic scattering at small (but not
too small) values of $t$ \cite{Donnachie:1983hf,Donnachie:1992ny}. In that case, fits to data suggest the
existence of a pomeron with intercept
\begin{equation}
\alpha_{\pom}(t) \approx \alpha_{\pom}(0) + \alpha'_{\pom}\;t \approx 1.08 + (0.25~{\mathrm{GeV}}^{-2}) \; t.
\label{eq:donlan}
\end{equation}
More recent analyses suggest that a global fit to all soft data from
the ISR, S$p\bar{p}$S and Tevatron may require a pomeron with a higher
intercept and substantial screening corrections (see 
for example Refs.~\cite{Luna:2008pp,Gotsman:2008tr,Ryskin:2009tj} and references therein).

\begin{figure}[htbp]
\begin{center}
\includegraphics[width=0.52\textwidth,angle=0]{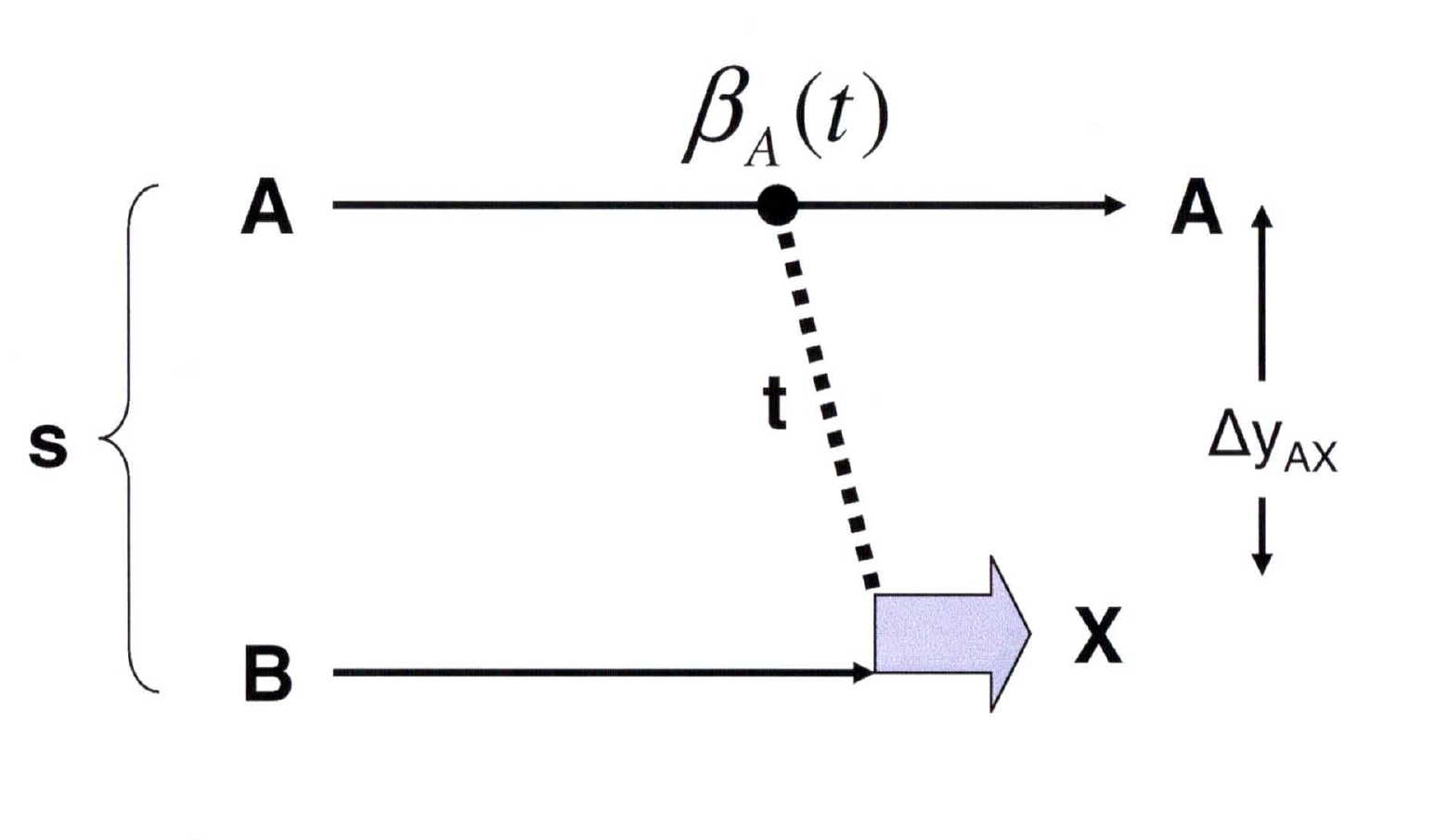}
\includegraphics[width=0.43\textwidth,angle=0]{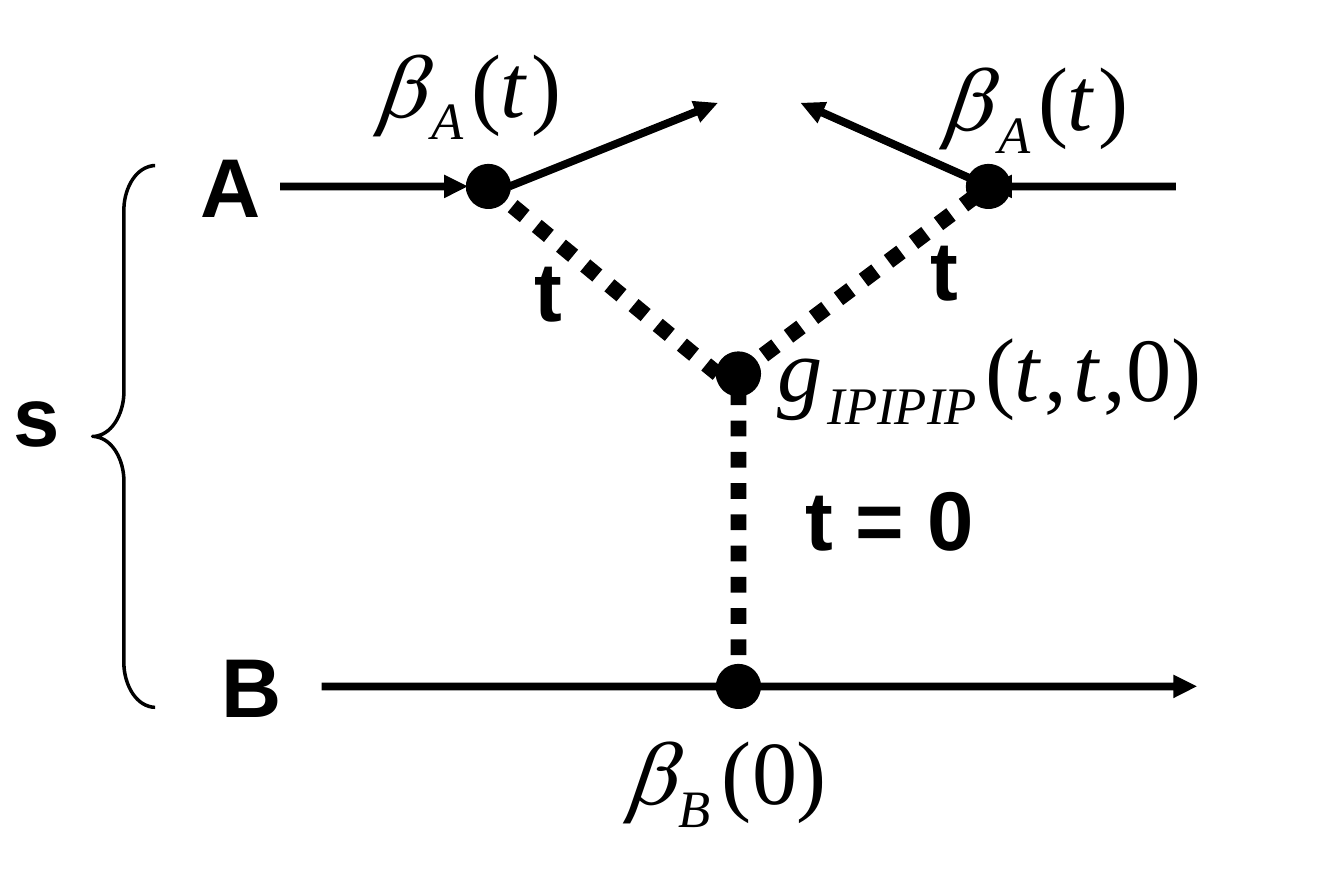}
\caption{(a) Diffractive excitation of particle $B$ to a state of mass $M_X$ by pomeron exchange.
(b) The corresponding cut diagram in the limit of large $M_X$.
\label{sdediag}}
\end{center}
\end{figure}

\begin{figure}[htbp]
\begin{center}
\includegraphics[width=0.52\textwidth,angle=0]{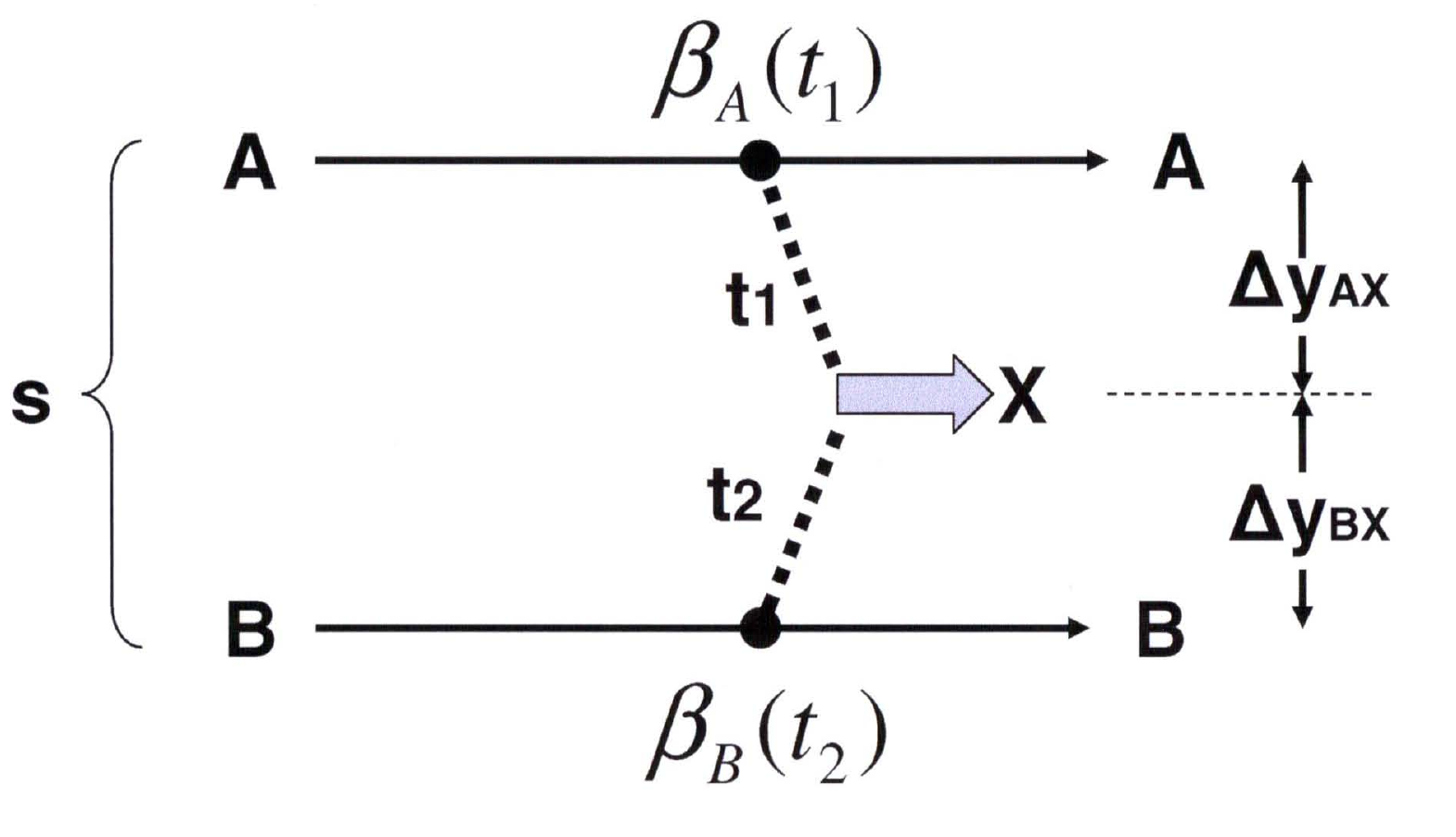}
\includegraphics[width=0.43\textwidth,angle=0]{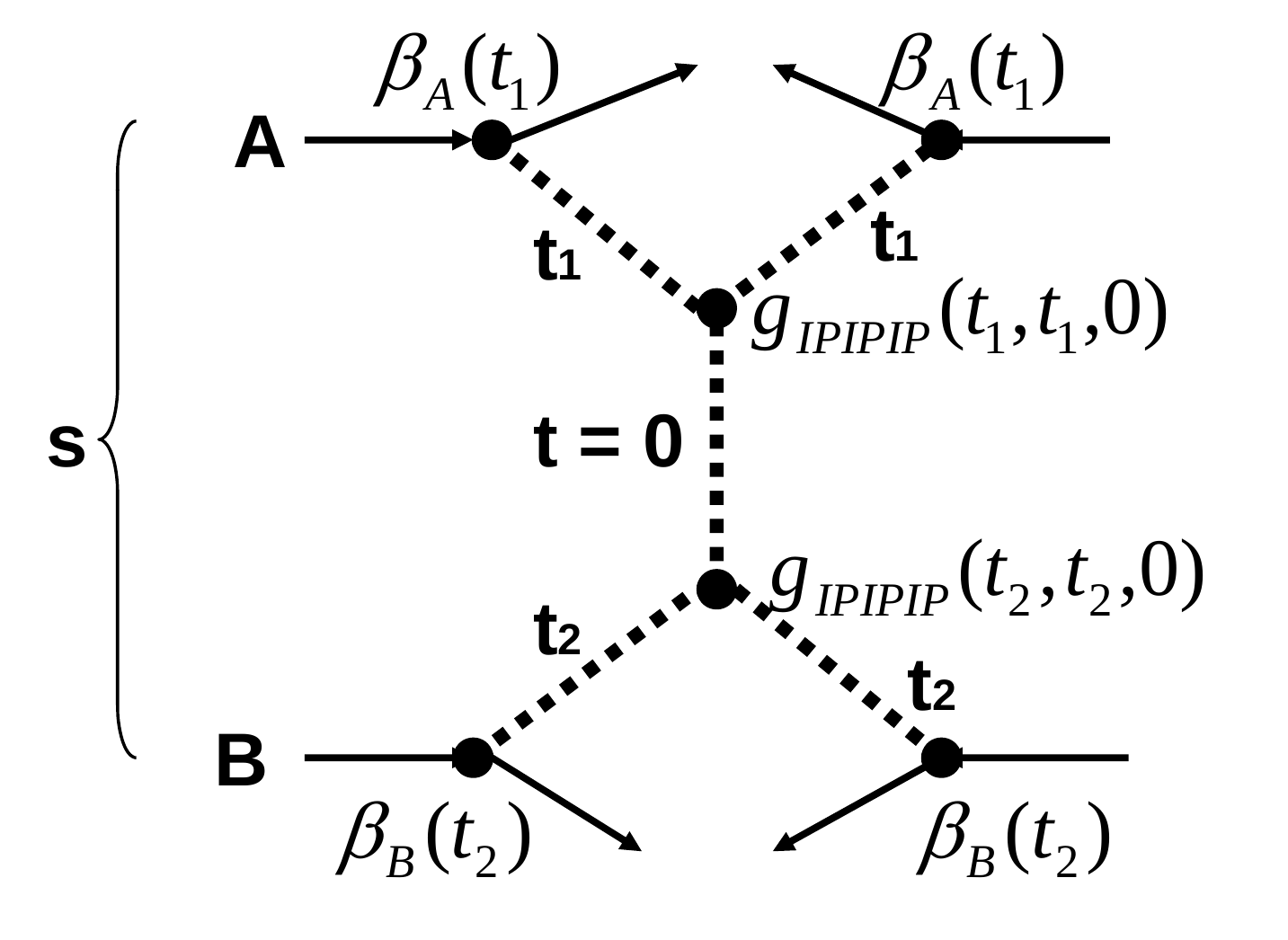}
\caption{(a) Diagram for double pomeron exchange. (b) The
  corresponding cut diagram in the limit of large $M_X$. 
\label{quint}}
\end{center}
\end{figure}

Regge theory is not restricted to the consideration of elastic
scattering amplitudes. A reggeon calculus can be developed, and used to tackle processes such as those illustrated in
Fig.~\ref{sdediag}(a) and Fig.~\ref{quint}(a). Again the dotted lines
represent pomerons, and pomeron dominance is presumed to pertain if the
relevant sub-energies are large enough, i.e. $\Delta y_{AX} \gsim 3$ in
Fig.~\ref{sdediag}(a) and $\Delta y_{AX},\Delta y_{BX} \gsim 3$ in
Fig.~\ref{quint}(a), where $\Delta y_{AX}$ is the rapidity interval
between $A$ and $X$ (and similarly for $\Delta y_{BX}$). For the
single diffractive dissociation process represented in
Fig.~\ref{sdediag}(a) one may write \cite{muellerOT,CandM}
\begin{equation}
M_X^2 \frac{\d\sigma}{\d t \, \d M_X^2} = \beta_{A}(t)^2 \, |\eta(t)|^2
\left(\frac{s}{M_X^2}\right)^{2\alpha_{I\!\!P}(t)-2} \; \sigma_{B\pom}(M_X^2,t)
\label{eq:triplepom}
\end{equation}
and we are invited to think of $\sigma_{B\pom}(M_X^2,t)$ as the total
cross section for $B\pom$ scattering at energy $M_X$. It is to be
noted that the normalization of $\sigma_{B\pom}(M_X^2,t)$ is a matter
of convention. Provided $M_X$
is sufficiently large, we expect that it is itself driven by pomeron
exchange and $\sigma_{B\pom}(M_X^2,t)\propto
(M_X^2)^{\alpha_{\pom}(0)-1}$. This is shown in Fig.~\ref{sdediag}(b) which
illustrates the $M_X^2$ discontinuity in the relevant three-body
amplitude.  We should stress that the pomeron is not a real particle and 
pomeron-induced cross sections are not directly measurable; however they are useful
constructs. Going one step further, we can rewrite Eq.~(\ref{eq:triplepom}) as
\begin{equation}
\frac{\d\sigma}{\d t \, \d\xi} =  f_{\pom/A}(\xi,t)  \; \sigma_{B\pom}(M_X^2,t)
\label{eq:triplepom1}
\end{equation}
where we define a pomeron ``flux''
\begin{equation}
f_{\pom/A}(\xi,t)= \beta_A(t)^2 \, |\eta(t)|^2 \left( \frac{1}{\xi} \right)^{2
  \alpha_{\pom}(t)-1}~ \label{eq:cep-regge}
\end{equation}
and $\xi$ is the fractional energy lost by the beam particle
$A$, i.e. $M_X^2 = \xi  s$. This approach describes very well
the HERA data on single diffraction dissociation, albeit with a
pomeron trajectory that differs from that in Eq.~(\ref{eq:donlan}).
In particular the $t$-dependence is consistent with a flat trajectory \cite{Newman:2009bq}. 

The study of double pomeron exchange ($D \pom E$), illustrated in Fig.~\ref{quint}, has a long history~\cite{Kaidalov:1979jz,Ganguli:1980mp,Kibble:1963zz,Lipes:1969ed,TerMartirosyan:1972ib,shankar,Kaidalov:1974qi,Chew:1974vu,Azimov:1974fa,Pumplin:1976dm,Desai:1978rh,Streng:1985ju}. In the Regge framework, such exchanges are responsible for the CEP process, and we may write the cross section for $A+B\to A+X+B$ in terms of the total cross section for two pomerons to fuse, producing the central system $X$, $\sigma_{\pom\pom}$:
\begin{equation}
\frac{\d\sigma}{\d t_1 \d t_2 \, \d \xi_1 \, \d \xi_2} = f_{\pom/A}(\xi_1,t_1)
  f_{\pom/B}(\xi_2,t_2) \; \sigma_{\pom \pom}(M_X^2,t_1,t_2)~.
\label{eq:reggecep}
\end{equation}
Again, the $\xi_i$ are the fractional energy losses, and
kinematics fixes $M_X^2 = \xi_1 \xi_2 s$. We shall return to this
formula for $D \pom E$ in Sections 4--6.

Eq.~(\ref{eq:triplepom1}) and Eq.~(\ref{eq:reggecep}) clearly exhibit
Regge factorization and the similarity
to the two-photon production case is striking - the pomeron flux
playing the role of the Weisz\"acker-Williams flux in the case of
photons (e.g. see Ref.~\cite{Budnev}).  Unfortunately
Regge theory does not tell us how to compute the cross section
$\sigma_{\pom \pom}(M_X,t_1,t_2)$, although it does predict the behaviour
for large $M_X$. So, although we have a model for the rapidity
dependence of the central system, we are not able to predict the
overall production rate without further model dependence. Furthermore, we should bear in mind that there is
no \emph{a priori} reason why more sophisticated pomeron diagrams should not
be relevant. 

It is time now to turn our attention to a rather different approach to
CEP, namely an approach based firmly within perturbative QCD. As we
shall see, it does not give rise to Regge factorization and leads to
essentially different predictions for the CEP of systems for which
$M_X \gg \Lambda_{{\mathrm{QCD}}}$. Furthermore, it is an approach
that has recently had some striking success in predicting the CEP rate of
$\chi_{c0}$ and dijets at the Tevatron.

%% file: cexptheory.tex
\section{QCD Models of Central Production}
\label{sec:dyna}

\begin{figure}[h]
\centering
\includegraphics[width=0.35\textwidth]{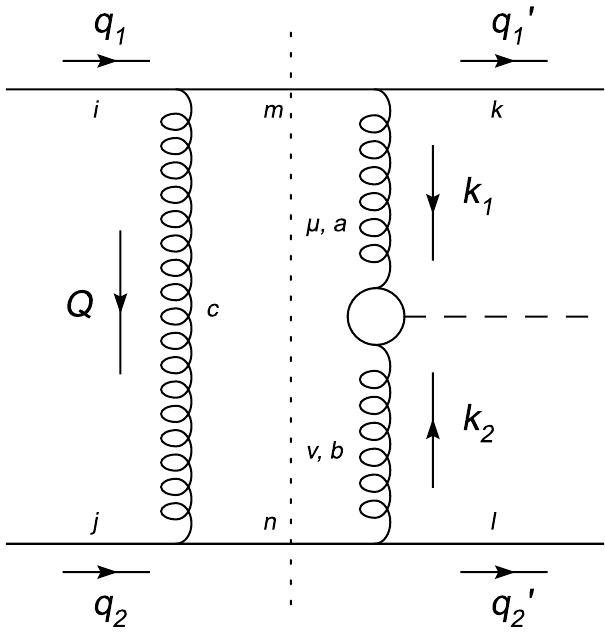}
\caption{The relevant lowest order Feynman diagram for $qq \to q+H+q$.}
\label{fig:qHq}
\end{figure}

\subsection{Perturbative QCD}\label{sec:Durham}

The perturbative approach to CEP was originally developed in
Refs.~\cite{jz0,Khoze:2001xm,kmrphptev,Khoze:1997dr,Kaidalov:2003ys}. We shall refer to the approach presented in 
those papers as
the Durham model, and it is our goal in this subsection
to review the calculation and comment upon its uncertainties. Our present
focus is on the central exclusive production 
of a Higgs, i.e. $pp \to p+H+p$, although the theory is not essentially different for other high mass systems 
provided they contain a suitable hard scale. In particular, we shall
use the same framework when it comes to comparing to Tevatron data on
CEP of dijets, di-photons and $\chi_c$ mesons.
The calculation starts by considering the process $qq \to q+H+q$ at lowest order in perturbation theory, as shown in
Fig.~\ref{fig:qHq}. The Higgs is produced via a top quark loop and a
minimum of two gluons needs to be exchanged in order that no colour be transferred between the
incoming and outgoing quarks. The outgoing quark momenta may be parametrised in terms of the momentum fraction each 
transfers to 
the Higgs, $x_i$, and their transverse momenta, $q'_{iT}$:
\begin{align}
	q'^\mu_1 &= (1-x_1) q_1^\mu + \frac{\B{q}'^\mu_{1T}}{(1-x_1)s} q_2^\mu + q'^\mu_{1T} \\
	q'^\mu_2 &= (1-x_2) q_2^\mu + \frac{\B{q}'^\mu_{2T}}{(1-x_2)s} q_1^\mu + q'^\mu_{2T} 
\end{align}
with $s$ denoting the centre-of-mass energy squared. In the high energy limit the colour-singlet amplitude is 
dominated by its imaginary part, as expected from arguments based on Regge theory (see for example Ref.~\cite{forshaw}). 
It may therefore be computed by considering only the cut diagram of Fig.~\ref{fig:qHq}. Thus, for small $x_i$ and 
with the colour-singlet exchange contribution projected out, the amplitude is given by
\begin{align}
	\Im {\mathrm{m}} A &\approx \frac{C_F^2}{N^2-1} \int \! \frac{\d^4 Q}{(2\pi)^2} \delta_{(+)}((q_1-Q)^2) \delta_{(+)}((q_2+Q)^2) \nonumber\\
		& \qquad \qquad \times \frac{2gq_1^\alpha2g q_{2\alpha}}{Q^2} \frac{2gq_1^\mu}{k_1^2} \frac{2g q_2^\nu}{k_2^2}  V_{\mu\nu}  \;.
\end{align}
In the Standard Model, the Higgs production vertex is
\begin{equation}
V_{\mu \nu} =  \left( g_{\mu \nu} - \frac{k_{2 \mu} k_{1 \nu}}{k_1 \cdot k_2}
\right) V~,
\end{equation}
where $V = m_H^2 \alpha_s/(4 \pi v) F(m_H^2/m_t^2)$, $v$ is the Higgs
vacuum expectation value and $F \approx 2/3$ provided the Higgs is
not too heavy~\cite{Ellis:1975ap,Voloshin:1985tc,Shifman:1979eb}. The Durham group also include 
a next-to-leading order K-factor correction to this vertex~\cite{kmrphptev}, though this is extracted from 
calculations of inclusive Higgs production~\cite{Kunszt:1996yp,Spira:1995rr} and not from an explicit calculation 
of the next-to-leading order corrections in CEP.

To proceed further, we parametrise the loop momentum in terms of Sudakov variables, $Q=\alpha q_1 +\beta q_2 +Q_T$; 
the $\delta$-functions which fix the cut quark lines on-shell then set $\alpha \approx -\beta \approx \B{Q}_T^2/s \ll 1$ and $Q^2\approx Q_T^2 = -\B{Q}_T^2$. As always, we neglect terms that are suppressed in $s$, such as the product $\alpha \beta$. 

We can compute the contraction $q_1^{\mu} V^{ab}_{\mu \nu} q_2^{\nu}$ either directly or
by utilising gauge invariance, which requires that $k_1^{\mu} V^{ab}_{\mu \nu} = 
k_2^{\nu} V^{ab}_{\mu \nu} = 0$. Writing\footnote{We can do this because $x_i 
\sim m_H/\surd{s}$ whilst the other Sudakov components are $\sim Q_T^2/s$.} 
$k_i = x_i q_i + k_{i T}$ yields
\begin{equation}
q_1^{\mu} V^{ab}_{\mu \nu} q_2^{\nu} \approx 
\frac{k_{1T}^{\mu}}{x_1} \frac{k_{2T}^{\nu}}{x_2} V^{ab}_{\mu \nu} \approx 
\frac{s}{m_H^2} k_{1T}^{\mu} k_{2T}^{\nu} V^{ab}_{\mu \nu} 
\end{equation}
since $2 k_1 \cdot k_2 \approx x_1 x_2 s \approx m_H^2$. Note that it is
as if the gluons which fuse to produce the Higgs are transversely polarized, i.e. we have for their polarisation vectors:
$\epsilon_i \sim k_{iT}$. Moreover,
in the limiting case that the outgoing quarks carry no transverse momentum
$Q_T = -k_{1T} = k_{2T}$ and so $\epsilon_1 = -\epsilon_2$. This is
an important result; it generalizes to the statement that the centrally produced
system must have a vanishing $z$-component of angular momentum in the limit that
the protons scatter through zero angle (i.e. $~ q_{iT}'^{2} \ll
Q_T^2$)~\cite{jz0}. Since we are interested in  very small angle
scattering this selection rule is effective. One immediate consequence
is that the Higgs decay to $b$-quarks may now be detectable. This is because, for massless quarks, 
the lowest order $q \bar{q}$ background vanishes identically (it does not vanish at next-to-leading order). The
leading order exclusive $b \bar{b}$ dijet background is therefore suppressed by a factor $\sim m_b^2/m_H^2$.
The dominant background thus becomes CEP production of $gg$-dijets, which can be reduced by $b$-tagging both jets.

Returning to the task in hand, the differential cross section is given by
\begin{eqnarray}
\frac{\partial\sigma}{\partial^2 \B{q}_{1T}' \partial^2 \B{q}_{2T}' \partial y} &\approx&
\left(\frac{N_c^2-1}{N_c^2}\right)^2 \frac{\alpha_s^6}{(2 \pi)^5}
\frac{G_F}{\surd{2}} \times
\left[ \int \frac{\d^2 \B{Q}_T}{2 \pi} 
\frac{\B{k}_{1T} \cdot \B{k}_{2T}}{\B{Q}_T^2 \B{k}_{1T}^2  \B{k}_{2T}^2} 
\frac{2}{3} \right]^2~, \label{eq:qHq}
\end{eqnarray}
with $y$ the rapidity of the Higgs. We are mainly interested in the
forward scattering limit whence 
\begin{equation}
\frac{\B{k}_{1T} \cdot \B{k}_{2T}}{\B{Q}_T^2 \B{k}_{1T}^2  \B{k}_{2T}^2} 
\approx -\frac{1}{\B{Q}_T^4}.
\label{eq:fsl}
\end{equation}
As it stands, the integral over $\B{Q}_T$ diverges. However, this is because the lowest perturbative order is insufficient for this process. 
Not only must we include a convolution with non-perturbative parton
distribution functions (PDFs) but, due to the exclusive nature of the
process, the perturbative contribution to the amplitude is enhanced at
each order in $\alpha_s$ by large logarithms in the ratio
$m_H^2/\B{Q}_T^2$. These terms are due to virtual corrections
involving soft gluons or partons that travel collinear to the incoming
hadrons. They must be summed to all orders in order to give a reliable prediction.

\begin{figure}[htb]
\centering
\includegraphics[width=0.7\textwidth]{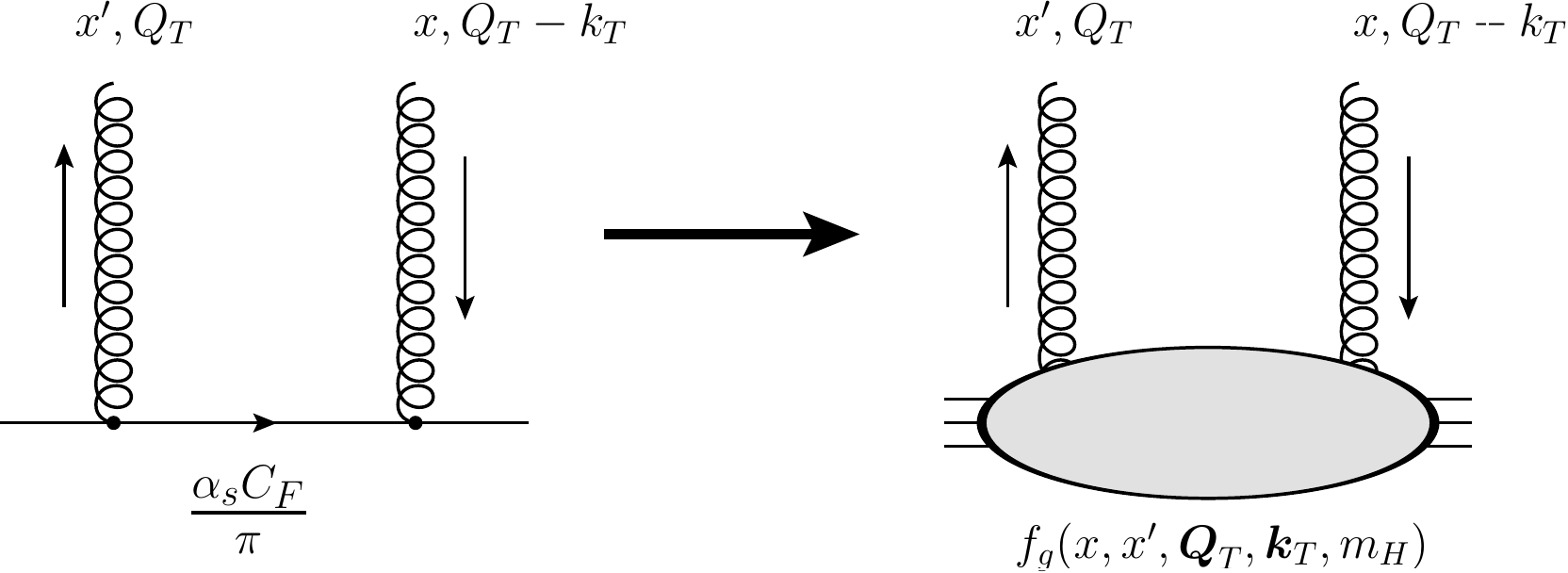}
\caption{The recipe for replacing the quark line (left) by a proton line (right).}
\label{fig:updf}
\end{figure}

The Durham prescription to account for these effects is shown in Fig.~\ref{fig:updf}. They replace $\alpha_s
C_F/\pi$, at the quark level, with a skewed, unintegrated, gluon PDF,
 $f_g(x,x',\B{Q}_T,\B{k}_T,m_H)$. The amplitude is dominated by the region $x\gg x'$, $\B{Q}_T^2\gg \B{k}_T^2$ and 
 in this limit one may relate the $f_g$ distribution to the standard, integrated, gluon distribution, 
 $g(x,\B{Q}_T^2)$~\cite{kmrphptev,Martin:2001ms}:
\begin{align}
	f_g(x,x',\B{Q}_T,\B{k}_T,m_H) &\approx e^{-b\B{k}_T^2/2} R_g \frac{\partial}{\partial \ln \B{Q}_T^2} \left(  \sqrt{T(\B{Q}_T,m_H)} xg(x,\B{Q}_T^2)  \right). \label{eq:pdfApprox}
\end{align} 
The $\B{k}_T$ dependence of the scattered proton is assumed to follow a Gaussian distribution, with slope parameter, $b\approx4~\textrm{GeV}^{-2}$, fixed by a fit to soft hadronic data~\cite{Khoze:2000wk} and cross-checked against the $\B{k}_T$-dependence of diffractive $J/\psi$ production at HERA~\cite{Levy:1997bh}. The factor $R_g$ is given by
\begin{align}
	R_g &= \frac{H_g\left( \frac{x}{2},\frac{x}{2};\B{Q}_T^2  \right)}{xg(x,\B{Q}_T^2)}
\end{align}
and accounts for the skewed effect ($x\neq x'$). Here $H_g$ is the skewed gluon distribution (see for example Ref.~\cite{Belitsky:2005qn}). In~\cite{Shuvaev:1999ce} $R_g$ was shown to be given approximately by
\begin{align}
	R_g &\approx \frac{2^{2\lambda +3}}{\sqrt{\pi}} \frac{\Gamma(\lambda+5/2)}{\Gamma(\lambda +4)}
\end{align}
if one assumes a simple power-law behaviour for the gluon density, $xg(x,\B{Q}_T^2)\sim x^{-\lambda}$. 
For the production of a 120~GeV$/c^2$ Higgs at the LHC, $R_g \approx 1.2$~\cite{Khoze:2001xm}; the off-diagonality therefore
provides an enhancement of $(1.2)^4 \approx 2$ to the cross section. Clearly the current 
lack of knowledge of the
off-diagonal gluon is one source of uncertainty in the calculation. 

Equation~(\ref{eq:pdfApprox}) also includes a Sudakov suppression factor, $T(\B{Q}_T,m_H)$. 
This is present because real emission from the process is 
forbidden\footnote{Note that, for a colour-singlet central system, a single real emission is 
forbidden regardless of experimental cuts, since it violates colour conservation.}. It is the 
Sudakov factor that collects terms in the perturbation series enhanced by 
logarithms in $m_H^2/\B{Q}_T^2$. Taking into account the leading and next-to-leading 
logarithms, i.e. all terms of order $\alpha_s^n\ln^m(m_H^2/\B{Q}_T^2)$ with $m=2n,2n-1$, the 
Sudakov factor has the form~\cite{kmrphptev,Dokshitzer:1978hw,Coughlin:2009tr}
\begin{align}
	T(\B{Q}_T,m_H) &= \textrm{exp}\left(  -\int_{\B{Q}_T^2}^{m_H^2/4} \! \frac{\d\B{q}_T^2}{\B{q}_T^2}\frac{\alpha_s(\B{q}_T^2)}{2\pi} \int_0^{1-|\B{q}_T|/m_H} \! \d z \left[  z P_{gg}(z) +n_f P_{qg}(z)  \right]  \right) \label{eq:SudakovFactor}
\end{align} 
where $P_{ij}$ are the usual DGLAP splitting functions \cite{Gribov:1972ri,Altarelli:1977zs,Dokshitzer:1977sg}. It is worth noting that, in order 
for this expression to collect the next-to-leading logarithms, the lower limit on the 
$\B{q}_T^2$-integral and the upper limit on the $z$-integral must be specified precisely. 
The upper $\B{q}_T^2$ limit in contrast, may be multiplied by any $\mathcal{O}(1)$ number 
without changing the leading or next-to-leading logarithms. This is because it corresponds 
to hard, non-collinear, emission and as such is not enhanced by a logarithm. The lower limit 
on the $\B{q}_T^2$-integral can be understood as follows. For hard collinear emissions, 
$1-z \sim \mathcal{O}(1)$, the region $\B{q}_T^2<\B{Q}_T^2$ is already included in the PDFs 
(see the discussion below). On the other hand, for the soft region, $1-z \sim \mathcal{O}(|\B{q}_T|/m_H)$ (relevant for the $1/(1-z)$ piece of $P_{gg}$), emissions with $\B{q}_T^2<\B{Q}_T^2$ cannot resolve the exchanged colour-singlet system (the size of which is of order $1/|\B{Q}_T|$) and so do not contribute.

The form of Eq.~(\ref{eq:SudakovFactor}) is somewhat different from that appearing in much of the literature. 
In particular, in Ref.~\cite{Kaidalov:2003ys} the upper limit on the $z$-integral was determined to be $|\B{q}_T|/(|\B{q}_T|+0.62 m_H)$. It was shown recently that this determination is incorrect and Eq.~(\ref{eq:SudakovFactor}) is in fact the correct 
form of the Sudakov factor~\cite{Coughlin:2009tr}. At the present time, the extent to which this 
modification alters previous predictions has not been fully assessed. However, it has been estimated 
to give approximately a factor two reduction in the cross section for Higgs masses in the range 
100--500~GeV$/c^2$, with the suppression increasing(decreasing) for larger(smaller) Higgs 
masses~\cite{Coughlin:2009tr}. As we shall see, this
uncertainty still lies within the range of the other uncertainties in the
calculation.

The Sudakov factor solves the $Q_T \to 0$ divergence problem we encountered in the 
lowest order calculation; the cross section, integrated over final state hadron momenta, 
is now given by
\begin{align}
	\frac{\partial \sigma}{\partial y} & \approx \frac{1}{256 \pi b^2} \frac{\alpha_s G_F \sqrt{2}}{9} 
	\left(   \int \! \frac{\d^2 \B{Q}_T}{\B{Q}^4_T} \frac{\partial}{\partial \ln \B{Q}_T^2} 
	\left(  \sqrt{T} x_1 g(x_1,\B{Q}_T^2)  \right)  \frac{\partial}{\partial \ln \B{Q}_T^2} 
	\left(  \sqrt{T} x_2 g(x_2,\B{Q}_T^2)  \right)    \right)^2 \label{eq:dsigdy}
\end{align}
and since the exponential Sudakov factor vanishes faster than any power of $Q_T$ as $Q_T \to 0$, 
this integral is finite. Furthermore, for a 120~GeV$/c^2$ Higgs, the 
typical $Q_T \sim 2~\textrm{GeV}/c$~\cite{Forshaw:2005qp}, justifying, \emph{a posteriori}, 
the use of perturbation theory. Note that the parton momentum fractions are, to a very good 
approximation, equal to the fractional momentum losses of the incident protons, i.e. 
$x_i \approx \xi_i$ and $y_X = \frac{1}{2} \ln(\xi_1/\xi_2)$.

\begin{figure}[h!]

	\includegraphics[height=0.3\textwidth]{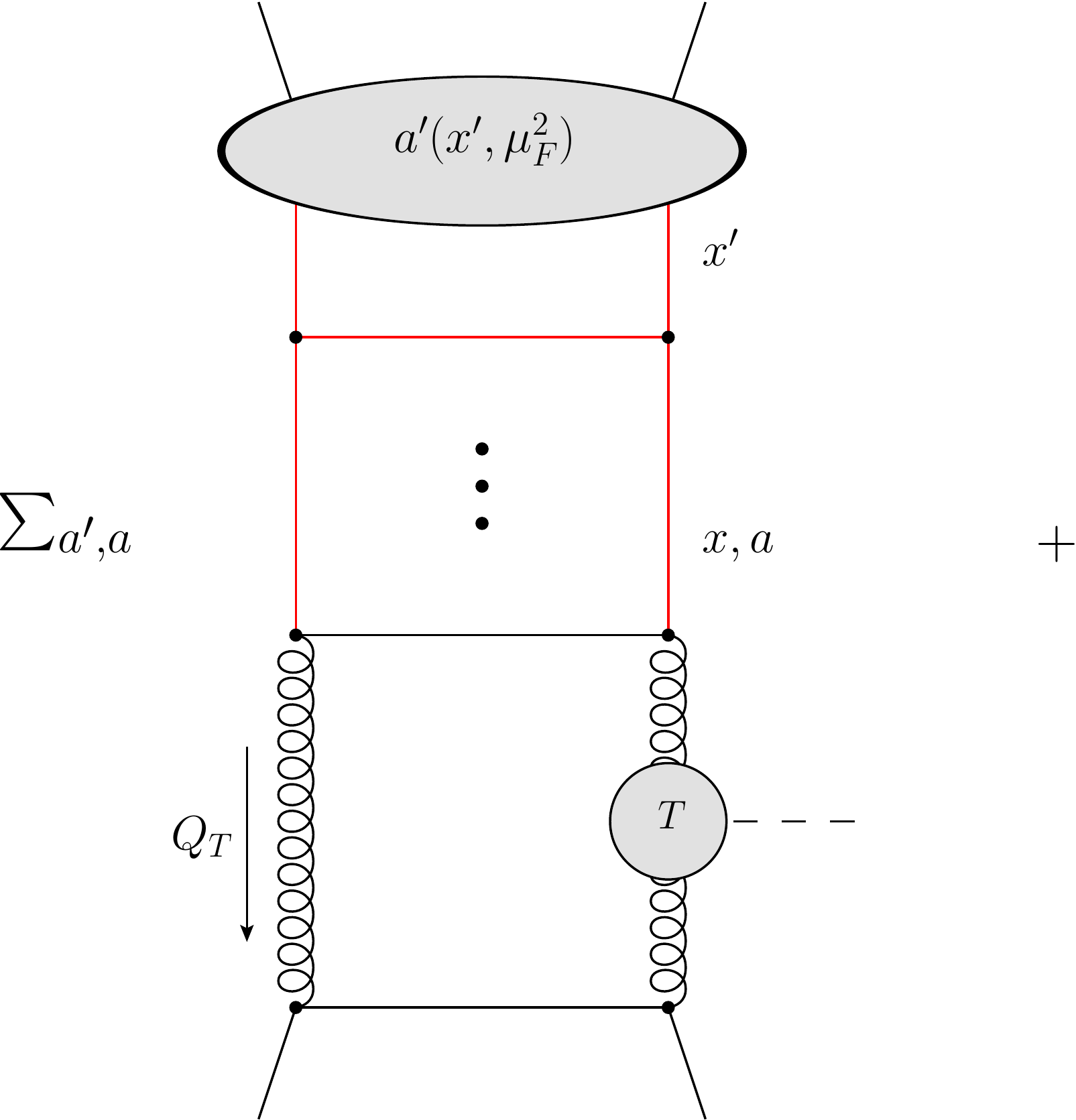} \hspace{0.05\textwidth}
	\includegraphics[height=0.3\textwidth]{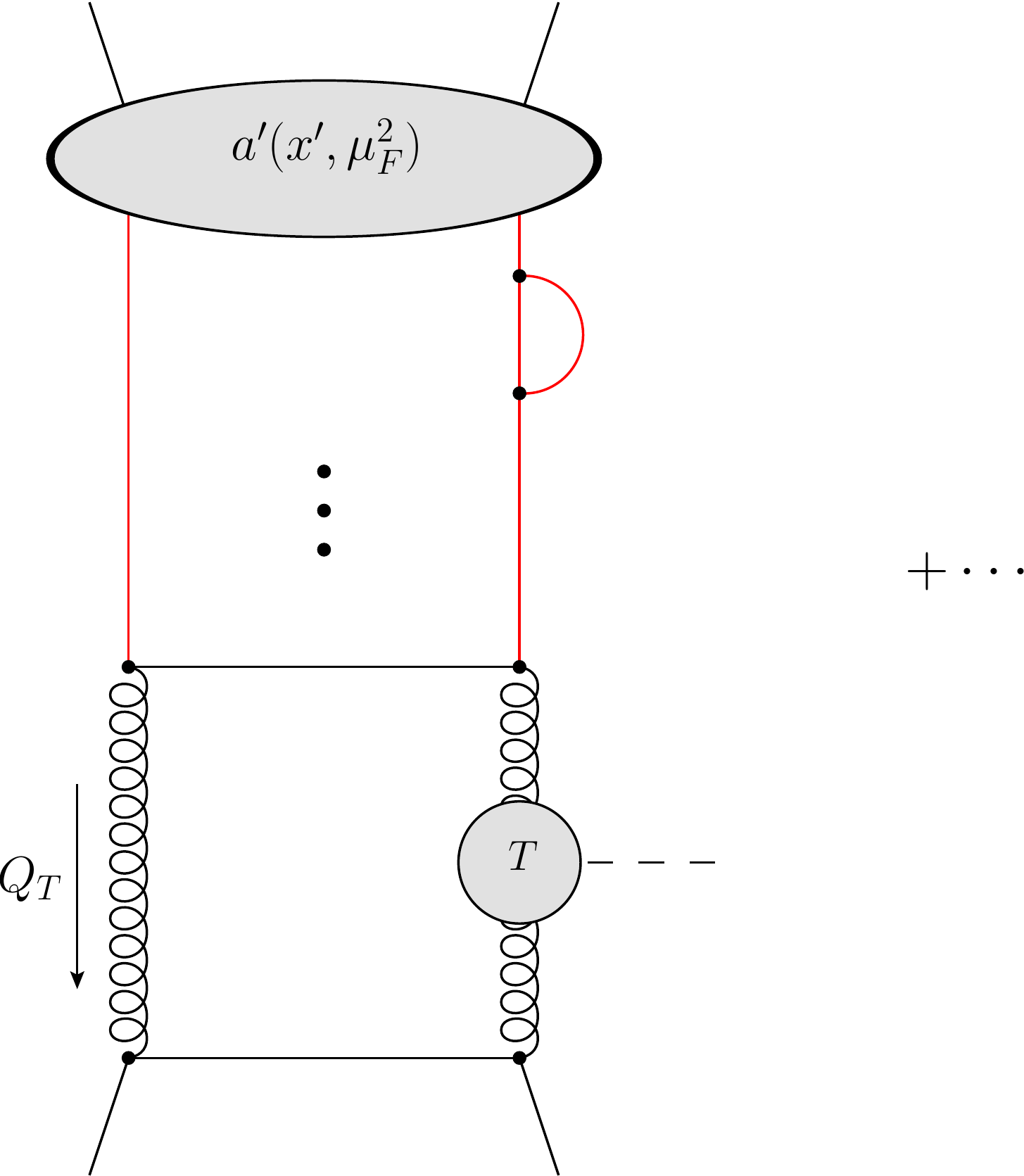} \hspace{0.05\textwidth}
	\includegraphics[height=0.2\textwidth,viewport= 60 -60 250 250]{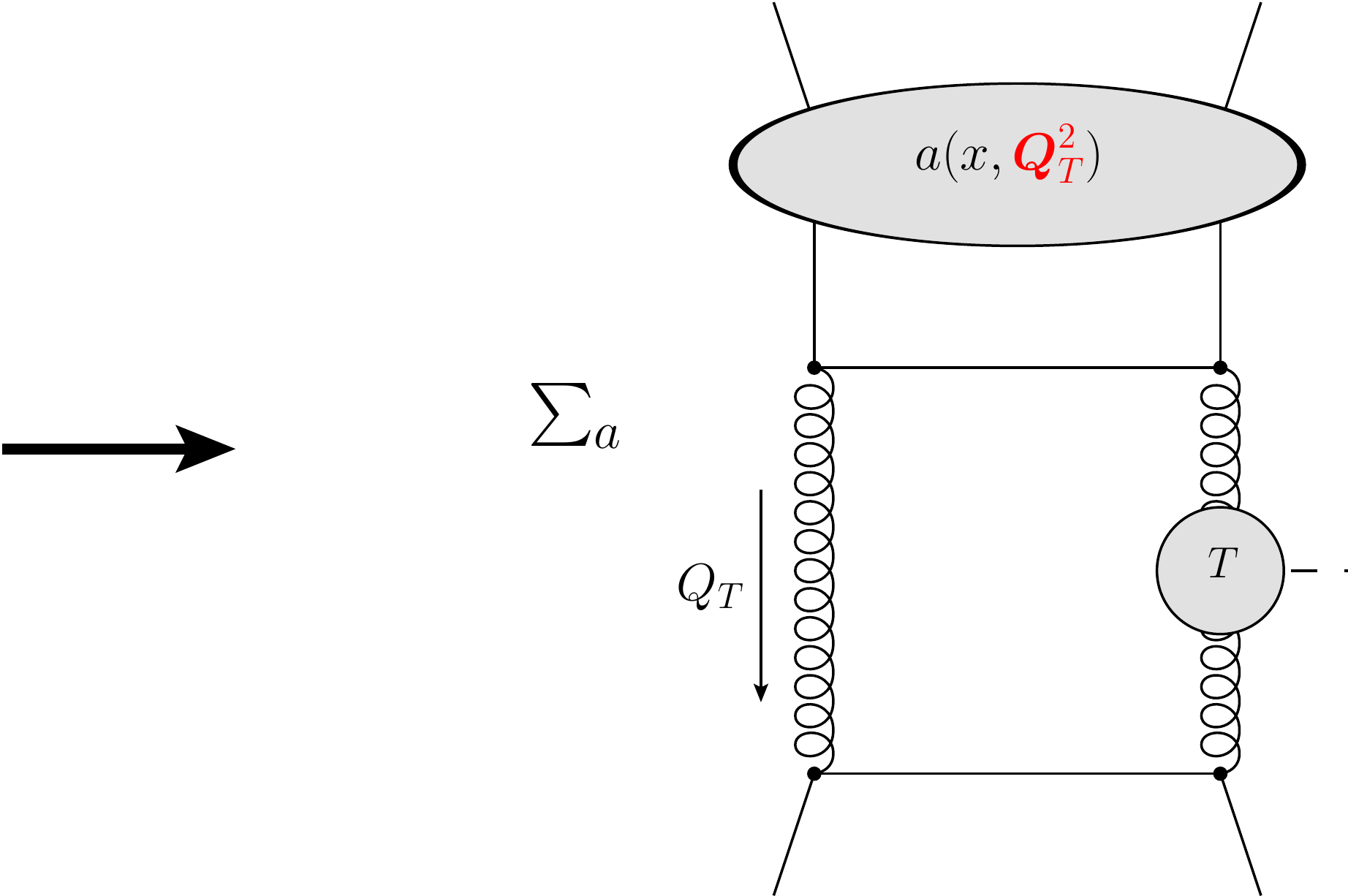}
	\caption{Schematic diagram of the transverse momentum ordered ladder corrections which evolve 
	the PDFs to the scale $\B{Q}_T^2$. Solid lines denote either quarks or gluons.}\label{fig:ladders}

\end{figure}

Returning to Eq.~(\ref{eq:pdfApprox}), the derivative structure can be understood 
as follows (see Ref.~\cite{Coughlin:2009tr} for more details). Transverse momentum ordered ladder diagrams, 
as shown schematically in Fig.~\ref{fig:ladders}, are 
enhanced\footnote{In a physical gauge the large logarithmic corrections are organised into 
ladders on a diagram-by-diagram basis. In contrast, the ladder structure only emerges in a covariant 
gauge after a sum over diagrams and application of the Ward identity 
(see for example Ref.~\cite{Dokshitzer:1991wu}).} by logarithms of $\B{Q}_T^2$. Summing up these contributions evolves the 
parton distributions associated with the upper and lower hadrons to the scale $\B{Q}_T^2$. 
The central hard process however, is sensitive to the transverse momenta of the final rung 
in each ladder. If the final rung has an energy much larger than its transverse momentum, 
$E\gg |\B{Q}_T|$, then it is included in the PDF and so we require the gluon PDF unintegrated 
over transverse momentum, $\partial g(x,\B{Q}_T^2)/\partial \B{Q}_T^2$. If, on the other hand, 
the final rung is soft, $E\sim |\B{Q}_T|$, then it may not be included in the PDF. This 
contribution is accounted for by the derivative of the Sudakov
factor. Shown also in 
Fig.~\ref{fig:ladders} (by the blob labelled $T$) are the corrections making up the Sudakov factor. These 
are virtual corrections to the $gg\to H$ sub-process with larger transverse momenta than the 
final ladder rungs. 


So far we have only discussed the Higgs production case, however the general features of the calculation are expected to remain the same for other central systems. The differential cross section for the production of a central system $X$, of invariant mass $\sqrt{\hat{s}}$, may therefore be written as
\begin{align}
	\frac{\partial \sigma}{\partial \hat{s} \partial y \partial \B{q}_{1T}'^2 \partial \B{q}_{2T}'^2}  &= \frac{\partial \mathcal{L}}{\partial \hat{s} \partial y \partial \B{q}_{1T}'^2 \partial \B{q}_{2T}'^2}  \d \hat{\sigma}(gg\to X) \;. \label{eq:generalCEPxsec}
\end{align}
The partonic cross section, $\hat{\sigma}$, is related to the matrix element for two on-shell gluons to produce the central system as
\begin{align}
	\d \hat{\sigma}(gg\to X) &= \frac{1}{2\hat{s}} \left|   \bar{\mathcal{M}}(gg\to X)  \right|^2 \d \textrm{PS}_X
\end{align}
where $\d \textrm{PS}_X$ is the phase-space of the final-state, $X$ and
\begin{align}
	\bar{\mathcal{M}}(gg\to X) &= \frac{1}{2} \frac{1}{N^2-1} \sum_{a_1a_2}\sum_{\lambda_1 \lambda_2} \delta_{a_1a_2}\delta_{\lambda_1\lambda_2} \mathcal{M}_{\lambda_1\lambda_2}^{a_1a_2}(gg\to X)
\end{align}
with $\mathcal{M}_{\lambda_1\lambda_2}^{a_1a_2}$ the amplitude for two on-shell gluons, with colours $a_i$ and helicities $\lambda_i$, to fuse to produce $X$. Finally, the effective luminosity is given by
\begin{align}
	\frac{\partial \mathcal{L}}{\partial \hat{s} \partial y \partial \B{q}_{1T}'^2 \partial \B{q}_{2T}'^2} &= 
	\frac{1}{\hat{s}} \left( \frac{\pi}{N^2-1} \int \! \frac{\d \B{Q}_T^2}{\B{Q}_T^4} f_g(x_1,x_1',\B{Q}_T,\B{q}'_{1T},
	\sqrt{\hat{s}}) f_g(x_2,x_2',\B{Q}_T,\B{q}'_{2T},\sqrt{\hat{s}})   \right)^2 \;.
\end{align}

Returning to Higgs production, before we can compute the cross section we need to
introduce the idea of rapidity gap survival. The Sudakov factor has allowed us
to ensure that the exclusive nature of the final state is not spoilt
by perturbative emission off the hard process. What about
non-perturbative particle production? The protons can in principle
interact independently of the perturbative process discussed above, and this interaction 
will lead to the production
of additional particles. We need to account for the probability that such emission
does not occur. Provided the hard process leading to the production of the
Higgs occurs on a short enough timescale, we may suppose that the physics which
generates extra particle production (including additional parton-parton interactions) factorizes, and that its 
effect can be accounted
for by an overall factor multiplying the cross section we just calculated. This
is the ``gap survival factor'', $S^2$, introduced in Ref.~\cite{bjorken} (see also~\cite{Dokshitzer:1991he}). It is defined by
\begin{equation}
 \d\sigma(p+H+p|\mathrm{no~soft~emission}) = \d\sigma(p+H+p) \times S^2 ~,
 \label{gaps}
 \end{equation}
where $\d\sigma(p+H+p)$ is the differential cross section computed above. The task is to
estimate $S^2$. Clearly this is not straightforward since we cannot utilize QCD
perturbation theory. However data on a variety
of processes observed at HERA, the Tevatron and the LHC can help improve our 
understanding of rapidity gap survival. We expect that $S^2 \sim 3\%$ for CEP
at the LHC (see for example \cite{Martin:2008nx} for an overview). Early measurements at the LHC
of rapidity gap processes will provide valuable information on $S^2$.  Note that the gap survival probability
has some process dependence: it will be higher for large impact
parameter, $r$, interactions. Thus we expect that
$S^2(\gamma\gamma) >  S^2(\gamma\pom) >
S^2(\pom \pom)$.

The issue of gap survival is frustrating to theorists. The
spacetime structure of the collision suggests that it may approximately factorize. Possible non-factorizing
contributions have been examined in Ref.~\cite{Bartels:2006ea}, although their
importance has been questioned in Refs.~\cite{Ryskin:2009tk,Khoze:2002py,Khoze:2006uj}. Assuming it to factorize,
we may use data to constrain it and accurate theoretical predictions
are possible. Here we shall present a simple model of gap survival which
should provide a good starting point for understanding the more
sophisticated treatments in
Refs.~\cite{Gotsman:2008tr,Ryskin:2009tk,Khoze:2006uj,Block,Strikman:2008er,Gotsman:2009bn}.

Dynamically, one
expects that the likelihood of extra particle production will be greater if the
incoming protons collide at small $r$. The simplest model which is capable of capturing this feature
is one which additionally assumes that there is a single soft particle production 
mechanism, let us call it a ``rescattering event'', and that rescattering events
are independent of each other for a collision between two protons at transverse
separation $r$. In such a model we can use Poisson statistics to model the
distribution in the number of rescattering events per proton-proton interaction:
\begin{equation}
P_n(r) = \frac{\chi(r)^n}{n!} \exp(-\chi(r))~. \label{eq:poisson}
\end{equation}
This is the probability of having $n$ rescattering events where
$\chi(r)$ is the mean number of such events for proton-proton collisions at
transverse separation $r$. Clearly the important dynamics resides in $\chi(r)$;
we expect it to fall monotonically as $r$ increases and to be
much smaller than unity for $r$ much greater than the QCD radius of the proton.
Let us for the moment assume we know $\chi(r)$, then we can determine $S^2$ via
\begin{equation}
S^2 = \frac{\int \d r ~ \d\sigma(r) ~ \exp(-\chi(r))}{\int \d r ~ \d\sigma(r)}
\end{equation}
where $\d\sigma(r)$ is the cross section for the hard process that produces the Higgs
expressed in terms of the transverse separation of the protons. Everything
except the $r$ dependence of $\d\sigma$ cancels when computing $S^2$ and so we
need focus only on the dependence of the hard process on the transverse momenta of
the scattered protons ($\B{q}_{iT}'$), these being Fourier conjugate to the transverse
position of the protons, i.e.
\begin{eqnarray}
\d\sigma(r) &\propto& [ (\int \d^2 \B{q}_{1T}' ~ e^{i \B{q}_{1T}' \cdot \B{r}/2} \; \exp(-b \B{q}_{1T}'^2/2))     
\times (\int \d^2 \B{q}_{2T}' ~ e^{-i \B{q}_{2T}' \cdot \B{r}/2} \; \exp(-b \B{q}_{2T}'^2/2) ) ]^2  \nonumber \\
&\propto& \exp\left( - \frac{r^2}{2 b} \right)~.
\end{eqnarray}
Notice that $b$ here is the same as that which enters into the denominator of
the expression for the total rate (Eq.~(\ref{eq:dsigdy})). There is thus a reduced sensitivity to $b$ 
when one includes the soft survival factor, due to the fact that as $b$ decreases so does $S^2$ (since the collisions are necessarily more central). Thus 
what matters is the ratio $S^2/b^2$.
   
It remains for us to determine the mean multiplicity $\chi(r)$. If there really is only
one type of rescattering event\footnote{Clearly this is not actually the case, but such
a ``single channel eikonal'' model has the benefit of being simple.} independent
of the hard scattering, then the inelastic scattering cross section can be written
\begin{equation}
\sigma_{\mathrm{inelastic}} = \int \d^2 \B{r} ( 1 - \exp(-\chi(r))),
\end{equation}
from which it follows that the elastic and total cross sections are
\begin{eqnarray}
\sigma_{\mathrm{elastic}} &=& \int \d^2 \B{r} ( 1 - \exp(-\chi(r)/2))^2, \\
\sigma_{\mathrm{total}}   &=& 2 \int \d^2 \B{r} ( 1 - \exp(-\chi(r)/2)).
\end{eqnarray}
There is an abundance of data which we can use to test this model and we can
proceed to perform a parametric fit to $\chi(r)$. This is essentially what is done in
the literature, sometimes going beyond a single channel approach. Suffice to say
that this approach works rather well. Moreover, it also underpins the models
of the underlying event currently implemented in the \textsc{pythia}~\cite{Pythia,Corke:2009tk} and \textsc{herwig}~\cite{Jimmy2, Jimmy1}
Monte Carlo event generators which have so far been quite successful in describing many of the
features of the underlying event, see Refs.~\cite{Borozan,Skands,Albrow:2006rt,Skands:2007zg,Gleisberg:2008ta,Bahr:2008dy,Bahr:2009ek}. 
Typically, models of gap survival predict
$S^2$ to be a few percent at the LHC. Although data
support the existing models of gap survival there is considerable room for improvement in testing
them further, thereby gaining greater control of what is perhaps the major
theoretical uncertainty in the computation of exclusive Higgs
production. Early diffractive studies at the LHC should give valuable information on $S^2$ (see for example~\cite{Khoze:2008cx}) and comparisons between different CEP channels at both the Tevatron and LHC will further constrain the theory (see Ref.~\cite{HarlandLang:2010ep}). 

\subsection{Other Models}  \label{sec:others}

Bialas and Landshoff (BL) developed~\cite{BL} a model inspired by a blend of QCD perturbation 
theory and Regge phenomenology to predict the rate of central Higgs production in $pp \to p+H+X+p$, i.e. 
central \emph{inclusive} Higgs production. Although their calculation looks ostensibly as if it is valid for 
exclusive production, BL were careful to emphasise that
``additional...interactions...will generate extra particles...Thus our calculation really is an inclusive one'', but with both
protons having $x_F > 0.95$.  
Inspired by this calculation, the Saclay group \cite{Saclay4} attempted to compute the rate for $pp \to p+H+p$ under 
the assumption that the exclusive rate can be obtained, modulo an overall gap survival factor, from the calculation 
in Ref.~\cite{BL}. We note in passing that the first attempt to compute
the central inclusive Higgs cross section, for proton-proton
collisions, appeared in \cite{Schafer:1990fz}. This was followed by a similar
analysis for nucleus-nucleus scattering in \cite{muller}, where the
first effort to estimate the exclusive Higgs cross section was
made. Other early studies of central inclusive and exclusive Higgs
production can be found in Refs.~\cite{Cudell:1995ki} and \cite{Lu:1994ys} respectively. 

We can understand the BL calculation starting from Eq.~(\ref{eq:qHq}). 
BL account for the coupling to the proton in a very simple
manner: They multiply the quark level amplitude by a factor of 9 (which corresponds to
assuming that there are three quarks in each proton able to scatter off each other). Exactly
like the Durham group they employ a form factor, $\exp(-b q_{iT}'^2)$, for each 
proton (at the cross section level), with $b = 4$~GeV$^{-2}$. Since BL are not 
interested in suppressing radiation, they do have a potential infrared problem since there
is no Sudakov factor. They dealt with this by following the earlier efforts of 
Landshoff and Nachtmann (LN) in replacing the perturbative gluon propagators with non-perturbative 
ones \cite{LN,DL}:
\begin{equation}
 \frac{g^2}{k^2} \to A \exp(-k^2/\mu^2). 
 \end{equation}
Rather arbitrarily, $g^2 = 4 \pi$ was assumed, except for the coupling of the gluons to
the top quark loop, where $\alpha_s = 0.1$ was used. 

Following LN, $\mu$ and $A$ are determined by assuming that the $p \bar{p}$ elastic
scattering cross section at high energy can be approximated by the exchange of two
of these non-perturbative gluons between the $3 \times 3$ constituent 
quarks. The imaginary part of this amplitude determines the total cross section for which
there are data to which they can fit. To carry out this procedure successfully,
one needs to recognize that a two-gluon exchange model is never going to yield the gentle
rise with increasing centre-of-mass energy characteristic of the total cross section. 
BL therefore also include an additional ``reggeization'' factor of $s^{\alpha_{\pom}(t)-1}$ in the
elastic scattering amplitude, where the pomeron trajectory is given by Eq.~(\ref{eq:donlan}).
In this way the two-gluon system models pomeron exchange.
They found that $\mu \approx 1$ GeV and $A \approx 30$~GeV$^{-2}$ gave a good fit to the data.
Similarly, the amplitude for central Higgs production picks up two reggeization factors.  

The inclusive production of a Higgs in association with two final state protons is clearly much more
infrared sensitive than the exclusive case where the Sudakov factor saves the day. Nevertheless, the Saclay model does not include the Sudakov suppression factor. Instead it relies
on the behaviour of the non-perturbative gluon propagators to render the $Q_T$ integral
finite. As a result, the typical $Q_T$ is much smaller than in the Durham case. 

Putting everything together, the Saclay model of the cross section for $pp \to p+H+p$ gives
\begin{eqnarray}
\frac{\partial\sigma}{\partial^2 \B{q}_{1T}' \partial^2 \B{q}_{2T}' \partial y} &\approx& S^2
\left(\frac{N_c^2-1}{N_c^2}\right)^2 \frac{\alpha_s^2}{(2 \pi)^5} \left(\frac{g^2}{4\pi}\right)^4
\frac{G_F}{\surd{2}} e^{-b q_{1T}'^2} e^{-b q_{2T}'^2} \nonumber \\ & & \hspace*{-2cm} \times ~
\xi_1^{2 - 2 \alpha_{\pom}(q_{1T}'^2)} \xi_2^{2 - 2 \alpha_{\pom}(q_{2T}'^2)}
\left[ 9 \int \frac{\d^2 \B{Q}_T}{2 \pi} \B{Q}_T^2 \; \left(\frac{A}{g^2}\right)^3 \exp(-3 \B{Q}_T^2/\mu^2) 
\frac{2}{3} \right]^2.\label{eq:pHp}
\end{eqnarray}
The only difference
between this and the original BL result is the factor of $S^2$. Integrating over the final
state transverse momenta and simplifying a little gives
\begin{eqnarray}
\frac{\partial\sigma}{\partial y}&\approx& S^2 \frac{\pi}{b+2 \alpha' \ln(1/\xi_1)} \frac{\pi}{b+2 \alpha' \ln(1/\xi_2)} \nonumber \\ && \times
\left(\frac{N_c^2-1}{N_c^2}\right)^2 \frac{G_F}{\surd{2}} \frac{\alpha_s^2}{(2 \pi)^5} \frac{1}{(4 \pi)^4} 
\left( \frac{s}{m_H^2} \right)^{2 \alpha_{\pom}(0)-2} \frac{1}{g^4}
\left[ \frac{A^3 \mu^4}{3}\right]^2.
\end{eqnarray}
A very similar formula was also presented by Bzdak \cite{Bzdak}, with the important difference that the Sudakov factor was reinstated, i.e. the integral over $\B{Q}_T$ in Eq.~(\ref{eq:pHp}) is replaced as follows
\begin{equation}
\int \d^2 \B{Q}_T \; \B{Q}_T^2 \; \exp(-3 \B{Q}_T^2/\mu^2)  \to
\int \d^2 \B{Q}_T \; \B{Q}_T^2 \; \exp(-3 \B{Q}_T^2/\mu^2)~T(\B{Q}_T,m_H)
\end{equation}
with $T(\B{Q}_T,m_H)$ given in Eq.~(\ref{eq:SudakovFactor}).

Not surprisingly, the Saclay and Bzdak approaches produce very different predictions: the Sudakov factor 
strongly suppresses the value of the predicted cross section. Broadly speaking, the Saclay prediction is 
similar to the Durham one for Higgs masses around 100~GeV$/c^2$ and (due to the absence of a Sudakov factor) overshoots 
it at higher masses. In contrast, the Bzdak calculation is typically much smaller than the Durham 
result. In both models, the choice of an exponentially falling gluon propagator means that 
there is little place for a perturbative component. However,
as the Durham calculation shows, there does not seem to be any good reason for neglecting
contributions from perturbatively large values of $Q_T$. In addition, neither calculation can lay claim to 
a systematic summation of the leading double and single logarithms, which is in contrast to the Durham calculation.
As we shall see in Section \ref{tevcep}, the CEP dijet data from
CDF~\cite{cdfjj} appear to exclude the Saclay model in favour of the
Durham one.

The perturbative Sudakov
factor is also included in the approach of Refs.~\cite{Petrov:2003yt,Petrov,Petrov:2007kn}.
This latter approach also uses perturbative gluons throughout the calculation
but Regge factors are included to determine the coupling of the gluons into the protons,
rather than the unintegrated partons of the Durham model. The results
are broadly consistent with those of the Durham model.

Finally, we mention the work of Szczurek and Lebiedowicz,
which takes the Durham model as the basis for (part of) their predictions for $pp
\to p+f_0(1500)+p$ \cite{Szczurek:2009yk} and $pp \to p+\chi_c+p$ \cite{Pasechnik:2007hm}.

In the following sections we shall survey existing results on CEP from
experiments performed up to the end of 2009. The reader most interested in the
LHC may skip Sections~\ref{isr}--\ref{spps}, but Section~\ref{tev} remains relevant.

%% file: cxpisr.tex
\section{CERN Intersecting Storage Rings (ISR)}
\label{isr}
 The ISR started in 1971, and provided $pp$ and $p\bar{p}$ (also $dd$
 and $\alpha\alpha$) collisions from $\sqrt{s}$ = 23 GeV to 63 GeV,
 greatly extending the energy range above the $\sqrt{s}$ = 7.6 GeV of the CERN PS and Brookhaven AGS. 
The higher energies allowed for a more direct study of pomeron
exchange. Ganguli and Roy gave an excellent review of Regge phenomenology as it stood after
     the first eight years of ISR operation~\cite{Ganguli:1980mp} . Before discussing the
     data, a note on nomenclature is in order. We will, especially in
     this section, follow the historical
terminology and use the term ``double pomeron exchange'' ($D \pom E$)
for inclusive studies of central particle production or
for exclusive particle production when we wish to emphasise its $\pom +\pom \to
X$ component. At the
low masses typical of the ISR, the central system will be dominated by
production of $\pi^+\pi^-$ making it in effect exclusive. We shall aim
to reserve ``CEP'' for strictly exclusive central systems.
It is now accepted that the pomeron is predominantly gluonic, which
means that $DI\!\!PE$ should be a good place to look for glueballs,
and at the ISR (and even at the lower $\sqrt{s}$ of the SPS) this became a strong motivation to study such processes.
This section provides some historical background, covering the ISR experiments that published results on 
$D I\!\!P E$ and CEP. 
     
    \subsection{Searches for double pomeron exchange}
\label{isrdpe}
At the start of the ISR era, light hadron spectroscopy was being
studied intensively and it was ``known'' that the
produced particles have low transverse momentum, $p_T$. Feynman had just invented the parton model and proposed that
Lorentz-invariant inclusive particle production cross sections should ``scale'', i.e. 
$E \d^3\sigma/\d^3p (p_T, x_F, s)$ should
become independent of $s$. Several single particle spectrometers
tested this and scaling was found to be
approximately true at small polar angles $\theta$, but was dramatically broken at $p_T \gsim$ 2 GeV$/c$ due to hard scattering of
partons, with the growing realisation that these are quarks and gluons, interacting according to the rules of QCD.

The ISR physics with which we are concerned in this review has two primary aspects: understanding the pomeron and
looking for glueballs. A full understanding of strong interaction physics must necessarily 
include an understanding of the pomeron/glueball sector. Yet, even after almost 40 years of QCD, the physics of these objects is still not well understood. 
The ISR enabled for the first time the study of pomerons beyond simple elastic scattering and low mass diffractive
excitation. One could study reactions such as $I\!\!P + p \rightarrow
X$ for $M_X >$ 10 GeV$/c^2$ and, our main concern
here, $I\!\!P + I\!\!P \rightarrow X$. At the ISR, the state $X$ could
be separated in rapidity from both outgoing coherently scattered
protons by large gaps $\Delta y \gsim 3$. 

In colliding proton beams the total rapidity coverage is given
by $\mathrm{ln}(s/m_p^2)$, which is 8.4 
at the top ISR energy of $\sqrt{s}$ = 63 GeV. 
It was found that single diffractive excitation, e.g. $p+p\rightarrow p+p^* \rightarrow p + (p\pi^+\pi^-)$ 
as studied at lower energies, now
extended well above the resonance ($N^*$) region. The inclusive forward proton spectrum has a 
distinct peak~\cite{Albrow:1976sv} for $x_F = 1-\xi > 0.95$,
 independent of $\sqrt{s}$ (Feynman scaling) and corresponding to $\pom$ exchange.
The diffractive mass calculated from $x_F = 1 - M_X^2/s$ extended up to
14 GeV/c$^2$, with this limit arising from the requirement of a rapidity gap adjacent to a leading proton 
of $\Delta y \geq 3$ units. For CEP, we similarly find that requiring
both protons to have $x_F > 0.95$ (or equivalently two gaps of $\Delta y
> 3$) one is restricted to $M_X \lsim 3$ GeV$/c^2$.

Earlier searches for $DI\!\!PE$ in fixed target experiments, such as that by the France-Soviet
Union (FSU) Collaboration at Serpukov~\cite{denegri}, were only able to put upper limits on the
cross section. FSU used a bubble chamber to study the exclusive reaction $p+p \rightarrow
p+\pi^+\pi^-+p$ at $p_{\mathrm{beam}}$ = 69 GeV$/c$, i.e. $\sqrt{s}$ =
11.5 GeV. All the events were found to have both pions close in
rapidity to one of the outgoing protons, consistent with single 
diffractive dissociation. A limit of $\sigma_{D\pom E} \lsim 20~\mu$b was determined
for events with $M_{\pi^+\pi^-} \lsim 0.7$ GeV$/c^2$ and
$|y_{\pi^+} + y_{\pi^-}| < 1.6$, a factor $\sim 20$ less than the dominant single
diffractive cross section $\sigma(p+p \rightarrow p +
(p\pi^+\pi^-))$. At this low value of $\sqrt{s}$ 
the beam rapidity is only $y_{\mathrm{beam}}$ = 2.5, too low for the
distinctive kinematics of $DI\!\!PE$. Another bubble chamber experiment, at Fermilab~\cite{derrick}, with the higher beam energy of
$p_{\mathrm{beam}}= 205$ GeV$/c$ ($\sqrt{s}$ = 19.7 GeV), placed an upper limit of 44 $\mu$b on the $DI\!\!PE$ cross section. 
It required the higher ISR energy for the $DI\!\!PE$ signal to emerge.

Chew and Chew~\cite{Chew:1974vu} made a prediction for the $DI\!\!PE$
signal, based on the single diffraction dissociation cross sections
for $A+B \rightarrow A+X$ ($\sigma^A$) and $A+B \rightarrow B+X$ ($\sigma^B$) and
an assumption of factorization, following Mueller~\cite{Mueller:1971ez}. They anticipated a differential $A+B\to A+X+B$
cross section given by
\begin{equation}
\frac{\d\sigma}{\d t_A \d t_B \d\xi_A \d\xi_B} =
\frac{1}{\sigma_T(AB)} \frac{\d\sigma^A}{\d t_A \d\xi_A}
\frac{\d\sigma^B}{\d t_B \d\xi_B}, \label{eq:chew}
\end{equation}
where $\xi_{A,B}$ are the fractional momentum losses of the colliding
particles and $\sigma_T(AB)$ is the total $AB$ cross section. Integrating over $t_{A,B}$ and over the $\xi_{A,B}$ range expected to be dominated by $DI\!\!PE$, they predicted
$\sigma_{DI\!\!PE}(\pi^+\pi^-) \approx 65 \; \mu$b at $\sqrt{s}$ = 63 GeV, mostly with
$M_X <$ 1 GeV$/c^2$.

\begin{figure}[htb]
\centering
\includegraphics[width=0.7\textwidth,angle=-90]{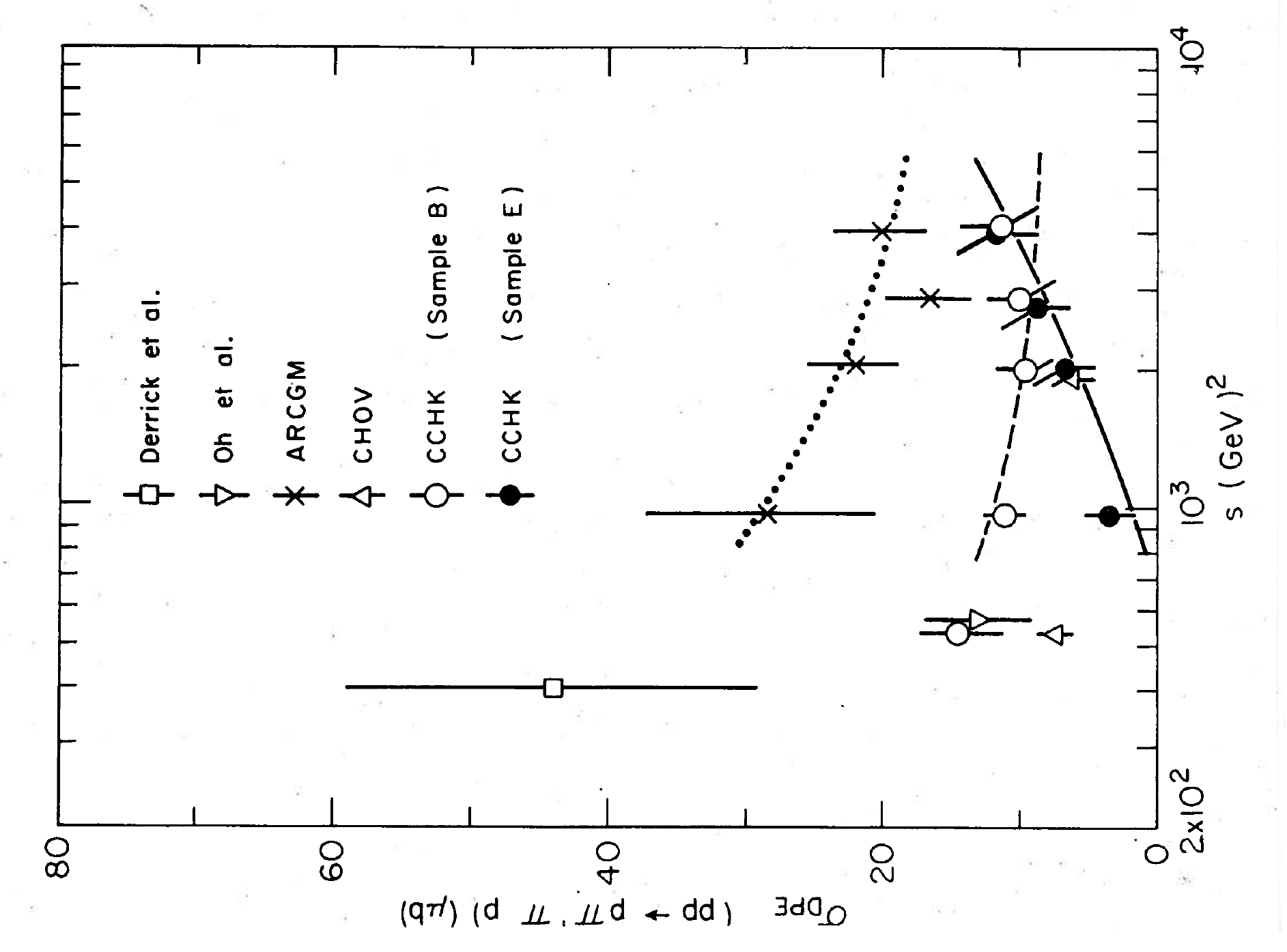}
\caption{Experimental $D I\!\!P E$ cross sections ($\mu$b) versus $s$ (GeV$^2$) at the ISR
together with the Regge calculations of Ref~\cite{Desai:1978rh}. The full circles and the rising solid line are
for two gaps with $\Delta y > 3$. The dashed line is for $|y_{\pi}| < 1.0$ and the dotted line
for $|y_{\pi}| < 1.5$. Figure from Ref.~\cite{Desai:1978rh}.
\label{fig:desai}}
\end{figure}

A more sophisticated calculation using Eq.~(\ref{eq:reggecep}), and
illustrated in Fig.~\ref{quint}, was performed by Desai \emph{et
  al.}~\cite{Desai:1978rh} (see also Ref.~\cite{Ganguli:1980mp}). To estimate $\sigma_{\pom\pom}$ they
assumed pomeron dominance, i.e. they took
\begin{equation}
\gamma_{A}^2(t_1)\sigma_{\pom \pom}(M_X^2,t_1,t_2)\gamma_{B}^2(t_1) = 
\frac{g_{\pom \pom \pom}(t_1) g_{\pom \pom \pom}(t_2)}{\sigma_{\pom}}
(M_X^2)^{\alpha_{\pom}(0) - 1} ,\end{equation}
where 
$$ \gamma_{A,B}^2(t) = \beta_{A,B}(t)^2 |\eta(t)|^2~, $$
$g_{\pom \pom \pom}(t)$ is the triple-pomeron coupling (e.g. see
Fig.~\ref{quint}(b)) and
$\sigma_{I\!\!P}$ is the pomeron contribution to the total cross section, i.e. 
\begin{equation}
	\sigma_T(AB) = \sigma_{I\!\!P} \left(\frac{s}{s_0}\right)^{\alpha_{I\!\!P}(0)-1} + \sigma_{I\!\!R} \left(\frac{s}{s_0}\right)^{\alpha_{I\!\!R}(0)-1} \;.
\end{equation} The values of $g_{\pom \pom \pom}(t)$ and the Regge
trajectory parameters were taken from the analysis of Field and
Fox~\cite{Field:1974fg} and secondary ($I\!\!R$) exchanges were
included as a correction (around 30--50\% at the lowest values
of $s$, falling to under 10\% at the higher values). 
They calculated cross sections both for two
gaps with $\Delta y > 3$ and for fixed ranges, $|y_{\pi}|<1.0$ or $|y_{\pi}|<1.5$, of the central pions, through the ISR energy range.
The predictions are shown together with the ISR data in
Fig.~\ref{fig:desai} and the agreement between the two is rather good. The predicted cross section for two gaps of fixed length rises with
$\sqrt{s}$, while for a fixed $|y_\pi| <
y_{max}$ cut the cross section falls with $\sqrt{s}$.
The predictions also indicate that the two protons should each have an exponential $t$-slope with $b_{D I\!\!P E} \sim
b_{\mathrm{elastic}}/2$ and they should be uncorrelated in azimuth and $t$. In the ISR energy range
$b_{\mathrm{elastic}} \sim$ 13 GeV$^{-2}$, with a small
$t$-dependence, rising like $b(s) = c + d \ln (s)$. Note that, under the
assumption that $\sigma_{\pom\pom}$ is independent of $M_X^2$, the
Desai \emph{et al.} approach is almost identical to that of Chew
and Chew, the only substantive difference being the treatment of
secondary exchanges. 

Desai \emph{et al.} also consider a pion exchange model for
$\sigma_{\pom\pom}$, and in this context they point out that
absorptive corrections may well be important. We refer to
Ref.~\cite{Desai:1978rh} and references therein for more detail. We shall
now turn to discuss the data, and note that Fig.~\ref{fig:desai}
contains the main results on the integrated cross section
measurements. Table~\ref{tab:isrex} summarises the ISR
$D\pom E$ cross section measurements.

The first experimental evidence for $DI\!\!PE$, at the ISR, was presented by the ARCGM Collaboration~\cite{baksay}  
which had forward proton tracking (but no momentum measurement) and full angular
coverage with scintillation counters. Candidate events had two leading non-collinear proton tracks with polar angle $6 < \theta <
10$ mrad, and exactly two hits in a scintillation counter hodoscope covering $|\eta| < 1.5$. Approximately 100 events were found at each ISR $\sqrt{s}$ value. Setting 
$t_{A,B} = - (p_{\mathrm{beam}} \theta_{A,B})^2$ they found no evidence
for any correlation between $t_A$ and $t_B$, and writing
$\d^2 \sigma/\d t_A \d t_B = a \; e^{-b(t_A+t_B)}$ they measured a slope $b = (9.9 \pm
1.8)$ GeV$^{-2}$, not significantly changing with energy. This is
somewhat higher than $ b_{\mathrm{elastic}}/2$ but is consistent at the 2$\sigma$ level. The cross section is
also approximately constant, from $\sigma_{D\pom E} = (28.4 \pm 8.1) \mu$b at
$\sqrt{s}$ = 31 GeV to $(20.2 \pm 3.3)\mu$b at $\sqrt{s}$ = 63 GeV. These are
expected characteristics of $DI\!\!PE$.

 The Split Field Magnet (SFM) facility at the ISR had nearly full angular coverage
   for charged particle tracks using proportional chambers, with dipole fields in the forward
   directions to analyse the momenta of the scattered protons. The central field was
   complicated (a quadrupole) and central particles were not identified but were assumed to be pions. A
   study by the CCHK Collaboration~\cite{dellanegra}
   selected events with two leading positive tracks ($x_F >0.9$) and forward rapidity
   gaps $\Delta y > 2$. 
   They observed the $DI\!\!PE$ characteristics of no correlations in $\phi$ and $t$, and measured $b = (5.5\pm0.9)$ GeV$^{-2}$.
    Extrapolating in $|t|$ they found $\sigma_{D\pom E} = (25\pm10)\mu$b, consistent with the subsequent Desai
   \emph{et al.}  calculation. 
   The mass distribution
   (assuming the central particles to be pions) had mostly $M_{\pi^+\pi^-} < 1$~GeV$/c^2$ with no significant $\rho\rightarrow \pi^+\pi^-$ signal, indicating $DI\!\!PE$ dominance, as $I\!\!P
   I\!\!P \rightarrow \rho$ is forbidden by isospin. 
   The authors point
   to the absence of an $f_0(980)$, but we will see later that it appears
   strongly in CEP as an edge, but not a peak. Au \emph{et al.}~\cite{Au:1986vs} later proposed that this is because 
   the $\pi^+\pi^- \rightarrow \pi^+\pi^-$ cross section is already as large as allowed by unitarity up to $M_{\pi\pi} \sim$ 1
   GeV$/c^2$.
   
   The CHOV Collaboration also studied $p+p \rightarrow p+\pi^+\pi^- + p$ at the SFM~\cite{dekerret}, selecting events with $x_F > 0.9$ for both protons and $|y_{\pi}| < 1.0$. After applying a
   four-constraint kinematic fit, the $\Delta\phi(pp)$ distribution showed two components,
   attributed to single diffraction and $D\pom E$. The inclusive proton spectra showed a diffractive peak for
   $x_F \gsim$ 0.95, while the region $0.90 < x_F < 0.95$ was dominated by non-diffraction (Reggeon $I\!\!R$ exchange). 
   They claimed, for $|y_{\pi}| < 1.0$,  $\sigma_{D\pom E} = (7.1\pm 1.0) \mu$b at 23 GeV and ($6.0\pm 1.5) \mu$b at 45
   GeV. Assuming the central particles to be pions, $M_{\pi^+\pi^-} < 1.5$ GeV$/c^2$.
   
   The CCHK Collaboration then followed the CHOV analysis with a more detailed
   study using five ISR energies and much higher statistics~\cite{drijard}.
   A sample was obtained with $x_F > 0.9$, predominantly $M_{\pi^+\pi^-} \lsim$
 1 GeV$/c^2$, with no observed $\rho$ resonance and with an indication of a dip around 1 GeV$/c^2$. 
 The helicity angle,
 defined as the angle in the $\pi^+\pi^-$
 centre-of-mass frame between the $\pi^+$ and the direction
 of the $\pi^+\pi^-$ system in the overall centre-of-mass frame, showed
 a flat distribution which suggests a dominantly $J=0$
 system. No correlation was seen in $\phi$ or $t$ between the protons, and the slope was measured to be $b =
   7.0\pm 0.5 (6.1\pm 0.8)$~GeV$^{-2}$ at $\sqrt{s} = 30.7(52.8)$~GeV. 
  Extrapolating in $|t|$, requiring both protons to have $x_F > 0.9$ and to
   have both rapidity gaps $\Delta y > 3$, the cross section
   $\sigma_{D\pom E}(\pi^+\pi^-)$ was found to rise from ($3.6\pm
   1.7)\mu$b to ($11.7 \pm 3.0)\mu$b
   as $\sqrt{s}$ increases from 30.7 to 62.3 GeV.
   
      The CHM Collaboration~\cite{armitage} studied events with two leading protons, measured with 
   single-arm magnetic spectrometers
   with $\sigma(p)/p \approx 0.6$\%, at $\theta \sim$ 30 mrad (and excluding elastic
   scattering by acollinearity). A scintillation counter hodoscope covered 98\% of the
   full solid angle and measured the directions of central charged
   particles, but without magnetic analysis. The mass and rapidity of the central hadronic system
   were determined entirely from the measured protons via
   \begin{equation}
   \frac{M^2_X}{s} \approx \xi_1 \xi_2; \;\; y = \frac{1}{2}\mathrm{ln}\left( \frac{\xi_1}{\xi_2} \right)~.
   \label{eq:massrap}
   \end{equation}
However, the study of
   centrally produced resonances was limited by the lack of tracking and the modest 
   mass resolution of $\sigma_{M_X}\approx 200$~MeV at $\sqrt{s}$ = 30 GeV.
  
 Table~\ref{tab:isrex} gives an at-a-glance summary of the above experiments; more details are given in the text and, of
   course, in the original publications.

   \begin{table*}
   \begin{center}
   \begin{tabular}{|c|c|c|c|c|c|c|c|}
   \hline
   Expt. & Collaboration. & Ref. & $\sqrt{s}$ & Forward $p$ & Central      &  Cuts: & $\sigma_{D\pom E}$ \\ 
            &                        &        &   (GeV)       & momenta&   momenta/ID   &  $x_p^{min}, y^{max}_{\pi}$     &  $(\mu$b) \\\hline \hline
R602/4  & ARCGM    & \cite{baksay}    &   31    &   &        & --, 1.5      &     28.4$\pm$8.1    \\
R602/4 & ARCGM    &   \cite{baksay}   &   45    &    &        &   --, 1.5    &     22.2$\pm$3.3   \\
R602/4  & ARCGM    & \cite{baksay}    &   53    &   &        & --, 1.5      &     16.8$\pm$3.1    \\
R602/4 & ARCGM    &   \cite{baksay}   &   62    &    &        &   --, 1.5    &     20.2$\pm$3.3   \\
 R407/8 & CCHK/SFM &\cite{dellanegra} &   31      &$\surd$ &  $\surd$  &   0.9, 1.5     &     25$\pm$10      \\   
 R401 & CHOV/SFM &\cite{dekerret}   &   23    &$\surd$ &  $\surd$  &  0.90, 1.0     &    7.1$\pm$1.0      \\ 
 R401 & CHOV/SFM &  \cite{dekerret} &   45  &$\surd$ & $\surd$  &0.90, 1.0& 6.0$\pm$1.5    \\ 
 R407/8 & CCHK/SFM &\cite{drijard}    &   23 &$\surd$ &  $\surd$  &$0.9,1.0$ & 14.4$\pm$3.1  \\ 
 R407/8 & CCHK/SFM &\cite{drijard}    &  31  &$\surd$ &  $\surd$  &$0.9,1.0 $& 11.0$\pm$1.2  \\ 
 R407/8 & CCHK/SFM &\cite{drijard}    &   45  &$\surd$ &  $\surd$  &$0.9,1.0$ & 9.4$\pm$1.9  \\ 
 R407/8 & CCHK/SFM &\cite{drijard}    &  53  &$\surd$ &  $\surd$  &$0.9,1.0$& 10.2$\pm$2.1  \\ 
 R407/8 & CCHK/SFM &\cite{drijard}    &   62  &$\surd$ &  $\surd$  &$0.9,1.0$ & 11.5$\pm$3.0  \\ 
 R407/8 & CCHK/SFM &\cite{drijard}    &  31 &$\surd$ &  $\surd$  &$0.9,1.0^d $& 3.6$\pm$1.7  \\ 
 R407/8 & CCHK/SFM &\cite{drijard}    &   45  &$\surd$ &  $\surd$  &$0.9,1.7^d$ & 6.8$\pm$2.7  \\ 
 R407/8 & CCHK/SFM &\cite{drijard}    &  53 &$\surd$ &  $\surd$  &$0.9,2.1^d $& 8.8$\pm$2.8  \\ 
 R407/8 & CCHK/SFM &\cite{drijard}    &   63  &$\surd$ &  $\surd$  &$0.9,2.4^d$ & 11.7$\pm$3.0  \\ 
R807 & AFS & \cite{Akesson:1983jz,Akesson:1985rn} &  63 &  & $\pi^+\pi^-$ & 0.95, 1.2 & 34$\pm$14$^{a,c}$\\
  R807 & AFS & \cite{Akesson:1983jz,Akesson:1985rn} &  63 &  & $\pi^+\pi^-$ & 0.95, 1.2 & 1.9$\pm$0.9$^{b,c}$\\
 R807 & AFS & \cite{Akesson:1983jz,Akesson:1985rn} &  63 &  & $K^+K^-$     &  0.95, 1.2        & 1.4$\pm$0.6$^c$\\
 R807 & AFS & \cite{Akesson:1983jz,Akesson:1985rn} &  63 & & $p\bar{p}$   &    0.95, 1.2         & 0.035$\pm$0.02$^c$\\
 R807 & AFS & \cite{Akesson:1983jz,Akesson:1985rn} &  63 &  &$\pi^+\pi^-\pi^+\pi^-$   & 0.95, 1.2  & 2.8$\pm$1.4$^c$ \\
 
 \hline
  \end{tabular}
  \end{center}
  \caption{ISR experiments reporting $D \pom E$ cross sections. See text and cited papers for details. 
  The forward protons are always tracked, but their momenta were not always
  measured. Only the AFS experiment identified the central particles,
  otherwise pions were assumed. $^a$For $M_{\pi\pi}
  <$ 1 GeV$/c^2$. $^b$For 1 GeV$/c^2 < M_{\pi\pi} <$ 2.3 GeV$/c^2$. $^c$Assuming slope $b$ = 6 GeV$^{-2}$. 
  $^d$For $\Delta y > 3$.}
  \label{tab:isrex}
  \end{table*}
  
\subsection{Searches for glueballs}
\label{gballisr}

Glueballs ($G$) are gluon bound states that are widely anticipated in
QCD. There is general agreement that the lightest glueball should be a
scalar with mass in the range $1-2$~GeV$/c^2$, with pseudoscalar and tensor
glueballs at higher mass. However, understanding the light scalars has
proven to be experimentally and theoretically very challenging,
and after thirty years of investigation the glueball sector is still
not well understood, due in no small part to the likely mixing of any
scalar glueball with $q\bar{q}$ (meson) bound states of the same quantum numbers. 
It is not our aim here to review the experimental status of glueballs;
excellent recent reviews can be found in
Refs.~\cite{Klempt:2007cp,crede}. Rather we will focus our attention
on data collected in $D\pom E$  that is of relevance to the search for
glueballs. It was Robson~\cite{robson} who, in 1977, first
suggested that $D\pom E$ should provide a gluon-dominated  channel that
favours central glueball production.

 Glueball searches in CEP are not only favoured by the gluon-dominated nature of $\pom$, but also
 the quantum numbers the central state must have: zero isospin, CP $= ++$ and even spin; it is thus a \emph{quantum number filter}. The lightest known states satisfying these constraints
are the very broad (and long controversial) $f_0(600)$ (or
$\sigma$) and the $f_0(980)$, which lies at the $K\bar{K}$
threshold. Note that scalar glueballs cannot lie on the pomeron
trajectory $\alpha_{\pom}(t)$ specified by Eq.~(\ref{eq:donlan}),
which instead anticipates a $J = 2$ tensor glueball with $M_G \approx 2.1$ GeV$/c^2$.
     
\begin{figure}[htb]
\centering
\includegraphics[width=0.7\textwidth,angle=-0.3]{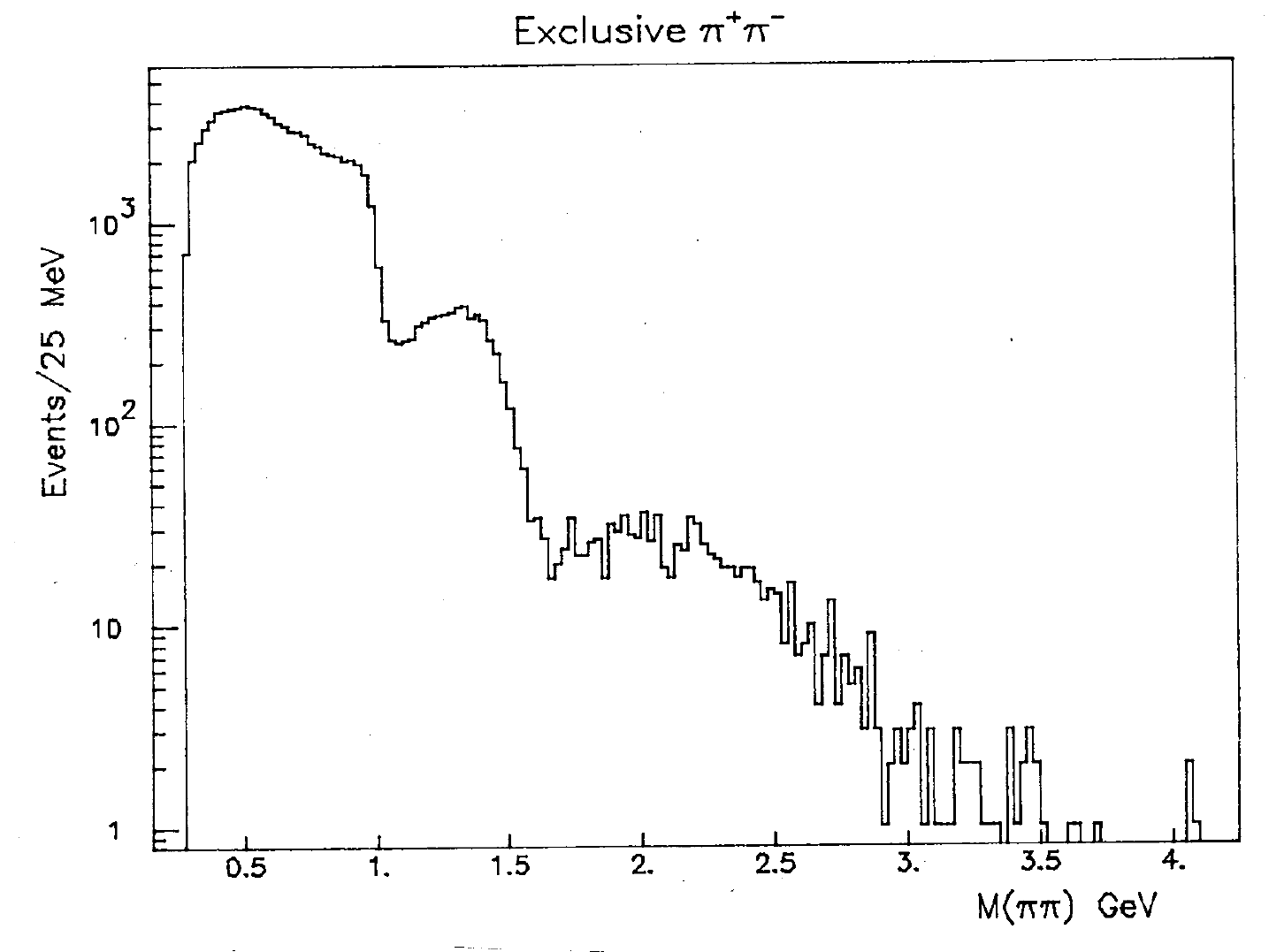}
\caption{$M_{\pi^+\pi^-}$ spectrum in $D I\!\! PE$ at the ISR (Axial
  Field Spectrometer, R807
  \cite{Akesson:1983jz,Akesson:1985rn}). Figure from Ref.~\cite{Akesson:1985rn}.
\label{cxafspipi}}
\end{figure}

  Following Robson's suggestion, Waldi, Schubert and Winter examined
  the $p+\pi^+\pi^- +p$ data, taken by the CHOV Collaboration,
 for resonant structure~\cite{waldi}. They observed a bump between 1100 and 1500
   MeV/c$^2$, which they incorrectly attributed to the $f_2(1270)$. Later the Axial Field Spectrometer (AFS) experiment
   (which we discuss in more detail in the next paragraph) showed that this bump, visible in Fig.~\ref{cxafspipi}, is predominantly S-wave
   and is broader than the $f_2(1270)$, with a small ($\lsim 10\%$)
   D-wave contribution from the $f_2(1270)$
   ~\cite{Akesson:1985rn}. Since the $f_2(1270)$ is also seen in
   $e^+e^- \rightarrow e^++\pi^+\pi^-+e^-$ (two-photon) production
   with a  cross section that favours a large {$q\bar{q}$}
   composition, it is not now considered to be a glue-dominated state.
 
   Compared to the CHOV analysis, the AFS Collaboration 
   benefitted from a factor 20 more
   statistics (89,000 exclusive events), better mass resolution, $\sigma_M$ = 10(25) MeV/c$^2$ at 1(2) GeV$/c^2$, 
   and particle identification~\cite{Akesson:1983jz,Akesson:1985rn}.
   Forward proton track detectors and
   rapidity gap detectors were added to the central spectrometer, 
   and CEP processes with $X = \pi^+\pi^-, K^+K^-, p\bar{p}$, and $\pi^+\pi^-\pi^+\pi^-$ were measured.
 The two-pion mass spectrum $M_{\pi^+\pi^-}$ shown in
   Fig.~\ref{cxafspipi} shows several interesting features.
   The cross section rises steeply from 
   threshold and there is 
   no sign of $\rho^0 \rightarrow \pi^+\pi^-$, which is forbidden in $D \pom E$. There is however an order-of-magnitude sharp drop just 
   below 1 GeV$/c^2$, followed by another bump-dip structure. The absence of a
   $\rho$, which is prominent in lower energy $p+p \rightarrow p+ \pi^+\pi^-
   +p$ experiments (e.g. at the SPS \cite{barberis}), is a sign
   that $I\!\!R$ exchange is unimportant in the AFS data. 
    Since photons can be exchanged across large rapidity gaps, they \emph{could}
   give rise to a centrally photoproduced $\rho$ via 
   $\gamma + I\!\!P \rightarrow \rho$. However the cross section is much lower than for $D I\!\!P E$, and protons with $|t| <$ 0.01
   GeV$^2$ were not detected, which further diminishes the rate. Despite the absence of a $\rho$ signal in the mass distribution, when the angular distributions were
   studied and a P-wave ($J$ = 1) component was selected, a small $\rho$ signal was seen, 
   demonstrating the sensitivity of such distributions\footnote{This method could in principle be applied to, e.g.,
   exclusive $p+p \rightarrow p+ \; W^+W^- + p$ (or $X = b\bar{b}$) at the LHC, if one had high enough statistics and good angular coverage. A scalar state (e.g. a Higgs) would show up
   only in the S-wave component.}.
   Conversely when the fixed-target experiment WA102~\cite{barberis} projected out the S-wave their
   $\rho$ peak disappeared and the $M_{\pi^+\pi^-}$ spectrum at $\sqrt{s}$ = 29 GeV had a similar 
   shape to the AFS spectrum. The partial wave 
  analysis of the AFS group showed that the ($J$=0) S-wave dominates
  as far up as 1700 MeV/c$^2$ ~\cite{Akesson:1985rn}.
  
  The striking drop at $M_{\pi^+\pi^-} \approx$ 1 GeV$/c^2$ in
   Fig.~\ref{cxafspipi} is the $f_0(980)$, the lowest mass
   narrow scalar state with full width $40 - 100$~MeV/c$^2$~\cite{pdg}. According to Au
   \emph{et al.}~\cite{Au:1986vs} the interaction can be understood as $I\!\!P + I\!\!P \rightarrow \pi\pi \rightarrow \pi^+\pi^-$ 
   with the final state
   interaction phase shift dominating the structure. Coupled channels such as intermediate 
   $K\bar{K}$ states contribute; similarly in the $K^+K^-$ final state intermediate $\pi\pi$ states
   $I\!\!P + I\!\!P \rightarrow \pi\pi/KK \rightarrow K^+K^-$ need to be taken into account.
    Au \emph{et al.} claimed that the coupled channel analysis could
    be interpreted as showing two narrow states 
   near 1 GeV$/c^2$: a $K\bar{K}$ ``molecule" and a glueball
   candidate. Narison~\cite{Narison:2002gv} suggested from an analysis
   of widths and couplings that both $f_0(600)$ and
   $f_0(980)$ have about 50\% $(gg)$ and 50\% $\frac{1}{\sqrt{2}} (\bar{u}u + \bar{d}d)$ in their wave functions.
   Mennesier, Minkowski, Narison and Ochs~\cite{Mennessier:2007wk} analysed $\pi^+\pi^-$ scattering
   below 700 MeV/c$^2$ in an analytic K-matrix model. This is the
   region of the broad $f_0(600)$. 
   They argue that the small direct coupling of this state to $\gamma\gamma$ is ``hidden" 
   by the $\pi\pi$, i.e. $f_0(600) \rightarrow \pi\pi
   \rightarrow \gamma\gamma$ and find that while
   $\Gamma(f_0(600)\rightarrow \gamma\gamma) \sim 3.9\pm 0.6$~keV, large
   enough to suggest a $q\bar{q}$ nature, after taking into account the rescattering, the direct
   width is only $\Gamma(f_0(600)\rightarrow \gamma\gamma) \sim
   0.13\pm0.05$~keV. This is considerably smaller than the width
   anticipated in the non-relativistic quark model for a
   $\frac{1}{\sqrt{2}} (u\bar{u}+d\bar{d})$ meson of similar
   mass. 
However, the interpretation of this state as a glueball is far from 
clear. For example, it is also commonly identified as the scalar partner 
of the pions under spontaneously broken chiral symmetry, especially in 
the context of NJL models \cite{Nambu:1961tp}. 
  
 Minkowski and Ochs in particular have interpreted the scalar $\pi\pi$ data from this and other 
 experiments in terms of a single very broad state
  extending from 400 MeV/c$^2$ to about 1700 MeV/c$^2$, with the $f_0(980)$
  and $f_0(1500)$ states superimposed, interfering destructively and
  therefore manifesting as dips in Fig.~\ref{cxafspipi}.  
  They find that in this energy range, after the $f_0(980)$
  and $f_0(1500)$ are subtracted, the $\pi\pi$ elastic amplitude describes a full loop in the 
  Argand diagram. They take this broad object, which they call the
  ``red dragon'', to be the lightest scalar glueball~\cite{Minkowski:1998mf}.
  This is a remarkable claim: The full $\pi^+\pi^-$ spectrum of
  Fig.~\ref{cxafspipi} from threshold to $\sim 1700$~MeV/c$^2$ is one
  very broad scalar glueball. 
  However any state with a width $\Gamma$ exceeding a few hundred MeV decays within $\sim\hbar c/\Gamma <$ 1 fm, and will not
  propagate beyond the interaction region as a free particle. The
  situation remains unclear however: Spanier, Tornqvist and Amsler reviewed
  the scalar mesons in the PDG~\cite{pdg} and concluded that ``The I = 0, 
  $J^{PC} = 0^{++}$ sector is the most complex one, both theoretically
  and experimentally.'' There is however something of a consensus that the scalar meson sector does contain a
  scalar glueball degree of freedom; the challenge is to understand
  its mixing with the quark states \cite{Close:2003tv}.

   The AFS experiment also measured central $K^+K^-$
   production~\cite{Akesson:1983jz,Akesson:1985rn}. The cross section rises very steeply from
   threshold,
   probably due to $f_0(980)$, but with no sign of exclusive $\phi \rightarrow K^+K^-$, which is 
   forbidden in $D I\!\!P E$.  The cross section ratio $\sigma(K^+K^-)/\sigma(\pi^+\pi^-)$ even exceeds
   1.0 above threshold, perhaps an indication of a gluonic component. Exclusive $p+(p\bar{p})+p$ with 
   mass $M_{p\bar{p}}$ from
   threshold to 2.8 GeV$/c^2$ was also seen (64 events), but with no evident structures. An extrapolation in $|t|$ gives an
   estimate of the total $\sigma(p+p \rightarrow p+p \bar{p}+p) \approx$ 35 nb (see Table~\ref{tab:isrex}). 
   We are not aware of a prediction for exclusive baryon pairs. One could expect a
   similar cross section, but scaled by $M^{-2}$, for other baryon pairs, e.g. $\Lambda\bar{\Lambda}$, as 
   the pomeron is flavour-blind. $D\pom E$ production of hyperon pairs was not studied at the ISR, but it would be
   possible at the Tevatron and the LHC. 

    The ISR also provided $\alpha\alpha$ collisions with $\sqrt{s}$ = 126 GeV. The forward drift chambers 
   installed for the CEP measurements could distinguish charge $Q=1$ and $Q=2$
   particles. The exclusive process
   $\alpha+\alpha \rightarrow \alpha + \pi^+\pi^- + \alpha$ was seen~\cite{Akesson:1985rn}, with the $M_{\pi^+\pi^-}$ spectrum having the same shape
   as in $pp$ collisions within limited statistics (395 events). The
   $\alpha$-particles must have been coherently scattered, making this
   is a very clean channel for $D I\!\!P E$, but the statistics were
   too low for serious spectroscopy.   
  
 The ISR was terminated in 1983, after the $Sp\bar{p}S$ collider was
  turned on. From a CEP perspective, what had we learnt?  Well, the
  existence of double
  pomeron exchange had been established and was in agreement with
  Regge predictions, but it had been a long
  struggle. 
  While CEP with large rapidity gaps in hadron-hadron collisions can
  come from $\gamma\gamma, \gamma I\!\!P,$ or $I\!\!P   I\!\!P$ interactions, only the latter had been observed. There was no evidence for the odderon, the C $= -1$ partner to the
  $I\!\!P$, except for a small difference in $pp$ and $p\bar{p}$ elastic scattering~\cite{isrppbar}
  at $t = - 1.2$ GeV$^2$ (the dip region). The reach of
  CEP was up to $M_X \approx 3$ GeV$/c^2$, which was favorable for hadron spectroscopy and especially 
  for glueball searches. The best detector for CEP, the AFS, was designed for high-$p_T$ physics 
  and added a CEP programme only in the final $2-3$ years of operation; the 
  analysis was completed after the ISR was turned off. The spectra for $ X = \pi^+\pi^-, K^+K^-, p\bar{p}$ and
  $\pi^+\pi^-\pi^+\pi^-$ showed structures (especially $\pi^+\pi^-$)
  but no unequivocal evidence for glueballs. The low mass  scalar
  sector is still unsettled, and
  the whole $M_{\pi\pi}$ region from $400-1700$~MeV/c$^2$ may even be a very broad scalar 
  glueball ``cut'' by the $f_0(980)$ and $f_0(1500)$.

%% file: cxpft.tex
\section{Fixed Target experiments}
\label{fte}
   After the ISR was closed in 1983, CEP of low
   multiplicity hadron states, for both hadron spectroscopy and for
   studies of production
   mechanisms, became an active field in fixed target experiments in
   the 1980s and 1990s. We shall restrict ourselves to a few brief
   comments since our focus is on high energy colliding beam experiments. 

Thanks to high 
   luminosities, dedicated  experiments in multiparticle spectrometers, such as the Omega facility at the CERN
   SPS experiments (WA76, WA91 and WA102) and experiment E690 at the
   Fermilab Tevatron, were highly productive. In particular, the
   experiments using the Omega spectrometer \cite{Barberis:1996tu,kirk} confirmed
   the earlier observation by the SFM collaboration (at the ISR)
   \cite{Breakstone:1990at} of a
   correlation between the properties of the central system and the
   directions of the outgoing protons.  Close \& Kirk
suggested that glueballs and $q\bar{q}$ mesons may have different
  dependencies on the difference in the transverse momenta of the
  outgoing protons~\cite{Close:1997pj,Close:2000dx} but this idea
  remains unproven. 
  Moreover, $\sqrt{s}$ 
   was only 29 GeV and so $I\!\!R I\!\!R$ and
  $I\!\!R I\!\!P$ exchanges were certainly not negligible. 
  Ref.~\cite{kirk} contains many relevant references to CEP at the
  Omega facility, with a large number of results that are
  of interest for meson spectroscopy.
  
  Experiment NA22 at the European Hybrid Spectrometer (EHS) also extracted $DI\!\!PE$ cross sections in $\pi^+p$ and $K^+p$
  collisions, although with $p_{\mathrm{beam}} = 250$ GeV$/c$ ($\sqrt{s}$ = 21.7 GeV) a large non-$DI\!\!PE$ background
  had to be subtracted~\cite{Agababyan:1993fb}. They found an inclusive (but mostly $\pi^+\pi^-$) $DI\!\!PE$ cross section of ($39\pm5\pm8)
  \mu$b for $K^+p$ and ($24\pm6\pm3)\mu$b for $\pi^+p$. Such a
  difference is qualitatively expected in Regge theory since $\sigma_T(K^+p) > \sigma_T(\pi^+p)$.
 
  At $p_{\mathrm{beam}}$ = 800 GeV$/c$, Fermilab Experiment E690
  studied the $X = K_S^0 K_S^0$ system, showing in particular
  a large $f_0(1500)$ peak~\cite{PhysRevLett.81.4079}.

%% file: cxspps.tex
\section{CERN Proton-antiproton Collider $\boldsymbol{(Sp\bar{p}S)}$}
\label{spps}
In 1981 the first $p\bar{p}$ collider, the $Sp\bar{p}S$, came into operation at
CERN. It provided an increase in $\sqrt{s}$ from 63 GeV at the ISR up to 630 GeV,
and even 900 GeV for some low luminosity ``ramping runs''. This
step-up in energy enabled the
discoveries of the $W$ and $Z$ bosons, and observations of dramatic 
high $E_T$ jets. Diffractive physics took a back seat during this
period, although experiment UA8, by adding forward Roman pots on 
both beam pipes to the
central UA2 experiment, observed diffractively produced
dijets~\cite{ua8jj} with $E_T$ up to $\sim$ 13 GeV. Ingelman and
Schlein had earlier proposed to describe the pomeron in hard processes as
having quark and gluon constituents, analogous to a hadron~\cite{ingel}. 
In this picture, the parton densities in the pomeron
can be measured in high mass diffraction. The jet kinematics gives the scattering parton
momenta and thus rather direct information on the parton distributions in the
$I\!\!P$. This approach led directly to the subsequent very precise measurement of
diffractive parton densities at the HERA collider (see for example Refs.~\cite{Wolf:2009jm,Newman:2009bq}). 

Experiment UA8 measured $p/\bar{p}$ tracks after (low-$\beta$) quadrupole magnets, and they studied
$p+X+\bar{p}$ with $X$ = ``anything'', using the UA2
central detector~\cite{ua8pxp}. They selected 107 events in the
$DI\!\!P E$-dominated region with $x_F >$ 0.95 (i.e. gaps $\Delta\eta
\gsim 3$ on both sides). Most of the events had $M_X < 8$~GeV$/c^2$ but no 
specific exclusive states $X$ could be resolved, as central particle momenta 
were not measured (UA2 had no magnetic field) and
the missing mass resolution was much worse than the mass derived using
the calorimeter alone (which was $\sigma_M\sim$ 1 GeV$/c^2$).
UA8 extracted the $D\pom E$ cross section,
$\sigma_{I\!\!P I\!\!P}$, assuming the validity of
Eq.~(\ref{eq:reggecep}) with a flux (see Eq.~(\ref{eq:cep-regge}))
given by
\begin{equation}
	f_{I\!\!P/p}(\xi,t) = \frac{K}{(1-t/a)^4}e^{bt} \left( \frac{1}{\xi}\right)^{2\alpha_{I\!\!P}(t)-1}
\end{equation}
and with parameters extracted from data on elastic scattering and single diffraction:
\begin{align}
	a &= 0.71~\textrm{GeV}^{2} \nonumber\\
	b &= 1.08~\textrm{GeV}^{-2} \nonumber \\
	K &= 0.74~\textrm{GeV}^{-2} \nonumber \\
	\alpha_{I\!\!P}(t) &= 1.035 +0.165 t+ 0.059 t^2 \;.
\end{align} 
UA8 inferred that
$\sigma_{I\!\!P I\!\!P} \approx 3$~mb for $2 \lsim M_X \lsim
8$~GeV$/c^2$, decreasing to $0.5 - 1.0$~mb 
for $10 \lsim M_X \lsim 25$~GeV$/c^2$. This is significantly larger than
expected assuming the simple factorization formula presented in Eq.~(\ref{eq:chew})
at the same total energy.
The UA8 paper also included a prediction for
${\d\sigma_{DI\!\!PE}}/{\d M_X}$ at the Tevatron and LHC assuming a
constant $\sigma_{I\!\!P I\!\!P} = 1$ mb. They predicted about  0.4 $\mu$b/GeV$/c^2$ at 
$M_X$ = 100 (200) GeV$/c^2$ at these colliders respectively. 

The largest experiment at the $Sp\bar{p}S$, UA1, was designed to
discover the $W$ and $Z$, but it also performed a CEP study at $\sqrt{s} = 630$~GeV
using forward rapidity gaps~\cite{ua1gxg}. They
required no energy in forward calorimeters with $3 < |\eta| < 6$ and some energy
in the central calorimeters ($5^\circ < \theta < 175^\circ$). A localized energy
deposition of $E_T > 1.5$~GeV and/or a ``jet'' with $E_T > 3$~GeV was
required to trigger, which excluded the possibility of
detecting low mass exclusive states. The data were compared to a non-diffractive sample, containing at least one
charged particle with $1.5 < |\eta| < 5.5$ in each forward direction. The mass $M_X$ of the central
hadrons had a mean $\langle M_X \rangle \sim$ 36 GeV$/c^2$ and extended
a little above 63 GeV$/c^2$ (above which at least one scattered proton must
have $x_F < 0.90$). 
At low $M_X$, the mean charged particle
multiplicity $\langle n_{ch} \rangle$ was measured to be similar to that in $pp$ or $e^+e^-$ collisions at
$\sqrt{s} \equiv M_X$, but it was found to rise faster and, by $M_X$ = 70 GeV$/c^2$, it was a factor of two higher. 
This effect has not been quantitatively explained, and hopefully it will be checked at the Tevatron or LHC.
Recall, however, that the final state in $e^+e^-$ collisions producing
dijets is $q\bar{q}$ and in $\pom\pom$ collisions
it is predominantly $gg$, so one might expect
$\langle n_{ch} \rangle_{\pom \pom} >\langle n_{ch} \rangle_{e^+e^-}$
as a result of the difference between quark and gluons jets, together with pomeron ``remnants''.
With the UA1 jet algorithm it was found that 5\% of the events have at
least one jet with $E_T > 10$~GeV, but this fraction would probably be
lower if the trigger had not required an energy deposition of $E_T >$
1.5 GeV. This class of central dijet events
is expected be a dominant background to searches such as $p+H+p \rightarrow
p+JJ+p$ at the LHC.

%% file: cxptev.tex
\section{Tevatron}
\label{tev}
\subsection{Introduction}
\label{tevdiffr}
Until the LHC began operation in December 2009, the Tevatron at Fermilab was the highest
energy hadron collider, making proton-antiproton collisions at
$\sqrt{s} = 1.96$~TeV. In Run I (1992-1996) it provided collisions at
$\sqrt{s} = 546$~GeV, 630 GeV (to equal the CERN $Sp\bar{p}S$
collider) and 1800 GeV.  The total delivered luminosity per experiment (CDF and D\O) was 110
pb$^{-1}$, compared to about 7000 pb$^{-1}$ so far in Run II (2004-2009).

In the first (Run I) phase of diffractive studies, installation of Roman pots by CDF, with silicon and drift
chamber track detectors along both beam pipes, allowed measurements of
elastically and diffractively scattered protons and antiprotons.
Elastic scattering was measured at $\sqrt{s} = 546$ (1800) GeV
for $-0.29 < t < -0.025$ GeV$^2$, where it was well described
by an exponential slope with $b = 15.28 \pm 0.58$
$(16.98\pm0.25)$~GeV$^{-2}$, and the integrated elastic cross sections were
$\sigma_{\mathrm{el}} = 12.9\pm 0.3$~mb and $19.7\pm 0.9$~mb, at the two energies~\cite{PhysRevD.50.5518}. By measuring the rates of
elastic scattering and inelastic collisions simultaneously, CDF
exploited the Optical Theorem to measure the total
cross section, $\sigma_T$, independent of the
luminosity~\cite{cdftot}. They found $\sigma_T = 61.26\pm0.93$
$(80.03\pm 2.24)$ mb, with
$\sigma_{\mathrm{el}}/\sigma_{\mathrm{tot}}$ rising from $0.210\pm
0.002$ to $0.246 \pm 0.004$ over this
$\sqrt{s}$ range. Two other experiments, E710~\cite{Amos:1990fw} and E811~\cite{Avila:2002bp} 
found significantly ($\sim 2.5\sigma$) lower values
for both $\sigma_{\mathrm{el}}$ and $\sigma_{\mathrm{tot}}$ (but with the same ratio) at $\sqrt{s} = 1800$~GeV. These results, together with LHC expectations, are discussed in
\cite{Khoze:2000wk} and \cite{Stenzel:2009cn}. CDF also measured the inelastic proton
spectra for $x_F \gsim 0.88$~\cite{cdfsd,Goulianos:1998wt}, where they observed
the expected dominance of the diffractive peak above $x_F \sim 0.95$.
This corresponds to diffractive masses $M_X = 400$ GeV$/c^2$; indeed
later Tevatron studies succeeded in observing $W, Z$ and jets with $E_T$ 
up to 80 GeV in single diffraction. Unfortunately the forward proton
detectors were removed without having measured CEP with both protons detected.

The second stage of diffractive physics at the Tevatron took place in
Run II, initially without any forward
$p/\bar{p}$ measurements. CDF and D\O\; both studied
hard single diffraction with forward rapidity gaps, after which CDF
installed a new set of Roman pot detectors 50 m downstream to measure
diffractive antiprotons. There were no detectors for forward
protons, so $p + X + \bar{p}$ could not be studied with both the $p$
and $\bar{p}$ tagged, however sets of scintillation counters with
$\theta < 3^\circ$ were added along the beam pipes. These ``beam
shower counters'' (BSC) acted as rapidity gap detectors, using showers
produced in the beam pipe by secondaries with $5.4 < |\eta| <
7.4$. When combined with ``miniplug calorimeters'' covering $3.6 <
|\eta| < 5.2$, gaps of $\Delta \eta > 3.8$ could be required on each
side, with or without a measured antiproton. Later D\O\; also installed
Roman pots with tracking on both sides, but has not yet presented any
CEP results. 

The single diffractive production of
dijets~\cite{Abbott:1999km,Affolder:2001zn},
$W$~\cite{Abe:1997jp,Abazov:2003ti} and $Z$~\cite{Goulianos:2009uj}
(preliminary) has been observed and the fraction of $W,Z$ or dijets
that are classed as diffractive, i.e.
either with a high-$x_F$ antiproton or a large forward gap, is around
1\%. This is to be contrasted with $ep$
collisions at HERA, where H1 and ZEUS found that
the corresponding fraction of diffractive to non-diffractive events
is $\sim 10\%$. This is in accord with the rough expectation that the
gap survival probability $S^2$ is lower by about an order of magnitude in hadron-hadron
collisions than in $\gamma^* p$ collisions.

CDF also studied the central production of dijet events
containing both a leading $\bar{p}$ with $0.035 < \xi_{\bar{p}} < 0.095$ and
a rapidity gap in the proton direction, corresponding to $0.01 <
\xi_{p} < 0.03$~\cite{PhysRevLett.85.4215}. Although the proton
is not observed, its $\xi$ can be computed using
\begin{equation}
\xi_p = \frac{1}{\sqrt{s}} \sum_i E_T^i e^{\eta^i},
\end{equation}
where the sum is over all the observed particles in the detector.
This constitutes an observation of $D\pom E$ dijet production, and a
comparison to the
theoretical predictions has been carried out using pomeron parton
densities extracted from the HERA data in Refs.~\cite{Appleby:2001xk,Cox:2005gr}. Again
a gap survival factor of around 10\% was needed to fit the data. Importantly,
it was noted that the rate of $D\pom E$ (two-gap) dijet events is
approximately 10\% of the rate of single diffractive (one-gap) dijet events , i.e.  
the price to pay for gap survival should not be applied twice.
This was also observed in a CDF study~\cite{cdfjgj} of \emph{inclusive} $D\pom E$, with a measured
antiproton with $0.035 < \xi_{\bar{p}} < 0.095$ and a rapidity gap on the proton side. 

\subsection{Central exclusive production}
\label{tevcep}

The study of CEP at the Tevatron is a recent activity, largely motivated by the 
possibility of studying exclusive $p+p \rightarrow p+H+p$ and related processes at the LHC. 
In 2001 there was a very large spread in predictions for this exclusive
cross section, some of which 
even suggested that a Higgs search may be possible at the Tevatron,
and a Letter of Intent to perform such a search by measuring both
final state protons
was submitted to the Fermilab Program Advisory Committee~\cite{Albrow:2001fw}. It did not proceed to the proposal stage however,
partly because of the large cross section uncertainty.
Although the exclusive cross section is much smaller than the inclusive one, if the protons are
measured precisely it was noted that the ``missing mass'' method can be
used to give a very good measurement of the central state, independent
of its nature~\cite{Albrow:2000na}\footnote{Ref.~\cite{Albrow:2000na}
  also pointed out that a measurement of the difference in the proton
  arrival times at the forward detectors could be used to greatly
  reduce the impact of pile-up background.}.
Although it is now understood that the Higgs is not likely to be
produced this way at the Tevatron, 
the measurement of other central systems, in particular those with the
same quantum numbers as the Higgs (such as $\chi_{c0}$ and
$\chi_{b0}$) constrains the predictions for $\sigma(p+H+p)$. The
diagrams for $\chi_{c0}$ and $\chi_{b0}$ production are similar to
those for exclusive Higgs production, but with $c$- and $b$-quark loops
instead of a top-quark loop. Of course the relevant scale is much lower for these
lighter states and this renders the applicability of perturbative QCD more uncertain.
The cleanest related process is $p + \gamma\gamma + \bar{p}$ (through
$\pom\pom \rightarrow \gamma\gamma$) but the cross section is small
for $E_T(\gamma) > 5$ GeV, where measurements were made. 
Exclusive dijet production has a much higher cross section but one
must then compromise on not having such a clean final state.

\begin{figure}[h]
\centering
\includegraphics[width=0.65\textwidth]{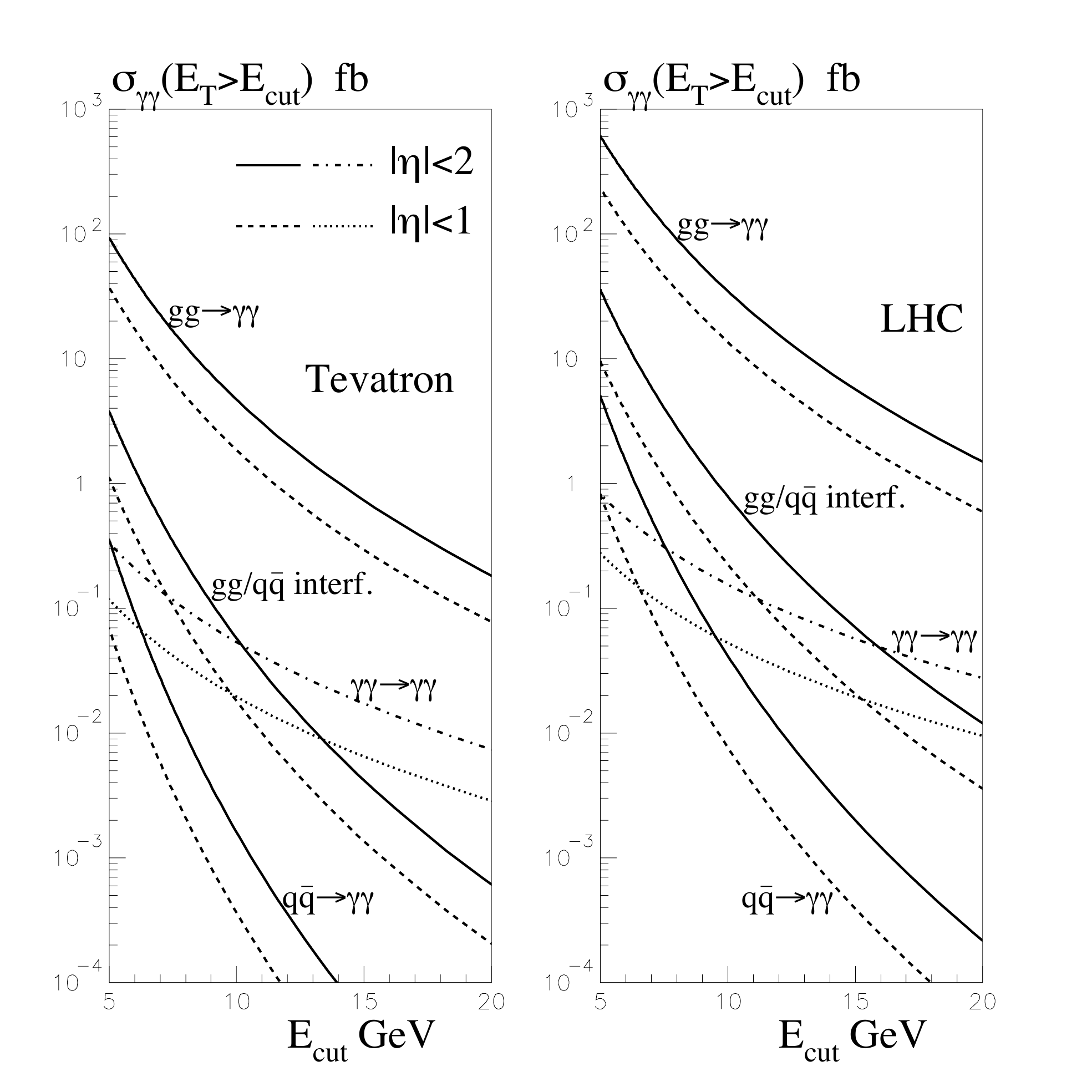}
\caption{Predictions for CEP di-photon production at the Tevatron and
  the LHC. Figure from Ref.~\cite{kmrsgg}.
\label{fig:kmrsgg} }
\end{figure}

Nevertheless, CDF has succeeded in measuring exclusive $\gamma\gamma$, $\chi_c$ and dijet
production, and we discuss each of these  in turn. We also
discuss exclusive $\gamma\gamma \rightarrow e^+e^-, \mu^+\mu^-$ and $\gamma\pom \rightarrow J/\psi,
\psi'(2S)$, which was recently seen for the first time in hadron-hadron
collisions. Apart from the dijet study, the forward $\bar{p}$ could not be detected, as the Roman pots (only on the $\bar{p}$ side) did not have acceptance for low
$M_X$. The analyses instead relied on finding events that contain just
the state $X$ in conjunction with an otherwise empty (i.e. consistent
with noise levels) detector. The miniplug calorimeters and BSC
counters, which have coverage out to $|\eta| =
7.4$, were crucial for this exclusivity requirement (the BSC was used as a veto in the trigger).

Firstly we shall discuss the exclusive $\gamma\gamma$ search~\cite{Aaltonen:2007na}, which was
combined with a CEP $e^+e^-$ search~\cite{Abulencia:2006nb}, as the trigger (and most of the analysis)
is identical. Only in the final step was the central tracking used to separate 16 $e^+e^-$ 
events from three with 
electromagnetic showers with $E_T >$ 5 GeV and no tracks. In all cases the showers had $\Delta \phi \sim \pi$ and $\sum \vec{E_T}$
small, and the $e^+e^-$ events agreed with the precise QED
expectation, providing a good control for the $\gamma\gamma$ candidates.
The gap survival probability is not an issue for the QED events; the impact parameter is large and 
$S^2 \sim 1$.
Also, the balance in $E_T$ and $\Delta\phi$ should make it possible to find QED events (especially $\mu^+\mu^-$)
in the presence of pile-up; this is now being studied in CDF.
Background, e.g. from $\pi^0\pi^0$ in the $\gamma\gamma$ candidate sample, could not be 
quantitively assessed \emph{a priori},
but two of the events had narrow single showers on each side and were
very unlikely to be background. The prediction using the Durham
model~\cite{kmrsgg} is shown in Fig.~\ref{fig:kmrsgg}. The
prediction of 36$^{\times 3}_{\div 3}$ fb for $E_T(\gamma) >$ 5 GeV and $|\eta(\gamma)| < 1$
would give 0.8$^{+1.6}_{-0.5}$ events, and the CDF data are in good
agreement with this. The two events correspond to $\sim 10^{-12}
\times \sigma_{\mathrm{inel}}$, showing that it is possible to find even very rare exclusive events. More CDF data has been taken with a lower threshold $E_T(\gamma) > 2.5$ GeV, and there
are plans to search for exclusive $\gamma\gamma$ events at the LHC. 

\begin{figure}[h]
\centering
\includegraphics[height=0.21\textwidth,angle=0]{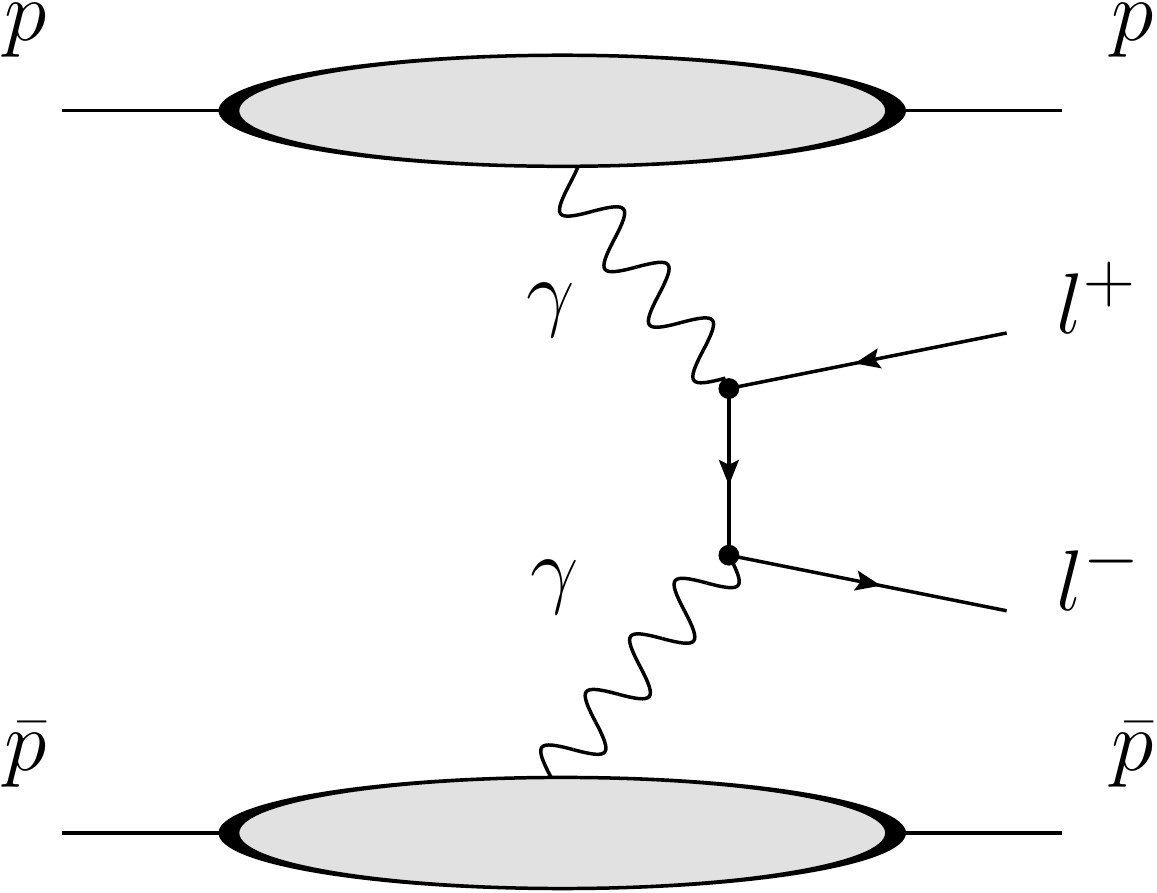} \hspace{0.02\textwidth}
\includegraphics[height=0.21\textwidth,angle=0]{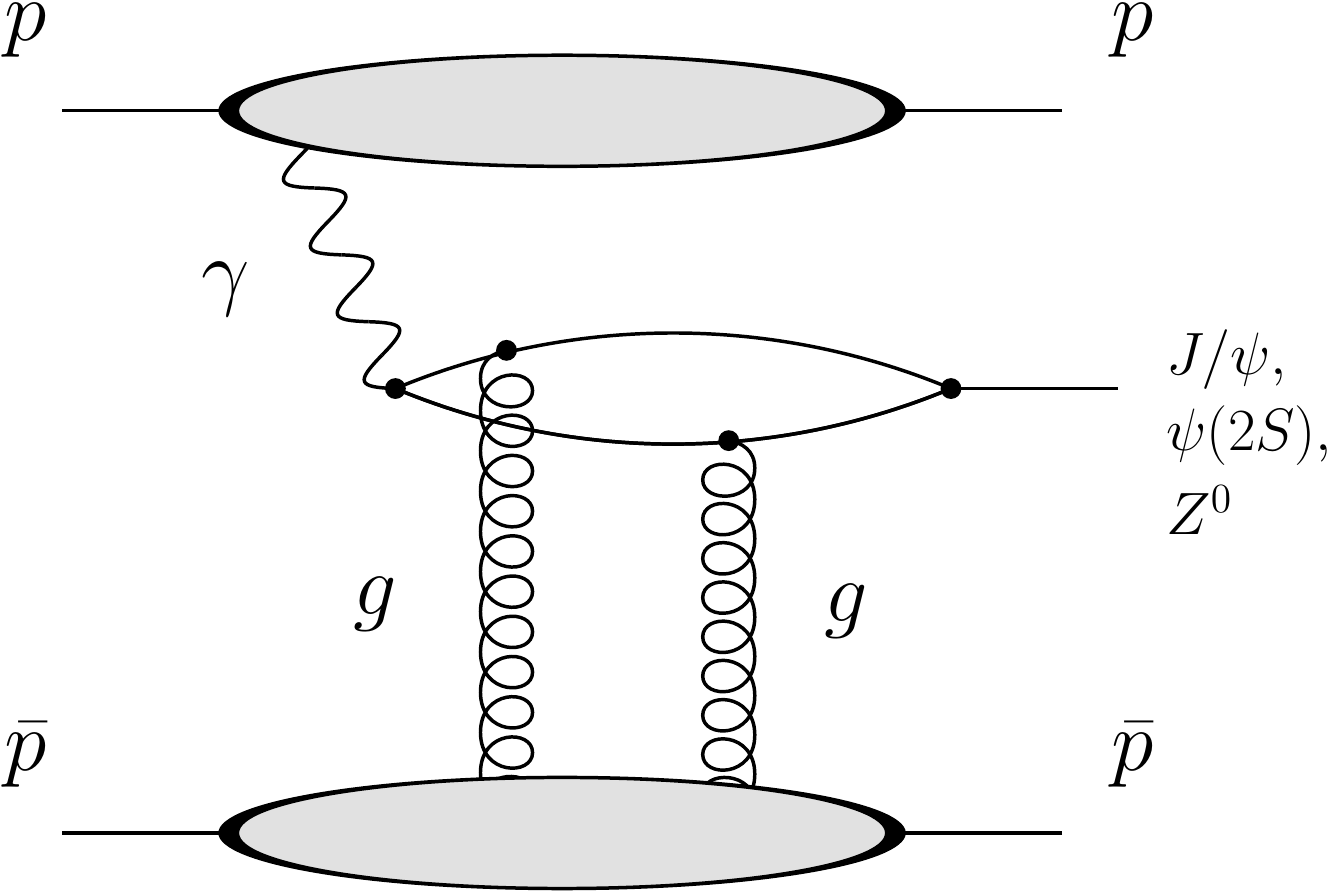}\hspace{0.02\textwidth}
\includegraphics[height=0.21\textwidth,angle=0]{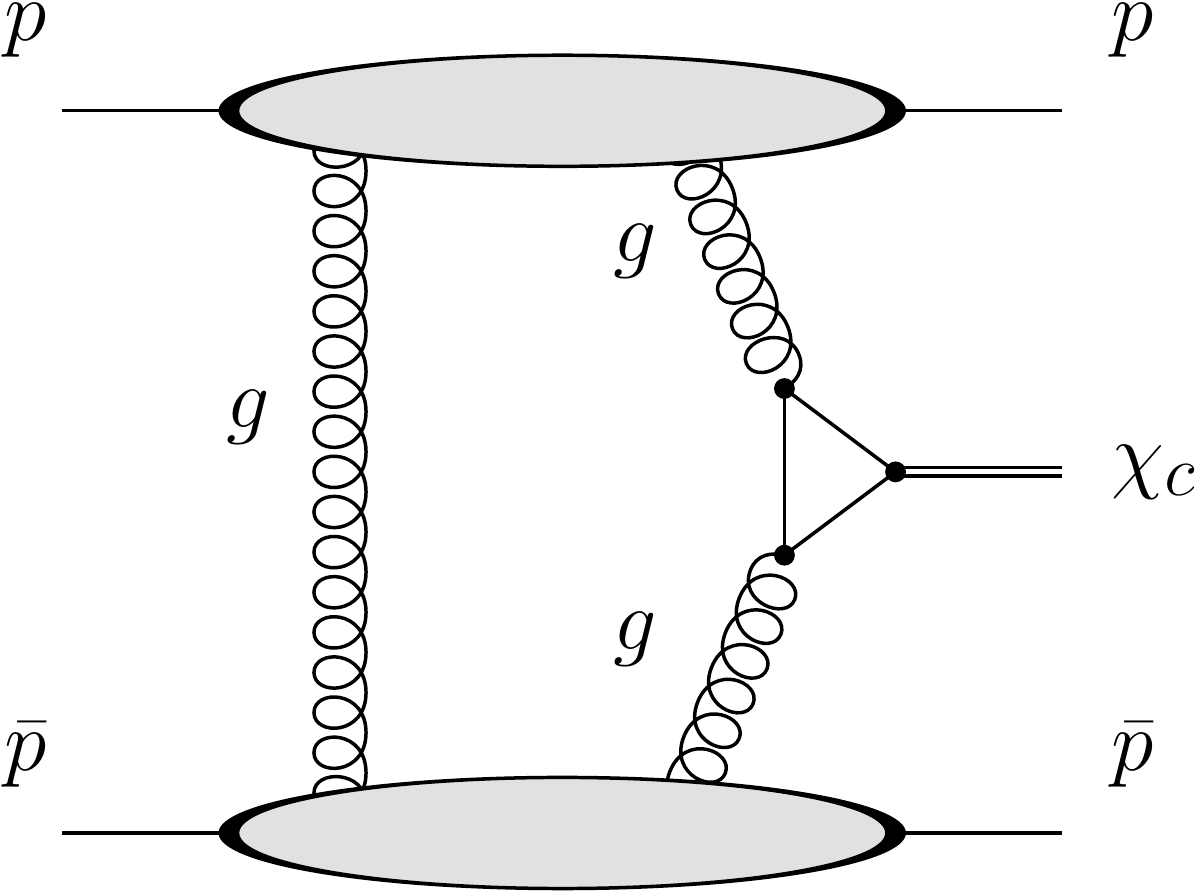}
\caption{Feynman diagrams for processes contributing to the exclusive di-lepton signal. (a) $\gamma\gamma\rightarrow l^+l^-$, (b) $\gamma\pom\rightarrow 
    J/\psi, \psi(2S), Z^0$, and
    (c) $I\!\!P I\!\!P \rightarrow\chi_{c0}$.
\label{fig:exclmmdiag} }
\end{figure}

In addition to the exclusive $\gamma\gamma$ search, CDF also studied the production of lepton pairs ($e^+e^-,\mu^+\mu^-$), either in association with no other particles or with one additional photon. Such exclusive leptons may be produced through several mechanisms, as shown in Fig.~\ref{fig:exclmmdiag}. We begin by discussing $\mu^+\mu^-$ production at low $M(\mu^+\mu^-)$, for which CDF used an exclusive di-muon trigger in the mass range $M(\mu^+\mu^-)\in [3.0,4.0]$~GeV$/c^2$ and $|\eta_\mu|<0.6$. The mass range was limited to $M(\mu^+\mu^-)>3$~GeV$/c^2$, as below $p_T$ = 1.5 GeV$/c$ muons range out in the
calorimeters.

\begin{figure}[h]
\centering
\includegraphics[width=0.70\textwidth]{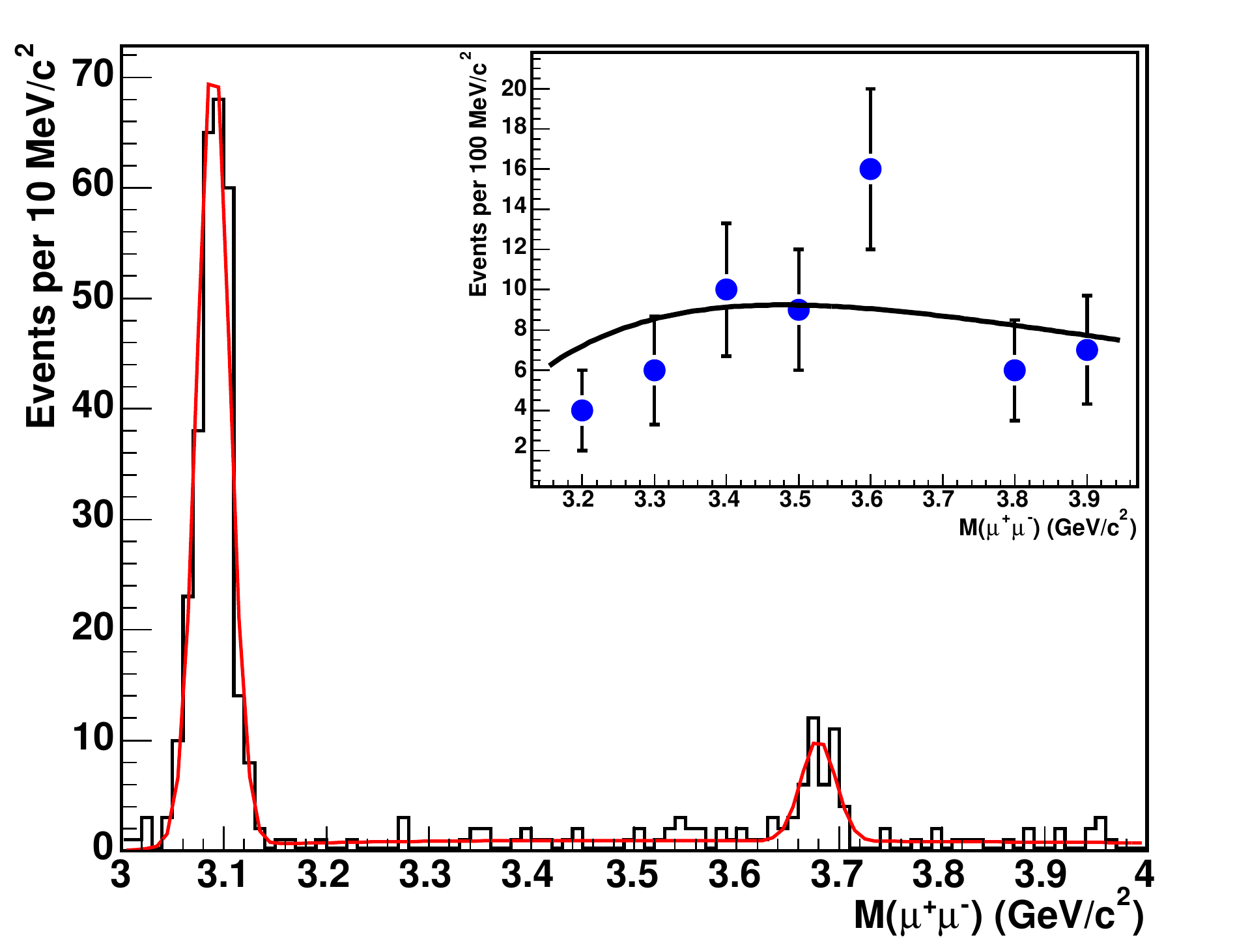}
   \caption{Di-muon mass distribution of exclusive $\mu^+\mu^-$ events in CDF, with no EM shower, (histogram) 
   together with a fit to two Gaussians for $J/\psi$
    and $\psi(2S)$, and a QED continuum. Inset: Data above the
    $J/\psi$ and excluding $3.65 < M(\mu^+\mu^-) < 3.75$ GeV$/c^2$ ($\psi(2S)$) with a fit to the QED spectrum. Figure from Ref.~\cite{Aaltonen:2009kg}.
\label{fig:mass} }
\end{figure}

Photoproduction of a vector meson, shown in
Fig.~\ref{fig:exclmmdiag}(b), is one source of exclusive muon pairs in
both $ep$ and $pp(\bar{p})$ collisions. The predictions for the
$p\bar{p}$ process are closely related to those for the $ep$ process,
bearing in mind the differing soft survival factors,
$S^2(p\bar{p})<S^2(ep)$. CDF recently observed the production of
$J/\psi$ and $\psi(2S)$, with subsequent decay to $\mu^+\mu^-$, via
this mechanism, the first such observation of vector meson
photoproduction in hadron-hadron
collisions~\cite{Aaltonen:2009kg}. Simply requiring rapidity gaps in
conjunction with two central muons results in an extremely clean spectrum, and any non-exclusive background (mostly undetected proton dissociation) is at most a few percent, see Fig.~\ref{fig:mass}. The cross sections agree with predictions~\cite{starlight,motykaz,schafer:2007mm}. Although these vector mesons cannot be produced in either $\gamma\gamma$- or
$\pom \pom$-collisions, they could be produced in odderon-pomeron interactions. 
While the odderon is required in QCD~\cite{ewerz}, the couplings are
expected to be suppressed compared to pomeron
couplings and there is at present no direct evidence for it.  

Another source of exclusive muon pairs, this time in association with a photon, is central exclusive $\chi_c$ production: $p+\bar{p} \to p+\chi_c+\bar{p}$, with subsequent decay: $\chi_c \to J/\psi + \gamma \to \mu^+\mu^- + \gamma$, see Fig.~\ref{fig:exclmmdiag}(c). This is a process much more closely related, at least theoretically, to CEP of a Higgs boson. Measurement of this at the Tevatron provides,
together with exclusive dijets which we shall discuss shortly, a
good test of the theory. CEP of $\gamma\gamma$ and $\chi_b$ are both
superior theoretically, but unfortunately their cross sections are
smaller. CDF has observed $p + (\chi_c\rightarrow
J/\psi+\gamma) + \bar{p}$  events, with $\d\sigma/\d y
|_{y=0}(\chi_c) = 76\pm 10 \pm 10$~nb, assuming the $\chi_{c0}$ dominates~\cite{Aaltonen:2009kg} (CDF could not resolve different $\chi_c$
states). The theoretical predictions of the $\chi_{c0}$
cross section are in agreement with the measured value. Ref.~\cite{kmrcharm} predicted
$\d\sigma/\d y|_{y=0}(\chi_{c0}) = 130$~nb, however since that paper 
the PDG value of the $\chi_{c0}$ width was lowered by 45\%~\cite{pdg}, correcting
their prediction to 90 nb and correcting Yuan's prediction~\cite{yuan} of 160 nb to 110 nb. Bzdak's prediction is 45
nb~\cite{bzdak1}. It should be noted however, that the assumed dominance of the $\chi_{c0}$ has been questioned. In
Ref.~\cite{jz0} it was suggested that higher spin $\chi_{c}$ states may contribute to the cross section, with their smaller 
production rates being compensated by larger branching ratios. Indeed, a recent re-analysis has been performed
by the Durham group and collaborators, including the effects of non-factorising, ``enhanced", contributions to the
gap survival
probability~\cite{HarlandLang:2009qe,HarlandLang:2010ep}. Their
studies suggest that the $\chi_{c0}$ production cross section is
somewhat lower than previous determinations, but that the reduction to
the total rate is compensated by the production of $\chi_{c1}$ and
$\chi_{c2}$ states. They find for the total production of all three
states, $\d \sigma/\d y|_{y=0} (\chi_c)=65~\textrm{nb}$ and the
calculation has been implemented in the Monte Carlo SuperCHIC~\cite{SuperCHIC}. The agreement
between theory and experiment is encouraging and indicates that the
predictions for CEP Higgs production at the LHC are unlikely to be far
off the mark.

Exclusive muon~\cite{Aaltonen:2009kg} (and
electron~\cite{Abulencia:2006nb}) pairs are also produced via the QED
$\gamma\gamma$ fusion process, depicted in
Fig.~\ref{fig:exclmmdiag}(a). This process generates a continuum in
the $M(\mu^+\mu^-)$ distribution over the which the $J/\psi$ and
$\psi(2S)$ peaks are visible, as shown in Fig.~\ref{fig:mass}. The
continuum agrees in both shape and magnitude with the QED
predictions. Also, since the exclusive $p+ \: l^+l^- +p$ cross section is very well known (only Coulomb elastic scattering is as well known), it
is a good candidate for measuring the absolute luminosity of the LHC,
and thus for calibrating high-rate luminosity monitors. The limit to the precision ($\sim 1\%-2$\%)  with which this can be done 
will come from the knowledge of acceptances and efficiencies
(favouring $\mu^+\mu^-$ over $e^+e^-$). One must do this in the
presence of some pile-up, as the probability of vetoing on pile-up
will depend on the unknown inelastic cross section. Another difficulty is that the protons may
dissociate, e.g.  $p \rightarrow p^* \rightarrow N\pi\pi$. The CDF
BSCs, situated along both beam pipes,
were efficient at detecting such dissociations, which create forward showers in the pipes and surrounding
material. CMS and ATLAS do not yet have such
counters, but there is a proposal to add them~\cite{cxfsc}. 

At high mass, lepton pairs may be produced in exclusive $Z$ photoproduction, $\gamma \pom \to Z$. A nearly real photon is radiated by one of the incoming beam particles, fluctuates into a $q\bar{q}$ pair, which then scatters by $\pom$ exchange on the other proton, see Fig.~\ref{fig:exclmmdiag}(b). CDF performed a search for this reaction~\cite{Aaltonen:2009cj}, however, the cross section is predicted to be only 
0.21 fb~\cite{goncalves} or 0.3 fb~\cite{motykaz} at the Tevatron, far below the CDF sensitivity. In contrast, as we shall discuss in section~\ref{sec:GP}, the cross section at the LHC is expected to be about 6~fb for $|y(Z)|<2$ and the $Z$ decaying to $e^+e^-$ or $\mu^+\mu^-$~\cite{motykaz}. 
It would be very interesting to measure
exclusive $Z$ production, as BSM theories with additional particles in the $\gamma I\!\!P \rightarrow Z$ vertex loop 
change the cross section. For example, in White's theory of the
pomeron there should be a ``large'' (but not quantified) increase, due to colour-sextet quark loops~\cite{white}.
Exclusive $\gamma\gamma \rightarrow l^+l^-$ pairs have been seen
with pair masses up to 75 GeV/c$^2$ and no observed $Z$
candidates~\cite{Aaltonen:2009cj,Albrow:2009nj}. 

\begin{figure}[htb]
\centering
\includegraphics[width=0.65\textwidth,angle=0]{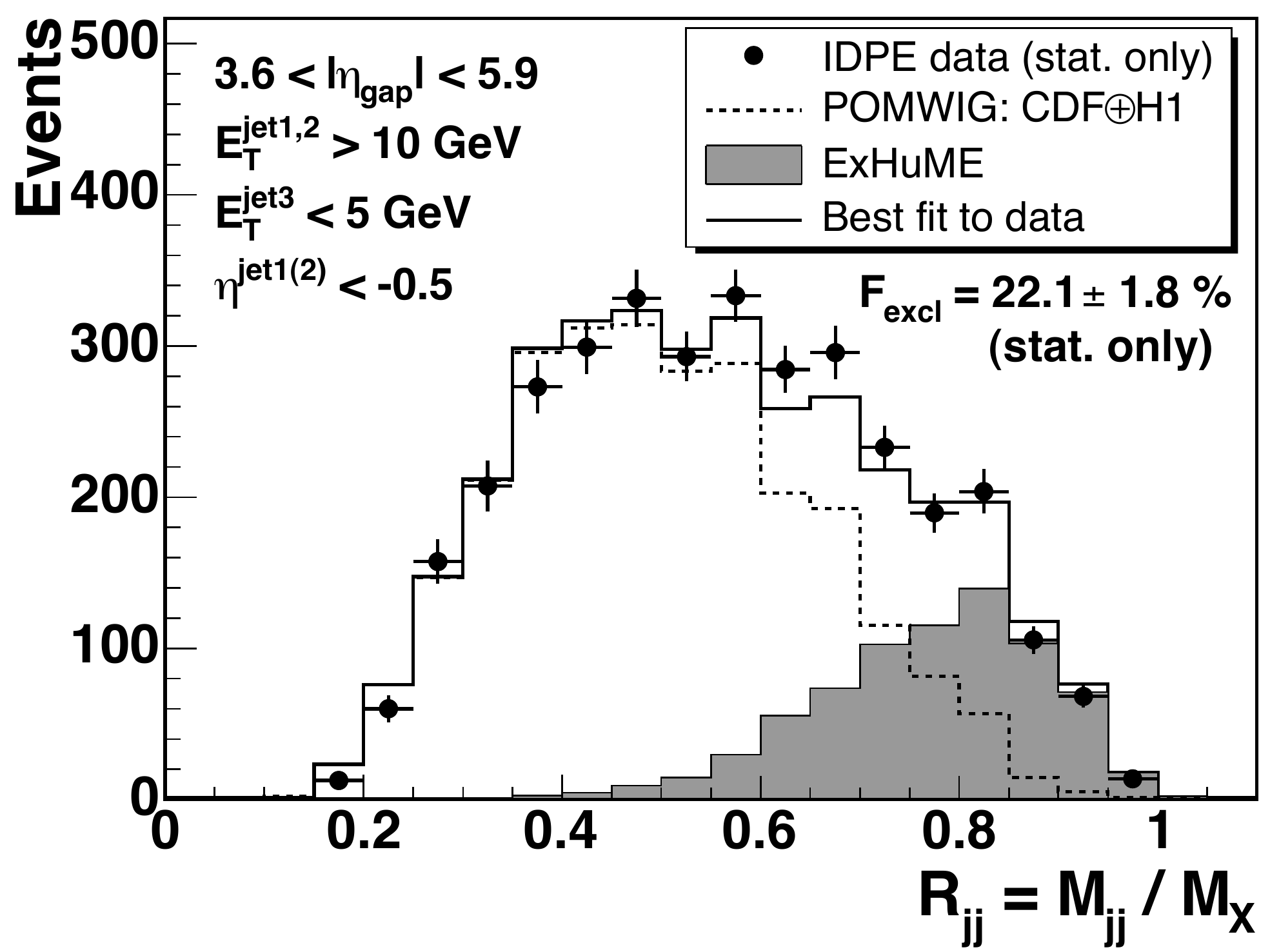}
   \caption{Dijet mass fraction for $D\pom E$ data (points) and best fit to the sum of two components:
   {\sc{pomwig}} $D\pom E$ + SD and ND background, and
   {\sc{E\textsc{x}H\textsc{u}ME}} (exclusive dijets). Figure from Ref.~\cite{cdfjj}. 
\label{fig:cdfrjj} }
\end{figure}

\begin{figure}[htb]
\centering
\includegraphics[width=0.65\textwidth,angle=0]{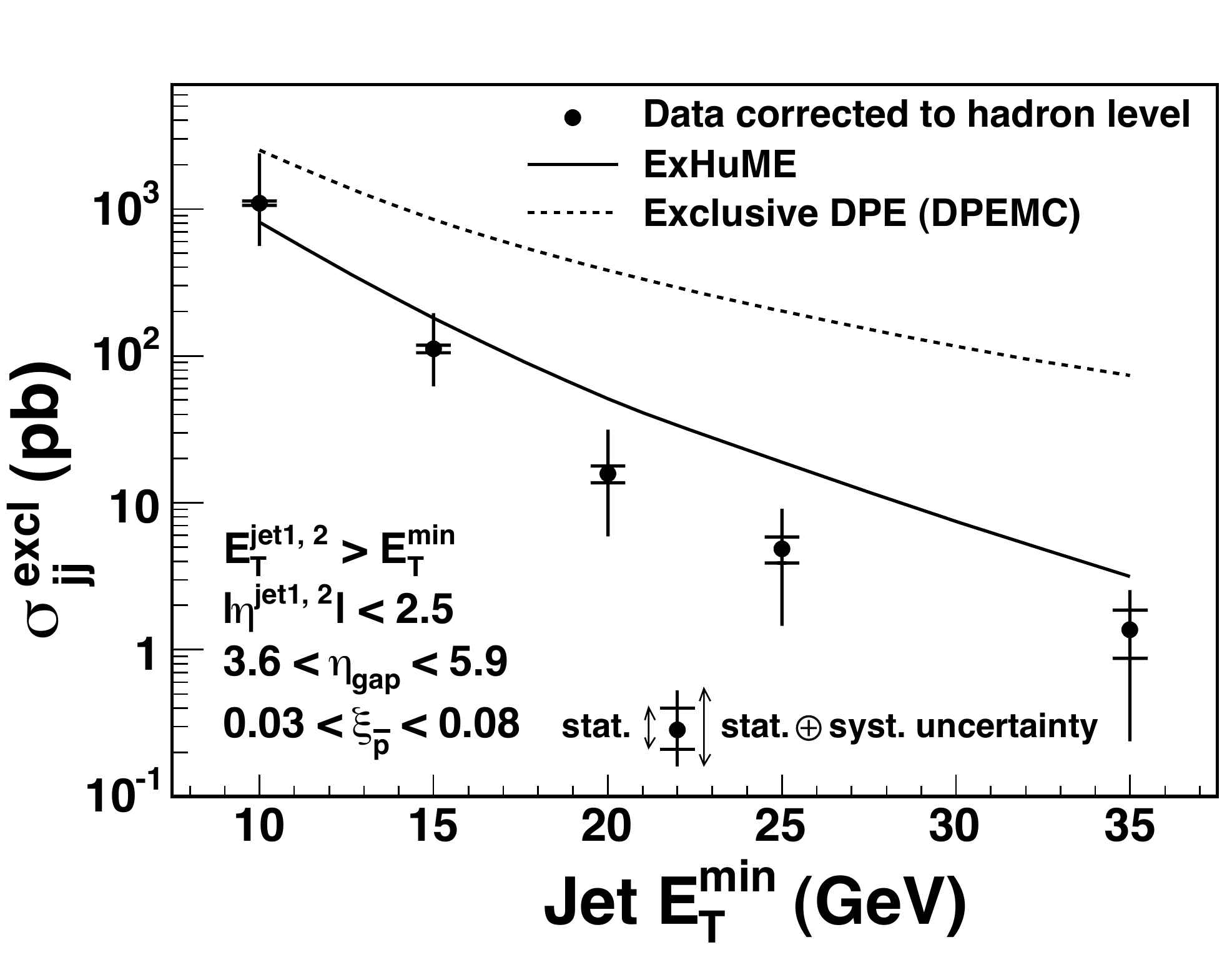}
   \caption{Exclusive dijet cross section \emph{vs.} the minimum $E_T$ of the two jets, compared to the
   \textsc{dpemc} and E\textsc{x}H\textsc{u}ME predictions. Figure from Ref.~\cite{cdfjj}.
\label{fig:cdfjjet} }
\end{figure}

Unlike exclusive lepton production, exclusive dijet production, $X = jj$, has the advantage of a
relatively high cross section but, 
unlike $X=\gamma\gamma$ or $\chi_c$,
it is subject to the usual uncertainties associated with defining jets. CDF measured it from a 
sample of $D \pom E$ dijet events, with a detected $\bar{p}$
together with a large gap on the $p$ side~\cite{cdfjj}. They
reconstructed the mass of the dijet system, $M_{jj}$, and the mass of all the particles in the central detectors, $M_X$.
Exclusive dijets would have the ratio $R_{jj} = M_{jj}/M_X$, shown in Fig.~\ref{fig:cdfrjj}, close to 1. 
Above $R_{jj} \sim 0.6$ the data show a significant excess over that
expected from \textsc{pomwig}, which is a Monte Carlo generator for diffractive processes, including an implementation of the Ingelman-Schlein model of $DI\!\!PE$~\cite{pomwig}. \textsc{pomwig} is tuned to CDF and H1 data and does not include an exclusive component. After combination with a full CDF detector simulation, the excess is well described by the predictions for exclusive dijet production given by the E\textsc{x}H\textsc{u}ME Monte Carlo generator~\cite{Monk:2005ji}, which implements the Durham model. Making such fits for several values of the minimum jet transverse energy, $E_T^{\textrm{min}}$, CDF derived the
exclusive dijet cross section shown in Fig.~\ref{fig:cdfjjet}. The
data are far below the prediction of the \textsc{dpemc} Monte Carlo generator~\cite{Boonekamp:2003ie}, which
implements the Saclay model discussed in Section \ref{sec:others}, and
although slightly below the E\textsc{x}H\textsc{u}ME prediction the
agreement is much better and certainly within the theoretical uncertainties\footnote{We note that better agreement between the Durham approach and the CDF dijet data was found in the analysis of Ref.~\cite{Khoze:2007td}.}.

Exclusive dijets
are expected to come primarily from $gg \rightarrow gg$ since quark pair
production, $gg \rightarrow q\bar{q}$, is suppressed due to the $J_z=0$ rule, by a 
factor $\sim m_q^2/M_{jj}^2$~\cite{jz0}. In particular (see for example~\cite{Khoze:2001xm})
\begin{equation}
 \frac{\d\hat{\sigma}_{excl}}{\d t}(gg \rightarrow gg) = \frac{9}{4}\frac{\pi \alpha^2_s}{E_T^4}
\label{eq:jz0gg}
\end{equation}
and
\begin{equation}
 \frac{\d\hat{\sigma}_{excl}}{\d t}(gg \rightarrow q\bar{q}) = \frac{\pi \alpha^2_s}{6 E_T^4} \frac{m_q^2}{M_{jj}^2}
\left( 1-\frac{4m_q^2}{M_{jj}^2}\right)
\label{eq:jz0qq}
\end{equation}
where these partonic cross sections are to be used in conjunction with Eq.~(\ref{eq:generalCEPxsec}). Light quark jets cannot easily be experimentally distinguished from
gluon jets, but some $c$-jets and $b$-jets can be tagged by secondary vertices or displaced tracks.
CDF found the that fraction of all exclusive dijets that are from $c$- or $b$-quark jets
is suppressed at $R_{jj} >$ 0.5 compared to lower $R_{jj}$ values. This lends
further support to the claim that the excess is due to
CEP dijet production. Note that $q\bar{q}$ dijet suppression will be very
important in reducing QCD backgrounds in exclusive $H\rightarrow
b\bar{b}$ searches at the LHC, as we discuss in Section~\ref{sec:higgs}. 

 Tevatron luminosities are now (in 2009) too high, even at the end of a store, to study events with no pile-up. A search is being
  made in CDF~\cite{Albrow:2009nj} for exclusive $\Upsilon \rightarrow
  \mu^+\mu^-$ and high mass di-muons, with pile-up, based on their
 distinctive kinematics (and no associated tracks on the $\mu^+\mu^-$ vertex). Unfortunately
  $\chi_b \rightarrow \Upsilon+\gamma$ is not likely to be visible
  since the cross section is expected to be much smaller than that of the
  $\chi_c$, and due
  to the difficulty in selecting the correct photon in events that are
  contaminated by pile-up.

%% file: cexplhc.tex
\section{The Large Hadron Collider}
\label{lhc}
\begin{figure}[h]
	\centering
	\includegraphics[width=0.4\textwidth]{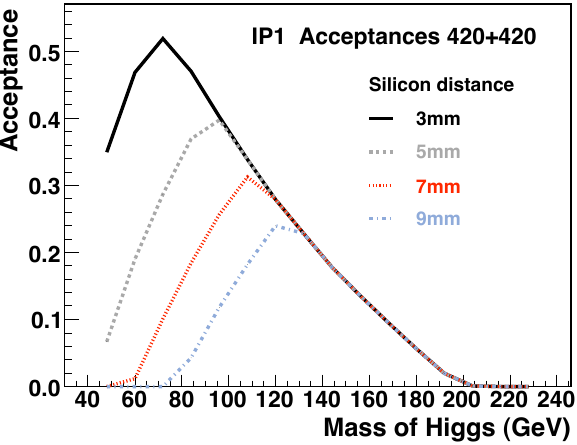} \hspace{0.05\textwidth}
	\includegraphics[width=0.4\textwidth]{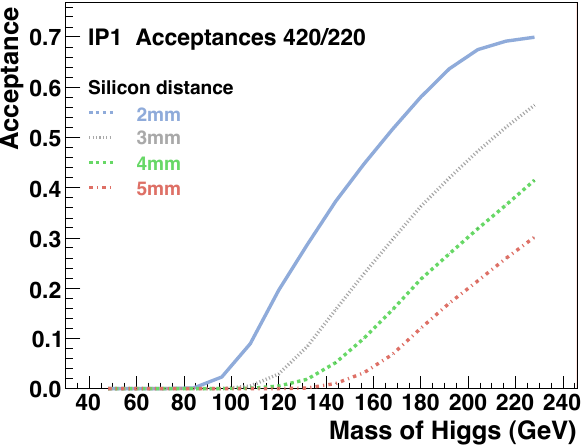}
	\caption{Forward proton detector acceptances, as a function of the 
	Higgs mass (mass of the central system), in the case of two protons detected at 420~m (left) and 
	in the case that one proton is detected at 420~m and one at 220~m (right). The different curves 
	correspond to differing positions of the active edge of the tracking detector relative to the beam. 
	Figure from Ref.~\cite{Albrow:2008pn}.}\label{fig:FP420acceptance}
\end{figure}

The idea to install detectors far from the interaction point at
CMS and/or ATLAS with the capacity to detect protons scattered through
small angles has gained a great deal of attention in recent
years, and the report presented in Ref.~\cite{Albrow:2008pn} constitutes a
significant milestone on the road to CEP physics at the LHC. 
Measurement of both protons by detectors located 420~m from the
interaction point at
ATLAS or CMS has the virtue that central systems with masses up to
200~GeV$/c^2$ can be measured with an event-by-event precision of
$\sigma(M_X)\approx 2-3$~GeV$/c^2$. Adding detectors at 220--240~m extends the acceptance to much higher central system masses, with the limiting factor becoming the rate of the production process. Fig.~\ref{fig:FP420acceptance} shows the forward detector acceptance as a function of 
central system mass for the case where both protons are detected at 420~m and when one is 
detected at 420~m and one at 220~m. The clean environment of CEP, combined with four-momentum constraints, makes for
reduced backgrounds (even in the presence of significant amounts of
pile-up), often aided by the fact that the centrally
produced system is predominantly in a $J_z=0$, C-even, P-even
state~\cite{jz0,Kaidalov:2003fw}. Of course having such a spin-parity filter also provides an
excellent handle on the nature of any new physics. For little
extra cost, forward detectors promise to significantly enhance the
physics potential of the LHC.

That said, CEP is not without its challenges. As we have seen, the theory is difficult,
triggering can be tricky, signal rates for new physics are often low,
high precision tracking and timing detectors need building and installing and pile-up needs to be
dealt with. As the last section showed, recent data on CEP from CDF at the
Tevatron gives confidence in the theoretical modelling, and
extensive studies have demonstrated that the other challenges
can be met, e.g. see Ref.~\cite{Albrow:2008pn}. Both in ATLAS and CMS, detectors at 420~m are situated too far from the central detector  to be included in the level 1 trigger,
although this should be possible in a future upgrade of the data
acquisition system. The data from the proton detectors are, of course, available for integration in a
higher level trigger. Note that, unlike the 420~m detectors, those at 220--240~m could be included in the level 1 trigger. 

\subsection{Higgs: SM and MSSM}
\label{sec:higgs}
The possibility of observing central exclusive Higgs boson production, $p+p \rightarrow p+H+p$, is 
largely responsible for the current interest in exclusive processes at hadron colliders. 

There are several important features of the CEP process which may allow one to extract, 
potentially unique, information on the structure of the Higgs sector. Firstly, precision measurements of the 
scattered proton momenta, in very forward detectors, offer the possibility of observing the Higgs as a bump in 
the missing mass distribution~\cite{Albrow:2008pn}. This is possible with a resolution $\sigma(M) \sim 2~\textrm{GeV}/c^2$ \emph{per event}, 
regardless of the final state (provided one can trigger on the central system). The precise calibration of the forward 
detectors necessary to achieve such a measurement can be obtained from data, using another exclusive (QED) 
process: $p+p \to p+ \mu^+ \mu^- +p$. The muons can be measured very well in the central 
detector, and then both proton momenta are known with an uncertainty limited by the incoming beam momentum
spread, $\Delta p/p\sim 10^{-4}$, i.e. $\Delta p \sim$ 700~MeV. To get sufficient rate for calibration 
on a reasonable time scale (e.g. one day), one must use muon pairs
with low mass, $M_{\mu^+\mu^-} \gsim$ 10 GeV$/c^2$. This is too small
to have any acceptance for detecting both protons but by
selecting same-side-dimuon events with $|\eta_{\mu^{\pm}}| \geq$ 2,
one proton can have large enough momentum loss to
be detected. 

In addition to obtaining the mass and width (provided it exceeds a few GeV) of any observed Higgs, one can also probe 
its spin and CP properties, due to the $J^{PC}=0^{++}$ selection rule. For non-zero angle scattering, the 
production of higher spin and odd parity states is allowed, though
they are strongly suppressed for high enough central masses by the
smallness of the final state proton transverse momenta \cite{Kaidalov:2003fw}. Given enough statistics one can distinguish the $J=0$ and higher spin states, 
by the azimuthal
correlations of the protons. Recall that at the ISR (Section~\ref{isr}), one could separate $X=\pi^+\pi^-$ data into 
S-wave ($J = 0$), 
P-wave ($J = 1$), and D-wave ($J = 2$) states, mass-bin by mass-bin, from the moments of the angular distributions of 
the pions, even finding a $\rho$ peak that was too small to see in the raw $M_{\pi^+\pi^-}$ spectrum.
This could be done at the ISR thanks to a high level of statistics
that is not likely to be available in $p+p \rightarrow p+ b\bar{b}+p$ data at the LHC, but 
the principle is the same. 

The spin-selection rule also has the advantage that it leads to a
leading-order suppression of the $b\bar{b}$ background in the CEP of
$H \to b\bar{b}$ (the suppression goes like $\sim m_q^2/M_{JJ}^2$, see Eq.(\ref{eq:jz0qq})).   
The production of exclusive dijets from light $q\bar{q}$ is likewise negligible, 
but CEP of gluon pairs receives no such suppression. Exclusive dijets,
$X = JJ$, are thus mostly $gg$ jets, 
with a modest admixture of $b\bar{b}$ jets. The $gg$ dijets can be discriminated against by heavy-flavour tagging, using a 
combination of displaced vertices and leptons in the jets from 
both $c$- and $b$-quark decays. In the exclusive dijet case one can
also profit from the fact that we 
can have $JJ = b\bar{b}$ or $gg$,
but pairs with net flavour (e.g. $bg$ or $bc$) are forbidden. 
Gluon splitting ($g \rightarrow b\bar{b}$) can be reduced by requiring both jets to be tagged.

\begin{figure}[htb]
\centering
\includegraphics[width=0.6\textwidth]{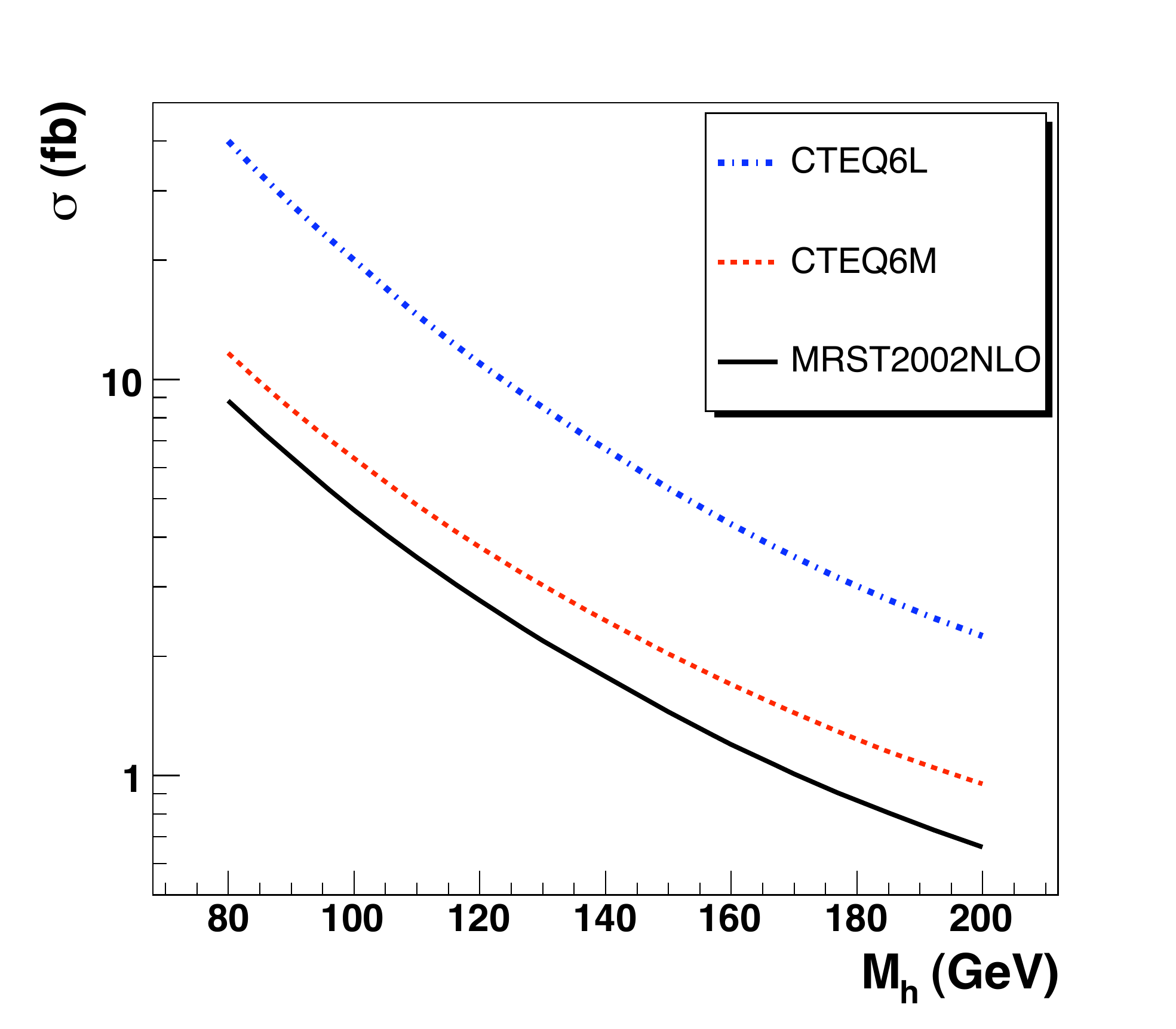}
\caption{CEP cross section for Standard Model Higgs production. Figure
  from Ref.~\cite{Albrow:2008pn}.}
\label{fig:SM}
\end{figure}

Fig.~\ref{fig:SM} shows how the cross section for producing a CEP SM Higgs
varies with Higgs mass for different gluon distribution
functions. The cross section is small and leads to low production
rates. In particular, the $b\bar{b}$ channel is very challenging, due to the aforementioned 
backgrounds and the additional background coming from pile up interactions at high 
luminosity, which we shall discuss in more detail later. Triggering in this case would certainly benefit from
having 220--240 m detectors in place, but even then one relies on optimistic
scenarios for the production cross section, detector acceptance and
trigger efficiency. 

In contrast, the observation of a SM Higgs in the $WW^{(*)}$ channel
is more promising. An initial study of the fully leptonic ($l^+l^- E\!\!\!\!/_T$) and
semi-leptonic ($l^- E\!\!\!\!/_T JJ$) decay modes, with basic experimental cuts, was performed
in~\cite{Cox:2005if}. The backgrounds considered in this study were
$\gamma\gamma$ fusion and exclusive $q\bar{q}$ production with
$W$-strahlung. The contribution of $gg\to W^+ W^-$ through a quark
loop was also considered, but found to be negligible. In order to
bring these backgrounds under control a range of cuts were
employed. The rapidities of the $W$ decay products, in the signal
case, are mostly central. In contrast, the backgrounds favour a
more forward distribution, hence a cut of $|\eta|<2.5$ was
imposed. The $\gamma\gamma$ process is dominated by the region in
which the outgoing protons carry very small transverse momenta, and a cut
was therefore made to remove this region. Finally, the dijet invariant
mass was constrained to lie within a window around the $W$ mass, thus
reducing the background from $q\bar{q}$ events not due to $W$
decay. After imposition of these cuts one could expect, for Higgs
masses in the range $140~\textrm{GeV}/c^2<m_H<200~\textrm{GeV}/c^2$,
$\sim 2-3$ events in 300~fb$^{-1}$ of data, with no appreciable background. One 
of these would be in the gold-plated fully leptonic channel, producing the 
striking signature of an event containing two leptons with no other tracks 
on the vertex and a large missing $E_T$. A subsequent analysis of the $WW^{(*)}$ channel 
with a fast detector simulation, taking into account pile-up background and using a more 
refined triggering strategy, found that the signal should be observable, with signal to 
background ratio of 1 or better, for $m_H \gsim 120~\textrm{GeV}/c^2$ and 300~fb$^{-1}$ 
of data~\cite{Albrow:2008pn}. 

\begin{figure}
\begin{center}
\includegraphics[width=0.75\textwidth]
                {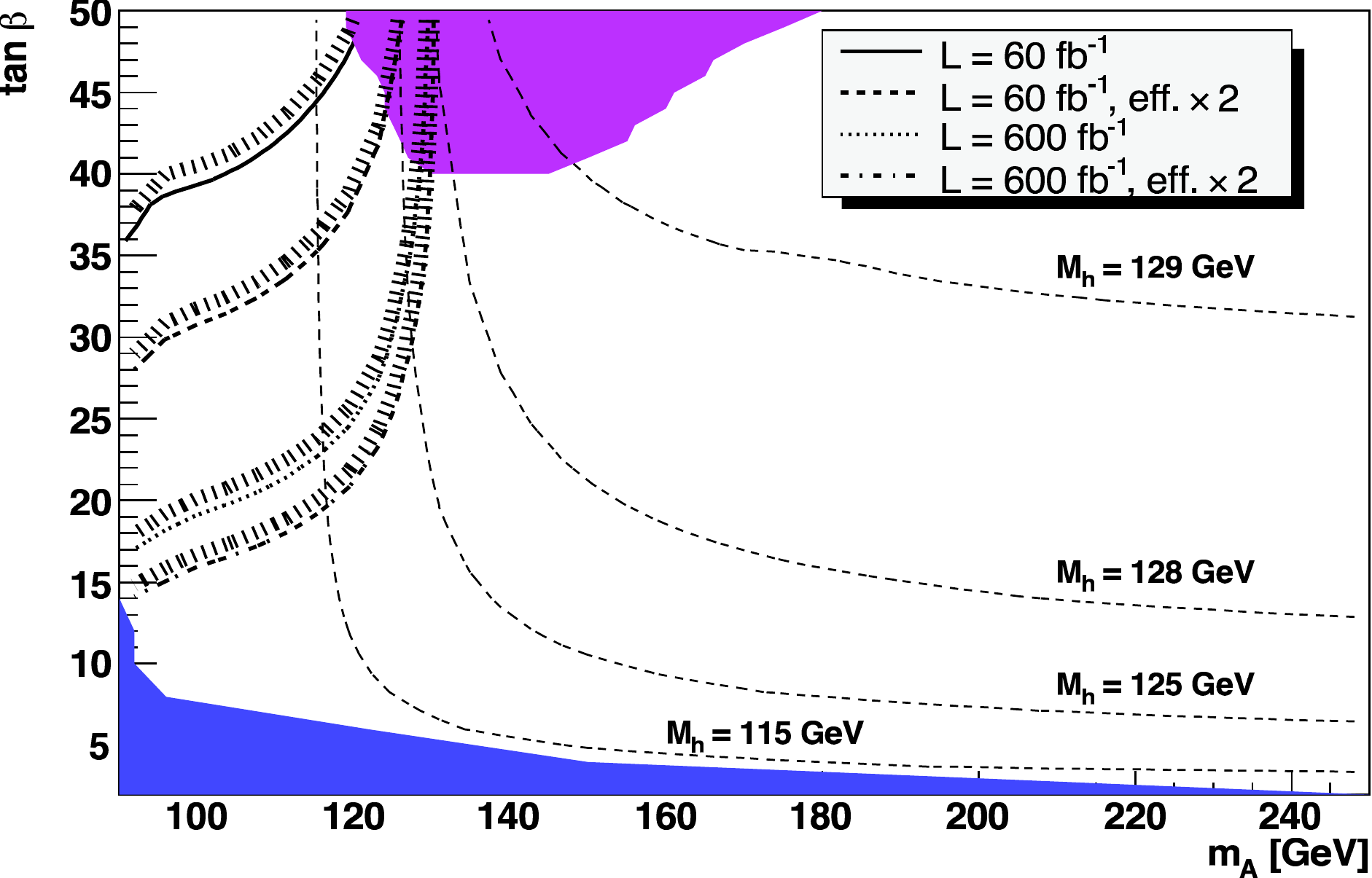}\\
\includegraphics[width=0.75\textwidth]
                {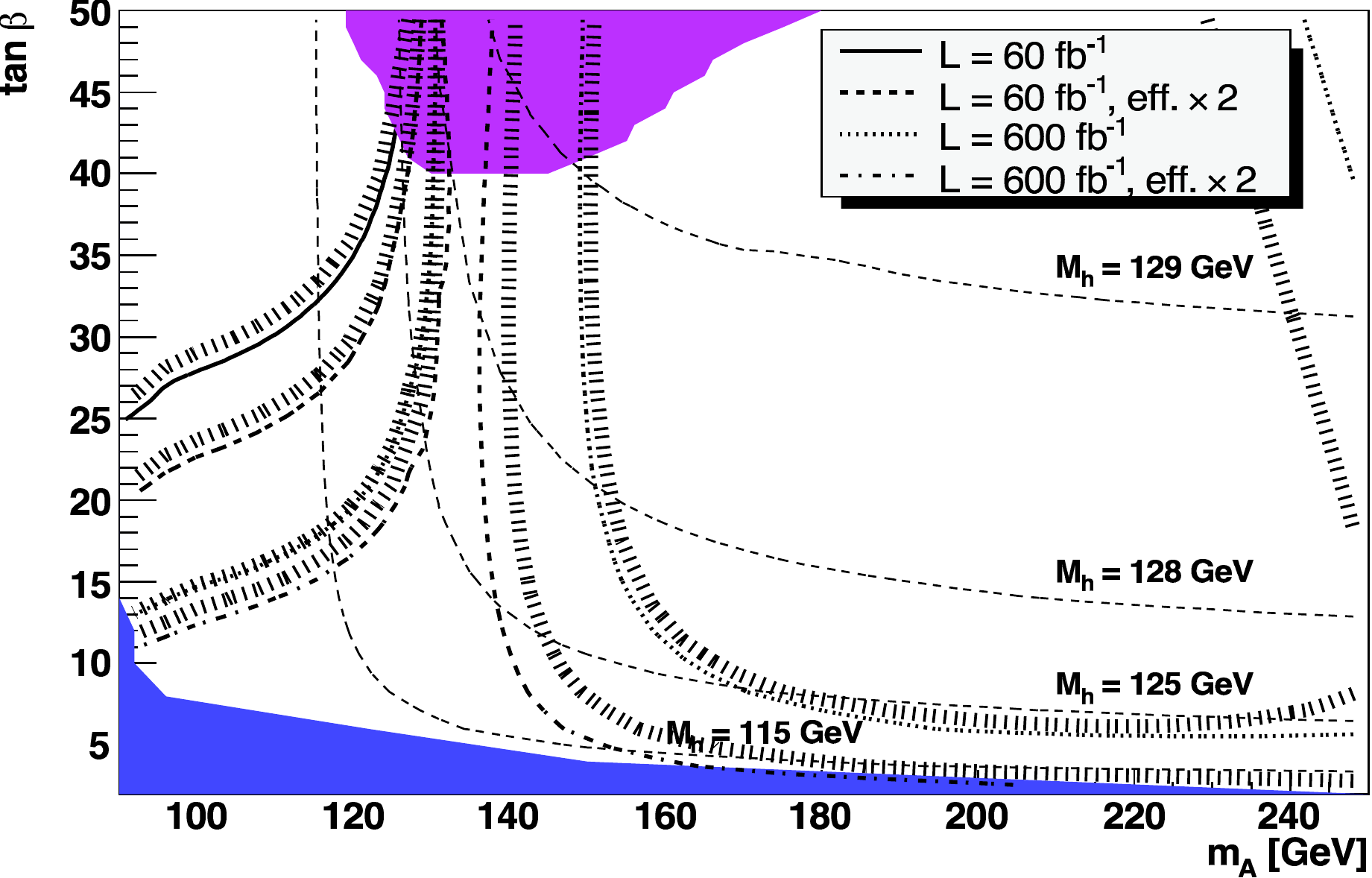}
\caption{
$5 \sigma$ discovery contours (upper plot) and contours of $3 \sigma$
statistical significance (lower plot) for the $h \to b \bar b$ channel in
CEP in the $M_A$ - \rm{tan}$\beta$ plane of the MSSM within the $\Mhmax$
benchmark scenario ($\mu=+200~\textrm{GeV}$) for different luminosity scenarios.  
The values of the mass of the light CP-even Higgs boson, $M_h$, are
indicated by contour lines.
The blue shaded region corresponds to the parameter region that
is excluded by the LEP Higgs boson
searches, while the purple shaded region is that excluded by searches at the Tevatron. Figure from Ref.~\cite{Heinemeyer:2009nj}.
}
\label{fig:heinemeyer2}
\end{center}
\end{figure}

\begin{figure}
\centering
\includegraphics[width=.65\textwidth]{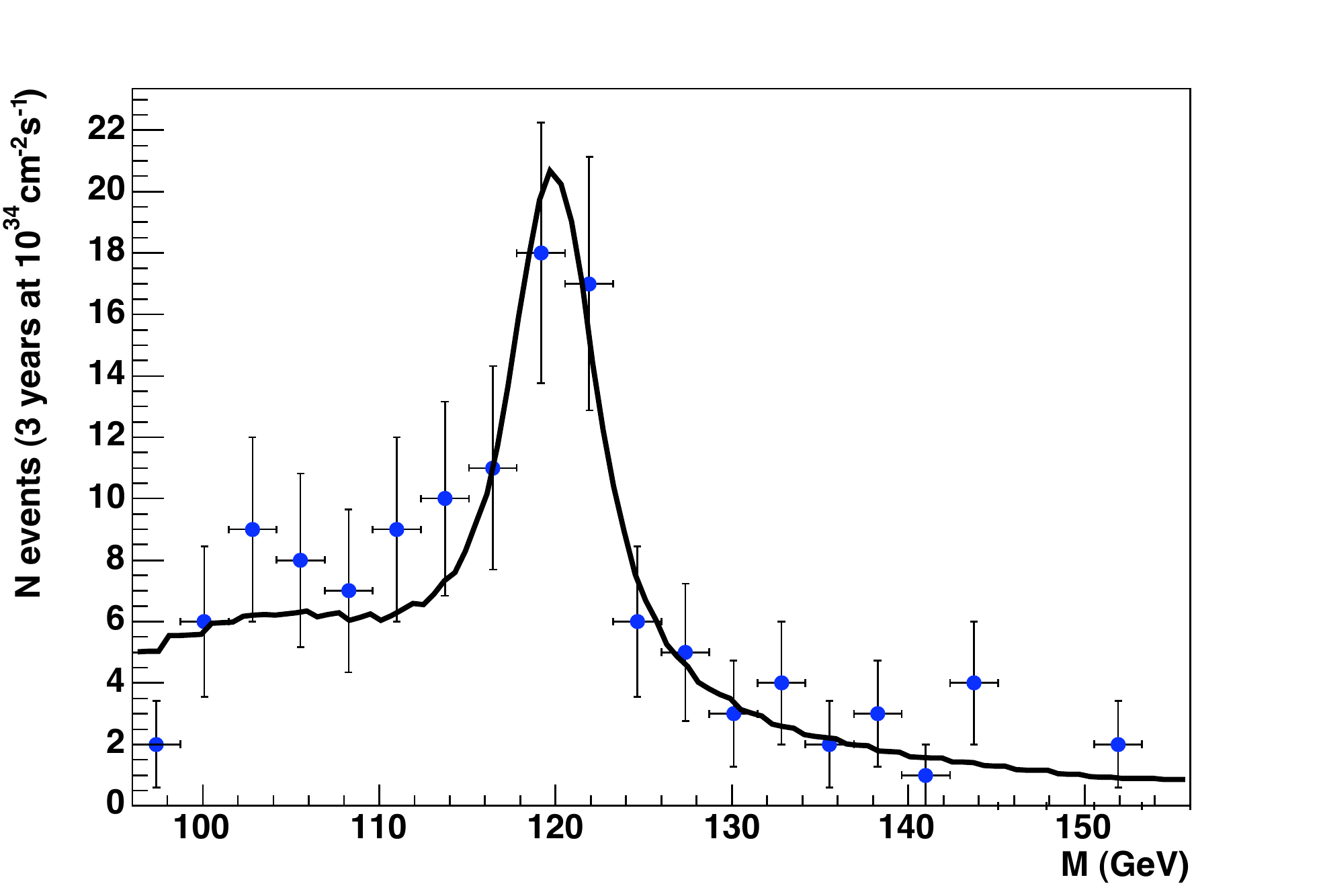}\\
\includegraphics[width=.65\textwidth]{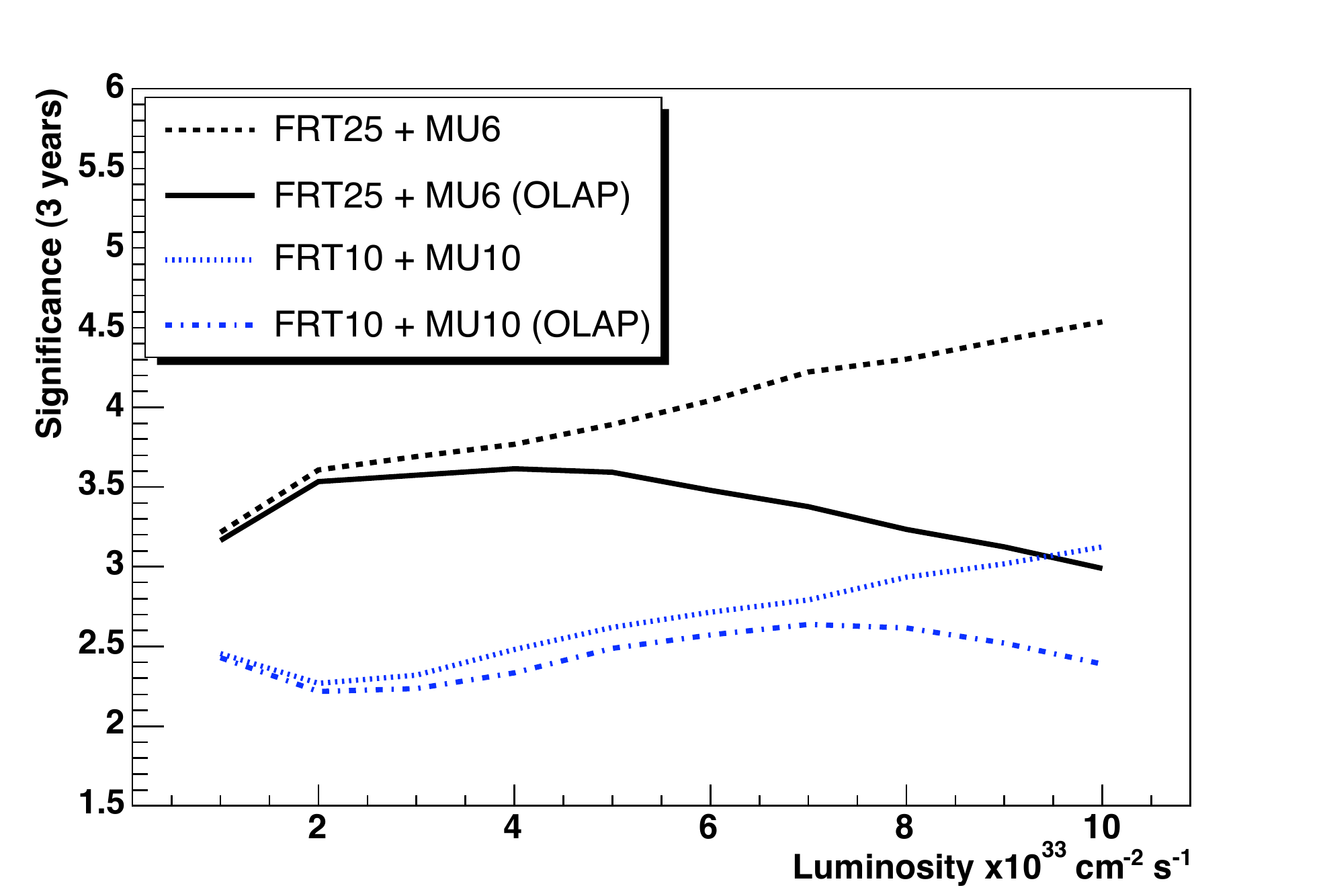}
\caption{Upper: A typical mass fit for the 120~GeV$/c^2$ MSSM $h\rightarrow b\bar{b}$ 
after 3 years of data taking at $10^{34}~$cm$^{-2}~$s$^{-1}$ and after removing the overlap 
background contribution completely (e.g. using improved timing detectors). The significance is 
$5 \sigma$ for these data. Lower: Significance of the measurement of a
120 GeV$/c^2$ MSSM Higgs boson versus luminosity, for two different combinations of muon
(MU6, MU10) and fixed-jet-rate (FRT25, FRT10) triggers. In each case
the curves correspond to the baseline (lower) and
improved-timing (upper) scenarios, i.e. the curves labelled ``OLAP'' include
pile-up background. Figure from Ref.~\cite{Albrow:2008pn}.}
\label{fig:peak2}
\end{figure}

Unlike the SM case, the $b\bar{b}$ channel can be rather
easier to explore in
certain MSSM scenarios. The neutral part of the MSSM Higgs sector consists of one light and one heavy CP-even scalar, 
the $h$ and $H$ respectively, and one CP-odd scalar, the $A$. For large $\tan \beta$ and small $m_A$, the $h b \bar{b}$ coupling 
is strongly enhanced over the SM case. Fig.~\ref{fig:heinemeyer2} shows an example of the region of parameter 
space\footnote{In the $M_h^{\textrm{max}}$ scenario with $\mu = + 200~\textrm{GeV}$.} in which one could observe 
$h\to b\bar{b}$ using CEP, for different amounts of integrated luminosity. Fig.~\ref{fig:peak2} shows the result 
of an in-depth analysis of one
particular point in the $m_A-\tan\beta$ plane ($\tan\beta =40$ and
$m_A=120$~GeV$/c^2$) \cite{Cox:2007sw}. The details of the two analyses can
be found in Refs.~\cite{Albrow:2008pn,Cox:2007sw,Heinemeyer:2007tu,Heinemeyer:2009nj} and they are in broad agreement.

\begin{figure}
\begin{center}
\includegraphics[width=0.75\textwidth]
                {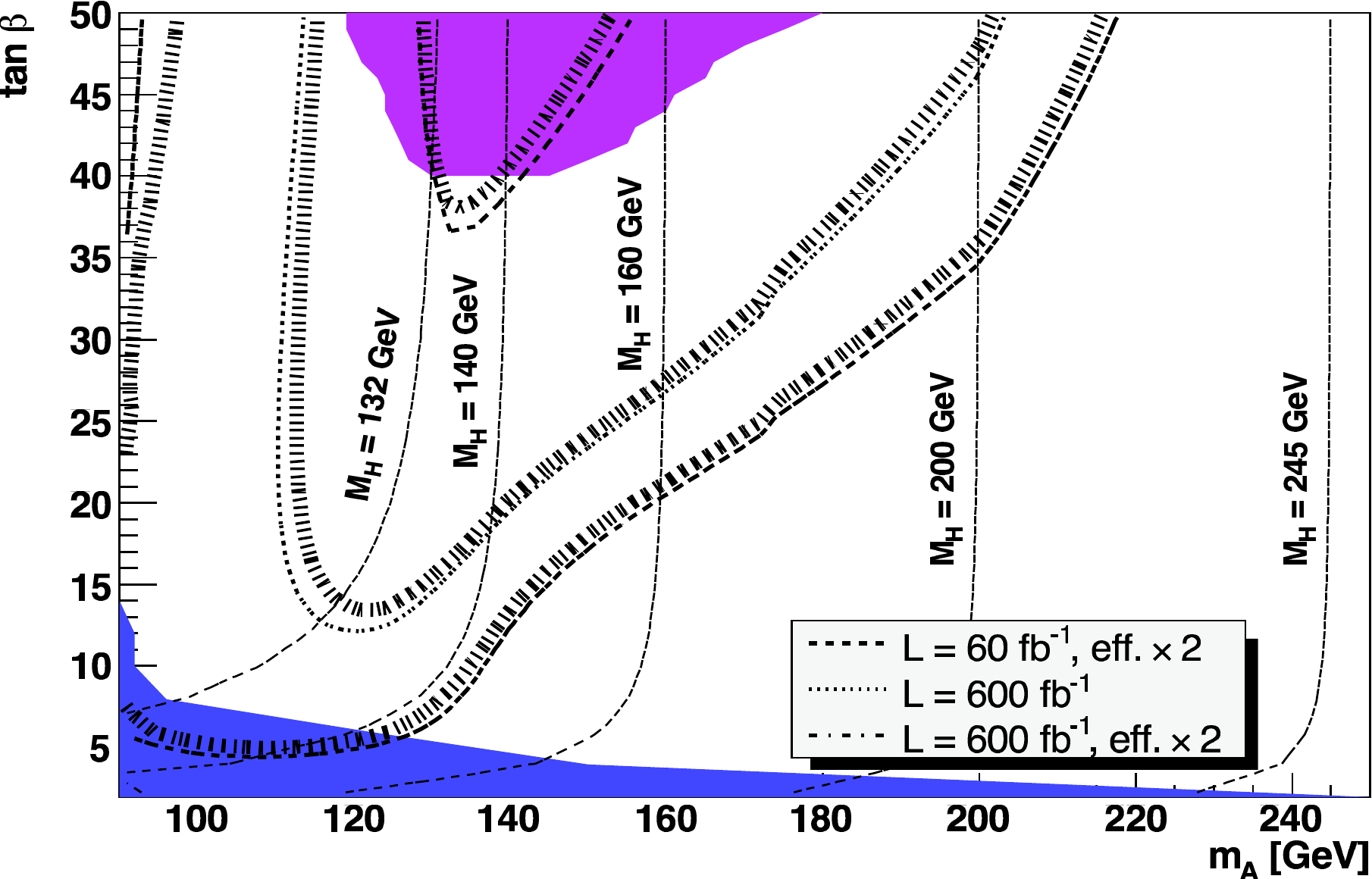}\\
\includegraphics[width=0.75\textwidth]
                {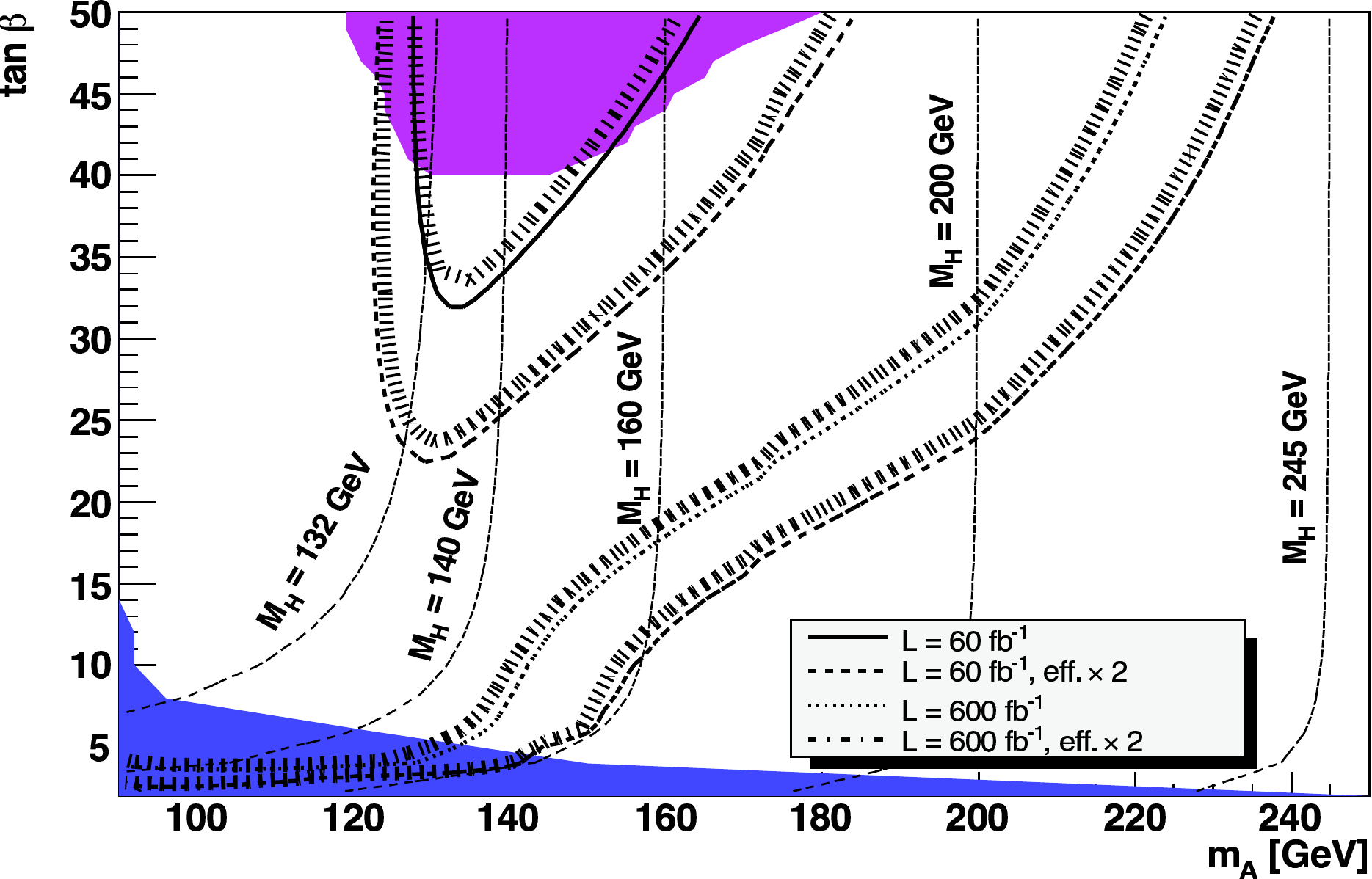}
\caption{
$5 \sigma$ discovery contours (upper plot) and contours of $3 \sigma$
statistical significance (lower plot) for the $H \to b \bar b$ channel in
CEP in the $M_A$ - \rm{tan}$\beta$ plane of the MSSM within the $\Mhmax$
benchmark scenario ($\mu=+200~\textrm{GeV}$) for different luminosity scenarios.  
The values of the mass of the heavy $CP$-even Higgs boson, $M_H$, are
indicated by contour lines. The blue shaded region corresponds to the parameter region that
is excluded by the LEP Higgs boson
searches, while the purple shaded region is that excluded by searches at the Tevatron. Figure from Ref.~\cite{Heinemeyer:2009nj}.
}
\label{fig:heinemeyer3}
\end{center}
\end{figure}

A similar pair of plots for $H\to b\bar{b}$ are shown in Fig~\ref{fig:heinemeyer3}. In addition to sensitivity in 
the low $m_A$ region, one can also hope to probe reasonably large values of $m_A$ (at large enough $\tan \beta$). 
If the MSSM parameters are in this region, an observation using CEP could be critical to understanding the Higgs 
sector at the LHC. This is because, in the large $m_A$ limit, referred to as the ``decoupling limit", the light CP-even 
Higgs becomes SM-like. In contrast, the coupling of the $H$ and $A$ (which are approximately mass degenerate) 
to SM gauge bosons is strongly suppressed. As a result, the dominant decay of the $H$ and $A$ is to $b\bar{b}$ ($\tau^+ \tau^-$ 
typically contributes approximately 10\%). This leads to the so called ``LHC wedge region"~\cite{:1999fr,Ball:2007zza,Gennai:2007ys}, where the $H$ and $A$ 
may escape detection in inclusive searches. Not only that, but current inclusive techniques, proposed to probe the spin 
and CP properties of the Higgs sector, rely on either the $ZZ$ decay mode at high mass~\cite{Buescher:2005re} or vector boson fusion at low mass~\cite{Plehn:2001nj,Hankele:2006ma,Ruwiedel:2007zz},
 both of which would be suppressed in this case (see also the
 discussion of this point in Ref.~\cite{Heinemeyer:2009nj}). It should also be noted that this decoupling of higher mass Higgs states 
 from the SM gauge bosons is a common feature of extended Higgs sectors, since such couplings are constrained by the 
 electroweak precision fits. Furthermore, the CEP production of the $A$ boson is strongly suppressed by the parity-even 
 selection rule. This means that any measurement of the $H$ mass and width using the forward proton detectors is 
 unaffected by a nearby $A$. This is not the case in inclusive production.

Returning to Fig.~\ref{fig:peak2}, the curves in the lower plot correspond to different trigger
scenarios. They also indicate the influence of pile-up background, which is of concern to any CEP analysis.
For brevity we denote a signal event with two detected protons and a
central state all arising from the same collision as [pXp]. Pile-up
backgrounds can originate when one or both protons and the central
system originate from different $pp$ collisions. If they originate from three different collisions
(triple pile-up) we write [p][X][p] and if they originate from two
different collisions (double pile-up) we write [pX][p] or [pp][X]. In the latter case,
since there is no acceptance for pure elastic scattering, the central
detector will contain particles from both the collision that led to
the two measured protons, as well as the particles from the second collision
(which has no detected protons). The double and triple pile-up rates both grow
strongly with luminosity and at $L\approx 10^{34}$ cm$^{-2}$s$^{-1}$
[p][X][p] typically dominates~\cite{Albrow:2008pn}. Fortunately there are many ways of
reducing the pile-up background to manageable
proportions. By measuring the four-momentum of the protons, one can
determine the mean
rapidity, $p_T$ and invariant mass of the central system. Consistency
between these values and those obtained using the central detector then
provide a strong constraint. Furthermore, a very large reduction in
the pile-up background can come from 
a precision measurement of the difference in time $\Delta t$ between
the arrival of the two protons in the forward detectors~\cite{Albrow:2008pn}.
If they came from the same collision, displaced a distance $z_{pp}$ from
the nominal vertex, then $\Delta t = 2 z_{pp}/c$
and one can require agreement between the measurements of $z_{pp}$ by the forward proton detectors and as determined from the 
central tracks.
The resolution is $\sigma(z_{pp})$ = 2.1~mm (3 mm/$\sqrt{2}$) for $\sigma(\Delta t)$ = 10~ps. As the distribution of interactions
has a spread  around 60~mm (at $1\sigma$) this leads to a large reduction in pile-up background.
 Detectors with time resolution $\sim$ 10 ps have been developed
for this purpose, and there is hope for further improvement (the spread in arrival
times due to different proton trajectories is only $\sim$ 2~ps). 

In
principle there are two other ways in which pile-up background can be
reduced by fast timing. Firstly, additional calibrations could permit the
actual collision time to be measured using the forward protons. The
collisions in a bunch crossing have a time spread $\sim 160$~ps. If the
individual collision times could also be measured using the detectors in
the central region, a match with the collision time from the protons
would give additional pile-up rejection. At present the
central detectors do not have sufficient time resolution to make this
practical. Secondly, one could place arrays of small ($\sim 1$~cm$^2$) fast-timing
detectors in the forward regions, e.g. $3 < |\eta| < 5$. Most inelastic collisions will produce
particles in both regions, but central exclusive events will
not. One can then veto events that contain extra particles
originating from the collision vertex (which is defined by the
measured protons). In the limit of the complete elimination of pile-up background, one
could make a $5\sigma$ discovery of an MSSM Higgs boson with 3 years of high
luminosity ($10^{34}$~cm$^{-2}$s$^{-1}$) data taking (the mass peak
is illustrated in the upper plot in Fig.~\ref{fig:peak2}).

\subsection{Higgs: NMSSM}
\label{sec:nmssm}

We now look at the
capabilities for CEP of Higgs boson production in the Next-to-Minimal
Supersymmetric Standard Model (NMSSM). More details can be found
in  \cite{Forshaw:2007ra}. Before discussing the specific details, let
us emphasise that this scenario is particularly interesting since it
deals with the possibility that a Higgs of mass $\sim 100$~GeV$/c^2$ decays
predominantly to light particles (in the case we shall review it is four taus). Such a
troublesome decay channel should be taken seriously and, as we shall see,
CEP can provide an excellent means to study decays of this type.

The NMSSM is an extension of the MSSM that
solves the $\mu$-problem, and also the little hierarchy problem. It achieves this by
adding a gauge-singlet superfield, $\hat{S}$, to the MSSM such that the $\mu$
term is now dynamical in origin, arising when the scalar member of
$\hat{S}$ aquires a vacuum expectation value (VEV). Since $\mu$ is no
longer fundamental and therefore is no longer naturally of order the GUT
scale (as would be the case if it were the only dimensionful parameter in the
superpotential), the $\mu$-problem is solved. The little hierarchy problem is also solved because a
lighter Higgs is allowed, thereby taking the pressure off the stop
mass, $m_{\tilde{t}}$. More specifically, the lightest scalar Higgs can decay
predominantly to two pseudoscalar Higgses and the branching ratio to
$b$-quarks is correspondingly suppressed, thereby evading the 114 GeV$/c^2$
bound from LEP (it drops to 86 GeV$/c^2$). Having a lighter Higgs
means that $m_{\tilde{t}}$ does not need to be so large, and that is preferred
by the $Z$ boson mass. 

The Higgs sector of the NMSSM extends that
of the MSSM by adding an extra pseudoscalar Higgs, the $a$ and an extra scalar
Higgs; crucially $\hat{S}$ is a gauge singlet and hence $h
\to aa$ can dominate with a light $a$ (i.e. below the threshold
for $a \to b\bar{b}$).
\begin{figure}[h]
\begin{center}
\includegraphics[width=0.55\textwidth]{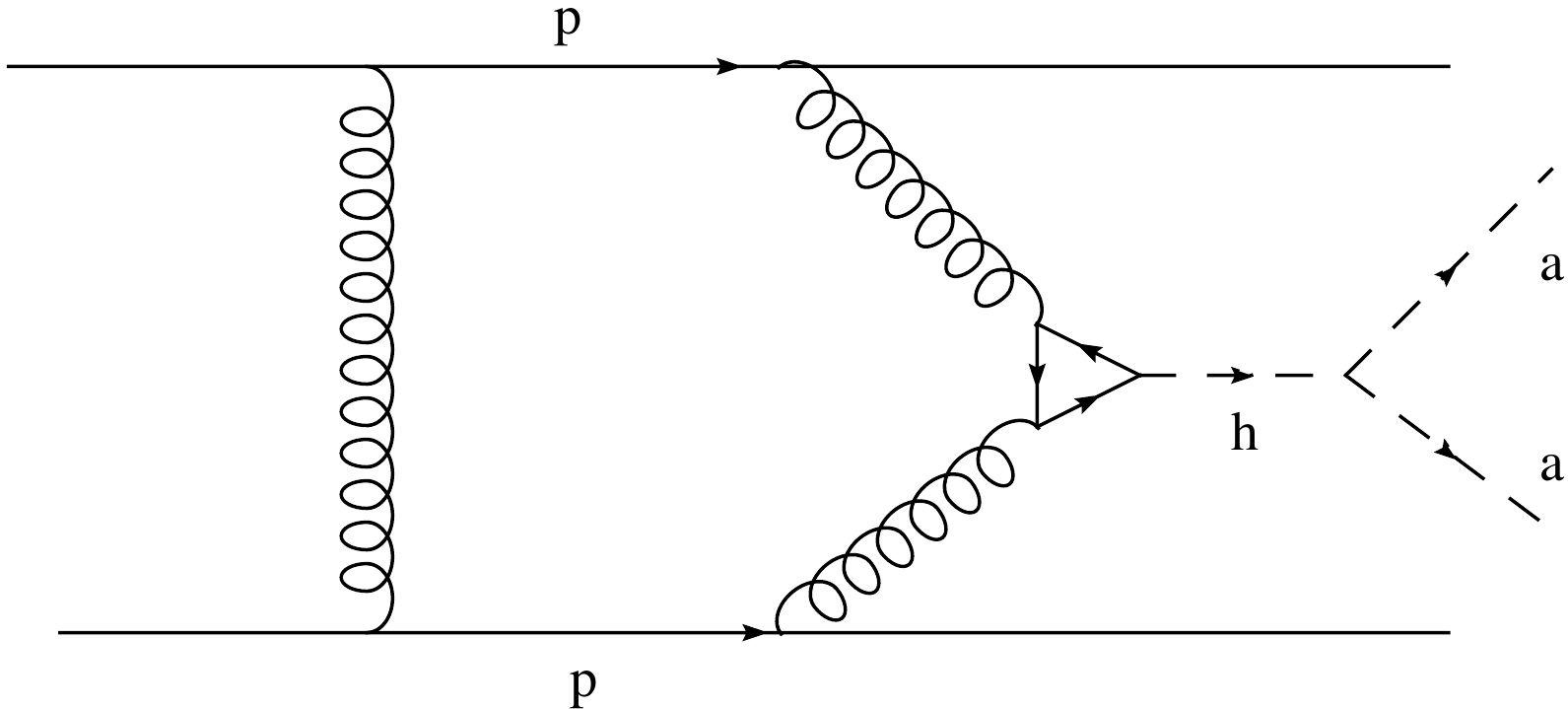}
\caption{CEP of an NMSSM Higgs.}
\label{fig:nmssm}
\end{center}
\end{figure}
Freed of the heavy $\tilde{t}$, it is natural to have a light Higgs with
a reducing branching ratio to $b$-quarks. Masses of the $h$ in the range 85--105
GeV$/c^2$ are most natural in this scenario and our attention will
focus on $m_h = 93$~GeV$/c^2$ and $m_a = 9.7$~GeV$/c^2$ with
BR$(h \to aa)=92\%$ and BR$(a\to \tau \tau)=81\%$ \cite{Ellwanger:2005dv}. The lightness of
the pseudoscalar $a$ means that the $h$ decays predominantly to 4$\tau$. Should such a decay mode be dominant at the LHC, standard
search strategies could fail and, as we shall see, CEP (as illustrated
in Fig.~\ref{fig:nmssm}) could provide the discovery channel.
This ``natural'' scenario of the NMSSM
has two additional bonus features that one might draw attention to:
(1) A light Higgs is preferred by the precision electroweak data
(recall that the best fit value is somewhat below 100 GeV$/c^2$), and (2) A $\sim$100 GeV$/c^2$
Higgs with a reduced ($\sim$10\%) branching ratio to $b$-quarks naturally
accommodates the existing 2.3 $\sigma$ LEP excess in $e^+e^- \to
Zb\bar{b}$ \cite{Dermisek:2005ar,Dermisek:2005gg}.

To detect the $4\tau$ decay of an NMSSM Higgs using CEP, we need
first to trigger on the event. In this regard, we require that at least one of
the $\tau$s decays to a sufficiently high $p_T$ muon (or two decay to lower $p_T$ muons). The detailed analysis is outlined in
\cite{Forshaw:2007ra}, here we shall just highlight the
key features. Table~\ref{tb:sigmas} shows how the signal (CEP) and backgrounds are
affected by the cuts imposed. The backgrounds are typical of CEP Higgs
studies: DPE refers to the $D\pom E$ production of a pair of $b$-quark
jets (which is the dominant $D\pom E$ background after $b$-tagging
both jets) and it was simulated
using the {\sc{pomwig}} Monte Carlo \cite{pomwig}; OLAP refers to
the background due to pile-up, in particular the three-fold
coincidence of two single-diffractive events with an inclusive $pp \to
X$ event. The QED background arises from $\gamma \gamma \to 4\tau$ and
is easily removed.

The top line of the
table is the cross section after requiring that there be at least one muon with
$p_T > 6$~GeV$/c$, which is the nominal minimum value to trigger at level 1\footnote{It will turn out that a 
higher cut of 10 GeV$/c$ is
  preferred in the subsequent analysis.}, 
and the condition that both protons be detected in the
420m detectors. There is also a loose cut on the invariant mass of the
central system. Of the remaining cuts, we single out the
``$N_{\mathrm{ch}}=4$ or 6'' cut for special mention. The charged track ($N_{\mathrm{ch}}$) cut is noteworthy because it can
be implemented at the highest LHC luminosities: we cut on 4 -
6 charged tracks that point back to the vertex defined by the
muon. Pile-up events add extra tracks (to both the signal and background), but they do not often
coincide with the primary vertex (e.g. within a 2.5~mm window) and do
not spoil the effectiveness of this cut. The ability to make such hard
cuts on charged tracks is of much
wider utility than in this analysis. The 4--6 track event is then
analysed in terms of its topology, exploiting the fact
that the charged tracks originate from four taus in the signal, which
themselves originate from two heavily boosted pseudoscalars. This analysis is purely
track-based with almost no reliance on the calorimeter, which is more affected by pile-up. 

\begin{figure}[t]
\begin{center}
\includegraphics[width=0.65\textwidth]{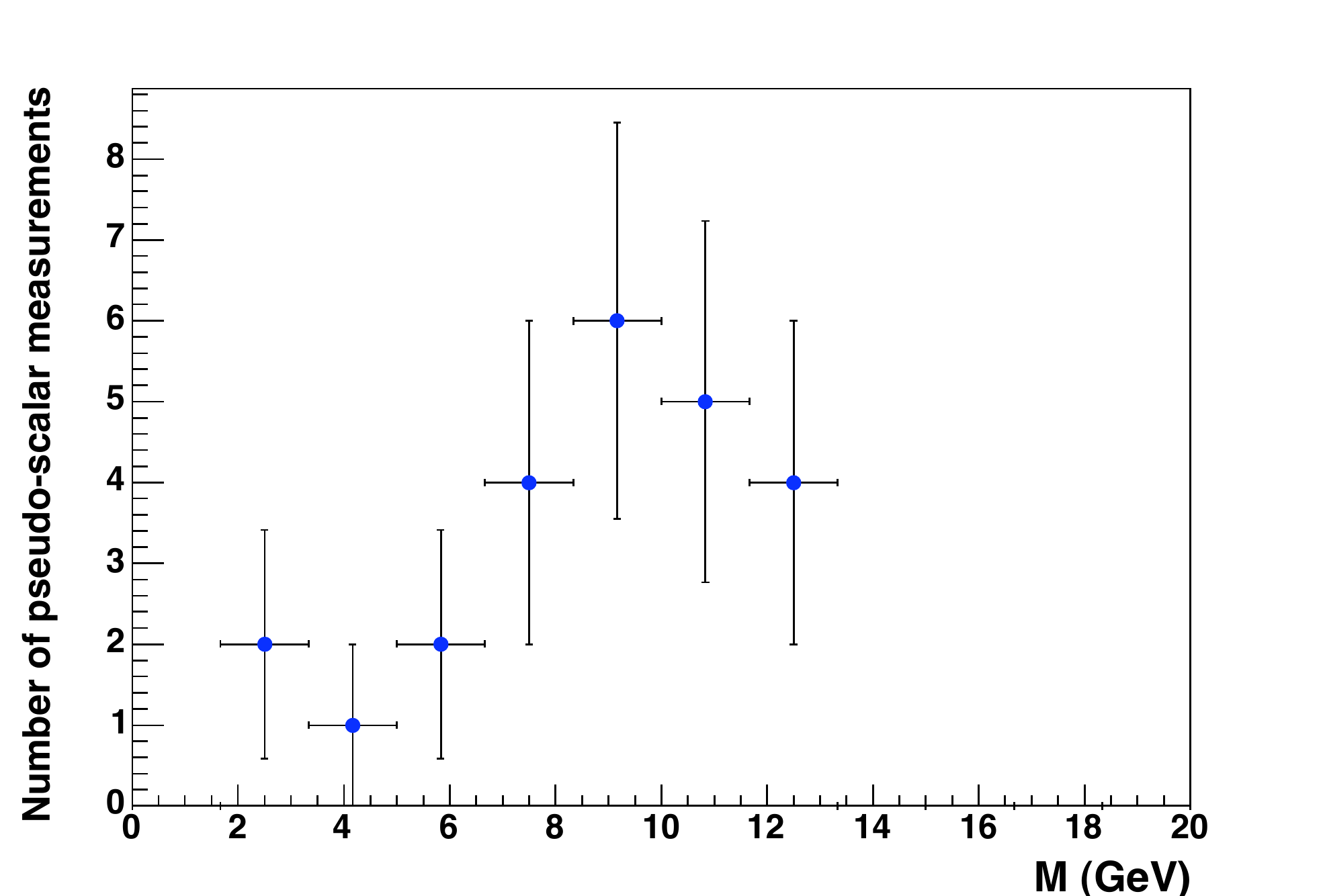}
\caption{A typical $a$ mass measurement. Figure from Ref.~\cite{Forshaw:2007ra}.}
\label{fig:amassdata}
\end{center}
\end{figure}

Accurate measurements of the proton momenta allow one to constrain
the kinematics of the central system (in particular its invariant mass, $p_T$,
and rapidity are known). We can also extract the masses of the
$h$ and $a$ on an event-by-event basis. The mass of the $h$ is
straightforward (it is measured directly by the forward proton
detectors) and a precision below 1 GeV$/c^2$ can be obtained with just a few events. The measurement 
of the pseudoscalar mass is more
interesting and potentially very important. The proton measurements
fix $p_z$, and $p_{x,y} \approx 0$ for the central system. In addition,
the $\tau$ pairs are highly boosted, which means they are collinear with
their parent pseudoscalars. Thus the four-momentum of each
pseudoscalar is approximately proportional to the observed (track)
four-momentum. The two unknown  proportionality constants (i.e. the
missing energy fractions) are overconstrained, since we have three
equations from the proton detectors. The result is that we can solve
for the pseudoscalar masses, with four measurements per
event. Fig.~\ref{fig:amassdata} shows a typical distribution of $a$
masses based on 180~fb$^{-1}$ of data collected at $3 \times 10^{33}$~cm$^{-2}$s$^{-1}$.

\begin{table*}
\centering
\begin{tabular}{|c|c|c|c|c|c|c|c|} \hline
 &  \multicolumn{3}{c|}{CEP} & DPE & OLAP &\multicolumn{2}{c|}{QED} \\ \hline
Cut & $H$ & $b\bar{b}$ &$gg$ & $b\bar{b}$& $b\bar{b}$ & $4\tau$ & $2\tau$~$2l$ \\ \hline
$p_{T0}^\mu$, $\xi_1$, $\xi_2$, $M$     & 0.442 & 25.14 & 1.51$\times10^{3}$ & 1.29$\times10^{3}$& 1.74$\times10^{6}$ & 0.014 & 0.467 \\ \hline
$N_{\mathrm{ch}} =$ 4 or 6 & 0.226 & 1.59 & 28.84 & 1.58$\times10^{2}$ & 1.44$\times10^{4}$& 0.003 & 0.056\\ \hline
$Q_h = 0$ & 0.198 & 0.207 & 3.77 & 18.69 & 1.29$\times10^{3}$ & 5$\times10^{-4}$& 0.010 \\ \hline
Topology & 0.143 & 0.036 & 0.432 & 0.209 & 1.84 & - & $<$0.001 \\ \hline
$p_T^{\mu}$, isolation & 0.083 & 0.001 & 0.008 & 0.003 & 0.082 & - & - \\ \hline
$p_T^{1, \not{\mu}}$ & 0.071  & 5$\times$10$^{-4}$ & 0.004 & 0.002 & 0.007  & -& - \\ \hline
$m_a > 2 m_\tau$ & 0.066 & 2$\times$10$^{-4}$ & 0.001 & 0.001 & 0.005 & - & - \\ \hline
\end{tabular} \caption{The table of cross sections for the signal and
  backgrounds for NMSSM Higgs production. All cross sections are in
  femtobarns. The pile-up (OLAP) background is computed at a
  luminosity of $10^{34}$~cm$^{-2}$~s$^{-1}$. From Ref.~\cite{Forshaw:2007ra}.}
\label{tb:sigmas}
\end{table*}

Although we can expect only a small number of signal events, the background is under control. One should be able to collect 4 signal events, on a background of 0.1 events, with 3 years running at an instantaneous luminosity of $5 \times 10^{33}$~cm$^{-2}$s$^{-1}$. Even with these few events, one may extract the masses of both the $a$ and the $h$, with a resolution below 2-3~GeV$/c^2$. The statistical significance of any discovery can exceed $5\sigma$ in a few years of running at luminosities in excess of $10^{33}$~cm$^{-2}$s$^{-1}$.

\subsection{Other new physics} \label{sec:onp}

In addition to the MSSM and NMSSM Higgs studies, the CEP of a number of other BSM 
central systems has been considered in the literature. 

For instance, in~\cite{Ellis:2005fp} a CP-violating version of the MSSM was studied. In such a model, the $h$, $H$ and $A$ may 
all mix to produce mass eigenstates, $H_1,H_2,H_3$, of indefinite CP. For large $\tan\beta$, the couplings of the $H_i$ 
to $b\bar{b}$ and $\tau^+\tau^-$ are enhanced, the states are naturally nearly mass degenerate (separated by a 
few~\textrm{GeV}$/c^2$) and strong mixing between the states leads to a complicated resonance structure. The CEP of the $H_i$, 
with decay to $b\bar{b}$ and $\tau^+\tau^-$ was studied and it was shown that the missing mass distribution could be 
used to perform a line-shape analysis of the resonance structure,
which is probably impossible in inclusive 
production. Also in the CP-violating MSSM, but in a different parameter region, the LEP bounds on the lightest 
Higgs mass become much weaker. In~\cite{Khoze:2004rc,Cox:2003xp}, CEP of such a light Higgs was studied, though in
this case the $b\bar{b}$ channel appears not to be observable and the $\tau^+\tau^- $ channel is marginal.

Another extension of the Higgs sector studied in relation to CEP is a
triplet Higgs model in which the 
Higgs sector consists of one complex Higgs doublet and two triplets
(one real and one complex)~\cite{Chaichian:2009ts}. In such a model, 
the neutral Higgs associated with the doublet, $H_1^0$, has an enhanced coupling to fermions\footnote{The $H_1^0$ becomes 
the Standard Model Higgs boson for vanishing doublet-triplet mixing.}. The observation, using forward proton detectors, 
in the $H_1^0\to b\bar{b}$ channel was considered and it was found to be observable, with a significance ranging 
from $\sim 4\sigma-14 \sigma$ over much of the parameter space with
60~fb$^{-1}$ of data. With such data, the mass could also be measured very accurately, with an expected 
error less than 2~GeV$/c^2$ over a significant region of parameter space (and in some regions better than 0.3~GeV$/c^2$).

In some extensions of the SM the Higgs decays predominantly to unobservable particles. 
The CEP of such an ``invisible Higgs" has been studied in~\cite{Belotsky:2004ex}. However, 
while a detection in the CEP channel would be especially useful, since it would not rely on 
the detection of the Higgs decay products, it is not clear that pile-up backgrounds can be 
brought sufficiently under control.

CEP Higgs production has also been studied in a model where the SM is complemented by a fourth
 generation of very heavy fermions~\cite{Heinemeyer:2009nj}. The Higgs coupling to gluons in 
 such a model is typically larger than in the SM. In the $H\to b\bar{b}$ channel, it was 
 found that a significance in excess of 3$\sigma$ could be obtained with 60~fb$^{-1}$ of 
 data over the range $114~\textrm{GeV}/c^2 \lsim m_H \lsim 145~\textrm{GeV}/c^2$ not excluded 
 by LEP or the Tevatron\footnote{The region $m_H\gsim 220~\textrm{GeV}/c^2$ is also experimentally 
 allowed, however the CEP rate for $H\to b\bar{b}$ is too low to be accessible in this region.}.

Away from the Higgs sector, the production of long-lived gluinos, expected in the ``Split Supersymmetry" 
model~\cite{ArkaniHamed:2004fb,Giudice:2004tc}, was studied in Ref.~\cite{Bussey:2006vx}. Such particles hadronise 
to produce so-called $R$-hadrons which look much like slow muons. The CEP process in this case is particularly 
clean, since cuts on the velocity of the $R$-hadrons reduce the
muon background to a negligible level. It was found that the measurements from the forward proton detectors could in this case give a measurement of the 
gluino mass to better than 1\%, for masses below 350~GeV$/c^2$ and with 300~$\textrm{fb}^{-1}$ of data. Although not 
a discovery channel, such a measurement would be highly complementary to any measurement in inclusive production, 
which suffers from large systematic uncertainties in this mass region~\cite{Kilian:2004uj}.

Another model relevant to CEP is White's theory of the critical pomeron~\cite{white}, which requires the existence of a pair
of massive colour sextet quarks, $Q_S$. Since $Q_S$ loops couple strongly to the pomeron and to weak bosons, processes
such as $\pom \pom \rightarrow W^+W^-$ and $ZZ$ (but not $WZ$) should occur with a much 
enhanced cross section. The threshold for this process is above the Tevatron energy but below the LHC energy. The
enhanced cross section may even be visible without forward proton tagging, but the latter will be needed to demonstrate
that $\pom \pom$ interactions are responsible. Photoproduction of $Z$, $\gamma\pom \rightarrow Z$, should also be
enhanced a factor $\geq$ 10 over the SM, and should be detectable.
  
\subsection{Two photon collisions}
\label{sec:ggcoll}

\begin{figure}[htb]
\centering
\includegraphics[width=0.55\textwidth]{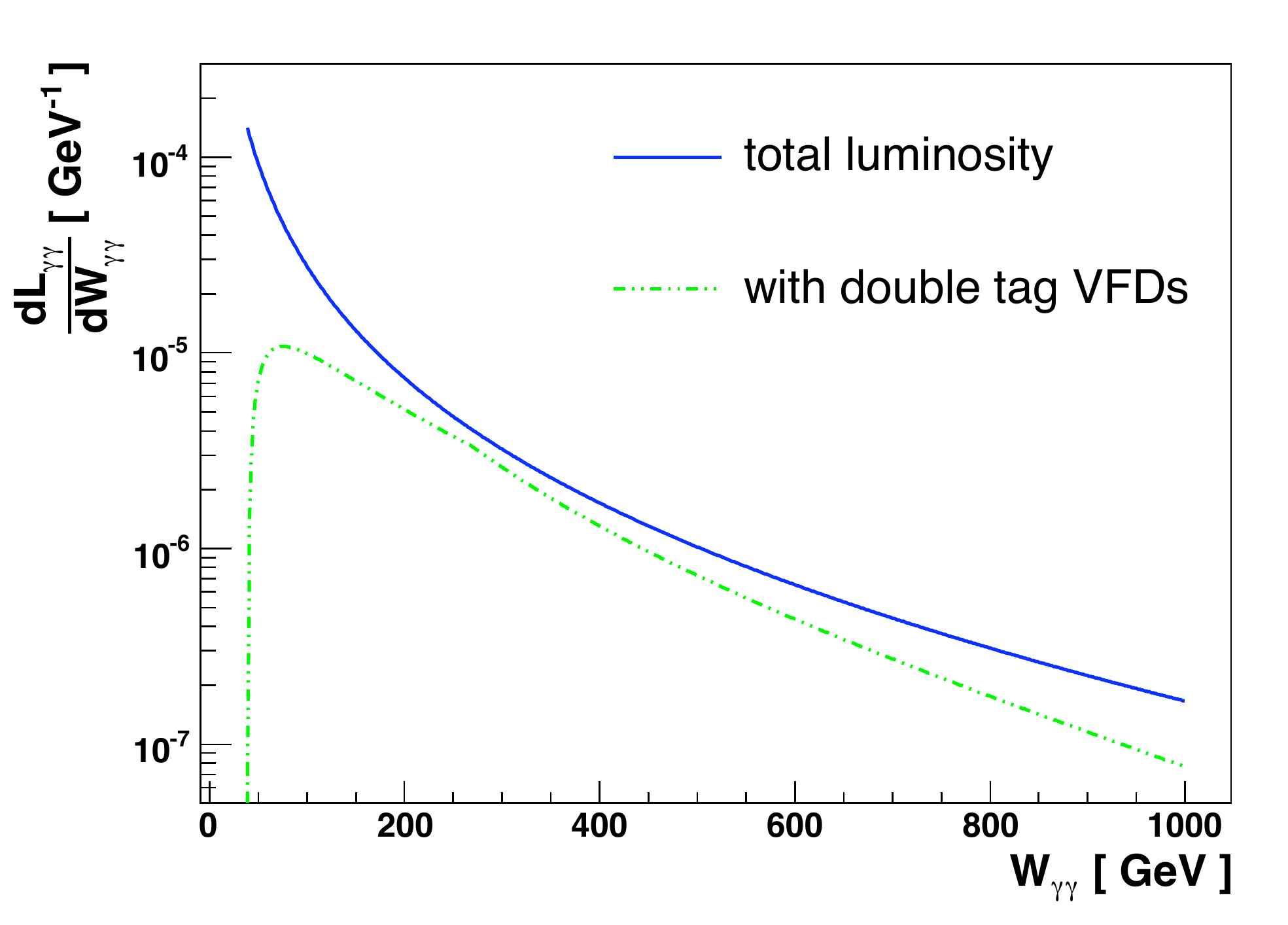}
\caption{Luminosity spectrum for photon-photon collisions at the LHC 
in the range $Q^2_{min}<Q^2<2$~GeV$^2$ (solid blue line) compared to the 
corresponding luminosity if the energy of each photon is restricted to
the forward detector tagging range  20 GeV $< E_\gamma <$ 900 GeV
(dashed green curve). Figure from Ref.~\cite{louvain}.}
\label{fig:GGlumi} 
\end{figure}

So far in this review our attention has focussed mainly on CEP via the strong
interaction. At the LHC, high proton luminosities will ensure that
the flux of bremsstrahlung photons off the colliding beams is
sufficiently large for CEP via $\gamma \gamma \to X$ and $\gamma \pom
\to X$ to be of interest. We focus in this section on the former.
The elastic photon flux is shown in
Fig.~\ref{fig:GGlumi}, from which the corresponding $pp$ cross
section, $\d\sigma_{pp}$, can be obtained:
\begin{equation}
\d\sigma_{pp} = \int \frac{\d L_{\gamma\gamma}}{\d
  W_{\gamma\gamma}}\d\sigma_{\gamma\gamma} \; \d W_{\gamma\gamma}~,
\end{equation}
where the integral is over the $\gamma\gamma$ centre-of-mass energy,
$W_{\gamma\gamma}$. To put this flux in context,
we note that it corresponds to a $\gamma\gamma$ luminosity that is 1\%
of the $pp$ luminosity for the production of central systems with mass
greater than 23~GeV$/c^2$, falling to 0.1\% for masses above
225~GeV$/c^2$. In addition, the photon virtualities are cut-off sharply by the
proton electromagnetic form factor and are small, i.e.
$\langle Q^2\rangle\approx0.01$~GeV$^2$; they can usually be assumed to
be zero.

Our discussion in this section follows
closely the presentation in \cite{Albrow:2008pn,louvain}. In turn, we
will consider CEP of: (i) lepton pairs; (ii) weak vector
boson pairs; (iii) supersymmetric (SUSY) pairs. The first of these
(especially using muon pairs) is a good candidate for measuring the
LHC absolute luminosity, the second is an accurate probe of the
$\gamma\gamma WW$ coupling and the latter could be used to make
accurate mass measurements of the SUSY particles from the forward protons.

\begin{table}[htbp]
\begin{center}
\begin{tabular}[htbp]{ll||c |c}
\hline
\multicolumn{2}{l||}{\textbf{Processes}}  & $\mathbf{\sigma}$ (fb) & \textbf{Generator}   \\\hline
$\gamma \gamma \rightarrow$ & $\mu^+ \mu^-$ ($p_T^{\mu}>$ 2 \GeVc, $|\eta^\mu|<3.1$) &  72 500 & { \sc lpair}~\cite{lpair}\\
& $W^+ W^-$ & 108.5 & { \sc mg/me}~\cite{mad,mad1} \\ 
& $F^+ F^-$ ($M = 100$ \GeVcc) & 4.06 & $\scriptscriptstyle{//}$ \\
& $F^+ F^-$ ($M = 200$ \GeVcc) & 0.40 & $\scriptscriptstyle{//}$ \\
& $S^+ S^-$ ($M = 100$ \GeVcc) & 0.68 & $\scriptscriptstyle{//}$ \\
& $S^+ S^-$ ($M = 200$ \GeVcc) & 0.07 & $\scriptscriptstyle{//}$ \\
& $H \rightarrow b\overline{b}$ ($M = 120$ \GeVcc) & 0.15 & { \sc mg/me}~\cite{mad,mad1} \\
\hline
\end{tabular}
\end{center}
\caption{Production cross sections for $pp \rightarrow p +X+ p $ (via $\gamma \gamma$ exchange) for various 
processes ($F$ = fermion, $S$ = scalar). $M$ is the mass of
the corresponding particle.~Table from Ref.~\cite{louvain}.}
\label{tab:backGG}
\end{table}

\begin{figure}[htb]
\centering
\includegraphics[width=8.5cm]{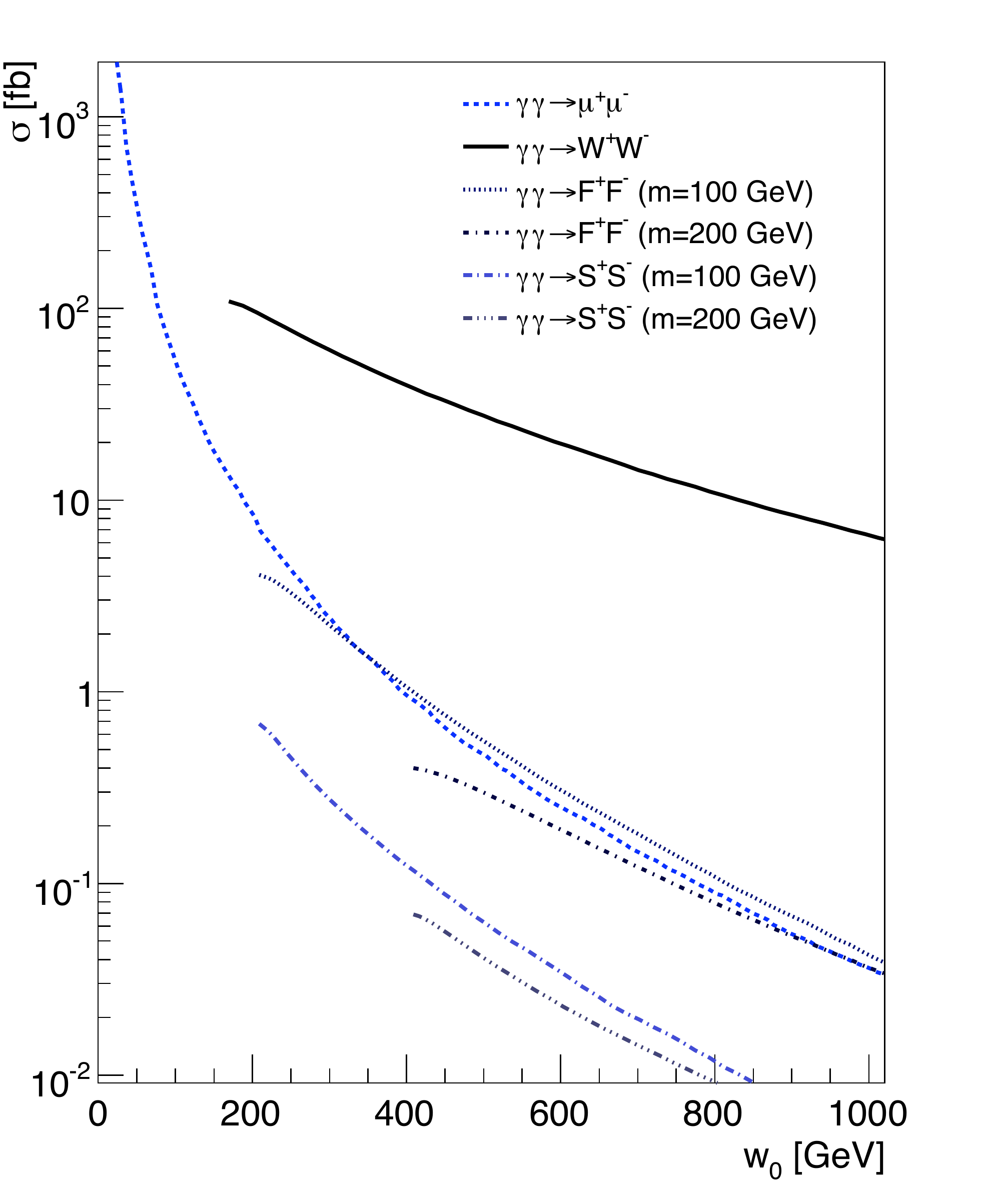}
\caption{Cross sections for various $\gamma \gamma$ processes at the LHC as a function
 of the minimal $\gamma \gamma$ centre-of-mass energy $W_0$. Figure
 from Ref.~\cite{louvain}.}
\label{fig:IntGG} 
\end{figure}

Using a modified version of 
{\sc madgraph}/{\sc madevent}~\cite{mad,mad1}, the Louvain
group~\cite{louvain} has computed the production cross sections
for the various processes of interest. These are tabulated in
Table~\ref{tab:backGG}. 
Since the cross sections for pair production depend 
only on the charge, spin and mass of the produced particles, the results
are shown for singly-charged, colourless fermions and scalars of two different masses. The cross
sections are also shown as 
a function of the minimal $\gamma \gamma$ centre-of-mass energy $W_0$ in Fig.~\ref{fig:IntGG}.
Note the hierarchy of cross sections, which is driven by the spin of
the relevant $t$-channel exchange. The cross sections are large enough
to merit further study. The two-photon production of SUSY pairs was first studied in
\cite{bib:Ohnemus} and the exclusive two-photon production of a
SM Higgs was studied in Refs.~\cite{higgs,kp}. The
rate is significantly smaller than that arising from strong
dynamics (see Fig.~\ref{fig:SM}) but it could become
particularly interesting in the case of an enhanced $H\gamma\gamma$ 
coupling. 

\subsubsection*{Lepton pairs}
\label{sec:muonpairs}

Two photon collisions can produce exclusive pairs of any charged
particles, but it will probably not be possible to observe $\gamma\gamma \rightarrow q \bar{q}$
because of overwhelming strong interaction backgrounds ($D \pom E$). However, lepton pairs ($e^+e^-, \mu^+\mu^-$
and also, to some extent, $\tau^+\tau^-$) should have very small
backgrounds, even in the presence of pile-up. One requires that there are no additional charged
particles on the dilepton vertex, and that $p_T(l^+l^-)$ is very
small or, equivalently, that the leptons are highly back-to-back ($|\pi - \Delta \phi| \lsim
1.0~$GeV$/M_{l^+l^-}$ rad) and
with similar $p_T$.  The $\Delta\phi$ cut is especially
powerful as it is forgiving of poor momentum resolution at high mass. The cross sections,
calculated by the {\sc lpair} Monte Carlo~\cite{lpair} at $\sqrt{s}$ = 14 TeV, integrated over
$M_{\mu\mu} > M_{\mu\mu}(min)$ and for different $\eta_{max}$,
are well approximated by the purely empirical fit: 

\begin{equation}
\sigma(M_{min},\eta_{max}) = (480 - 1.25 \times M_{min}) \times
\frac{\eta^{2.2}_{max}}{M_{min}^{2.4}} \mathrm{pb}~.
\label{lpairfit}
\end{equation}

This equation agrees with the predictions of LPAIR to within $\pm$4\% for $M_{min}$ from 8 GeV$/c^2$ to 100
GeV$/c^2$ and out to $\eta_{max}$ = 3. This pure QED cross section is so well known that it may 
provide the most precise measurement of the luminosity of the LHC~\cite{kp,lumi,Shamov:2002yi,Bocian:2004ev,
Krasny:2006xg,Vermaseren:1982cz,Khoze:2000db}
 (integrated over a period), and help calibrate the luminosity
monitors, which give a prompt rate measurement. 
For example the cross section for $M_{\mu^+\mu^-} >$ 10
GeV$/c^2$ and $|\eta|<2.0$, where one should have excellent and very well-known acceptance, 
is 8.55 pb. In 500 pb$^{-1}$, with an unprescaled trigger, one would therefore have $\sim 4275$ events, which translates into a
$\sim$ 1.5\% statistical uncertainty. A knowledge of the acceptance and efficiencies at this level is thus desirable. Without detecting both protons (and no LHC experiments will be able to do that
in normal low-$\beta$ (high luminosity) running at such low masses) a concern is that one or both protons may
dissociate. This probability can be measured in shower counters along the beam pipes at low luminosity, when pile-up can be neglected.

Although the QED calculation of the dilepton cross section is
accurately known, one still needs to worry about the issue of gap
survival $S^2$. We have already seen that this suppression is large for CEP
Higgs production, however $\gamma \gamma$ collisions are much more
peripheral and the gap survival factor is much larger. Nevertheless,
it must be understood at the percent level if we are to exploit CEP
dilepton production as a means to measure the luminosity. Fortunately,
studies indicate that the gap survival should be very close to 100\% for low-mass
dilepton production \cite{Khoze:2000db}.

How does $\gamma\gamma \rightarrow \tau^+\tau^-$ compare with exclusive $p+p \rightarrow p + H + p \rightarrow
p+\tau^+\tau^- +p$? Differentiating the above equation we get
$\d\sigma/\d M \approx$  0.4 fb/GeV$/c^2$ with $|\eta(\tau)| < 2.0$ at 120 GeV$/c^2$. 
The branching fraction for a SM Higgs (120) $H\rightarrow
\tau^+\tau^-$ is $\sim$ 0.07 and so, 
if $\sigma(p+H+p) \sim$ 3 fb~\cite{Khoze:2001xm},
the rates will be at least similar. For example, in a 4~GeV$/c^2$ mass window
(which is commensurate with the expected mass resolution of the
forward detectors) the cross sections would be around 1.6~fb and
0.2~fb respectively. Even assuming no other backgrounds and 100\%
detection efficiency, this translates into no more than $2\sigma$
evidence of $H \rightarrow \tau^+\tau^-$ with 100~fb$^{-1}$ of
data. It will therefore be difficult to see
a SM Higgs in this decay channel at the LHC, but it
may be feasible if the Higgs production rate is enhanced
(e.g. see \cite{Heinemeyer:2009nj}).  The $\gamma\gamma$ process is a continuum with well-known
cross section on which $H\rightarrow \tau^+\tau^-$ would (if the production cross
section is sufficiently large) be a narrow peak. By measuring the
protons, one could exploit the fact that the background is peaked at
smaller values of $|t|$ than the signal to improve the significance
of any observation.

As discussed earlier, exclusive $\mu^+\mu^-$ events are also very important for
the absolute calibration of the forward 
proton spectrometers. Note that the two-photon exclusive 
production of $e^+e^-$ pairs can also be studied at the LHC, though triggering of such events 
is more difficult, and the $\Delta\phi$ and $p_T(e^+e^-)$ measurements are worsened by bremsstrahlung. 

Finally, we note that the prospect of studying so called ``unparticles" using the exclusive $\gamma\gamma \to l^+ l^-$ channel has been studied in~\cite{Sahin:2009gq}. 

\subsubsection*{$W$ and $Z$ boson pairs}
\label{sec:WWZZ}

As illustrated in Fig.~\ref{fig:IntGG}, the rate for pair production of
$W$ bosons via $\gamma \gamma \to W^+W^-$ is large (around 100~fb) and this can be
exploited to investigate the corresponding quartic coupling. The signal is very clean
and offers the possibility to study this process out to $M_{WW} \sim 500$~GeV. The cross section for $p_T^{\mu} >$ 10 GeV$/c$ and $|\eta^{\mu}| < 2.5$
from $p+p \rightarrow p +(\gamma\gamma \rightarrow W^+W^- \rightarrow \mu^+\mu^- \nu \bar{\nu}) + p$ is 0.76~fb~\cite{louvain}.
The large reduction comes from the two $W \rightarrow \mu \nu$ decays. The cross section is only 
slightly reduced after adding the 
requirement of at least one forward proton tag.
The unique signature of $W$ pairs in the fully leptonic final state, no additional tracks on the $l^+l^-$ vertex, 
large lepton acoplanarity and large missing transverse momentum
strongly reduces the backgrounds. 
At low LHC luminosities (i.e.
$ \ll 10^{33}$~cm$^{-2}$s$^{-1}$), pile-up is not a problem and the events can be recognised even without
detecting the outgoing protons \cite{louvain,Chapon:2009ia}. However tagging
the protons would give improved sensitivity and a $M_{WW}$ measurement. At higher
luminosities, proton tagging is needed in order to suppress pile-up backgrounds.
The situation is similar for the production of $Z$ boson pairs,
assuming fully leptonic, or semi-leptonic decays. 
In the SM, $\gamma\gamma \rightarrow ZZ$ is negligible (it is absent at
tree level). Thus any observation would be a direct measurement of an anomalous
$\gamma \gamma ZZ$ coupling. 

In CEP of $W^+W^-$ and $ZZ$, the fully leptonic decays (e.g. $ WW \rightarrow e^- \mu^+ E\!\!\!\!/_T$) have practically no
background,
but constitute only 4.4\% and 0.45\% respectively of all decays.
However in CEP one may be able to use additional four-momentum constraints to perhaps use about 50\% of the events, i.e. all decays
except the (six-jet) fully hadronic. For example, in $p+p \rightarrow
p + WW +p \rightarrow p + l \nu J J + p$, take the
four-momenta of all detected objects $p,p,l,J,J$ and calculate the missing mass, which is $m_{\nu} \sim 0$.
Similarly if $X = ZZ \rightarrow l^+l^- \nu \bar{\nu}$ the missing mass to $p,p,l^+l^-$ is $m_Z$
(provided the missing $Z \rightarrow \nu \bar{\nu}$ is on-shell). CEP can therefore produce a mass peak in $Z \rightarrow \nu \bar{\nu}$.
With similar 4-momentum constraints as much as 50\% of all (on-shell) $W^+W^-$ and $ZZ$ events can be
used.

Anomalous $\gamma\gamma WW$ couplings were first studied, within the
context of a future $e^+e^-$ collider in~\cite{anom}. Under the
assumptions of electromagnetic gauge invariance, C and P conservation
and custodial symmetry (to keep the tree-level $\rho = M_W^2/(M_Z^2
\cos^2 \theta_W)$ parameter equal
to unity) there are only two relevant couplings to lowest order in an
expansion in the inverse of the scale of new physics, $\Lambda$,
i.e. the lowest non-renormalizable terms occur at dimension 6 in the
lagrangian~\cite{anom}:
\begin{equation}
L_6 = -\frac{e^2}{8}\frac{a_0}{\Lambda^2}
F_{\mu\nu}F^{\mu\nu}W^{+\alpha}W^-_{\alpha} - \frac{e^2}{16}\frac{a_c}{\Lambda^2}
F_{\mu\alpha}F^{\mu\beta}(W^{+\alpha}W^-_{\beta} +
W^{-\alpha}W^+_{\beta}). 
\end{equation}
The anomalous couplings are denoted by $a_{0,c}$ and they are zero in
the Standard Model. Sensitivity to these anomalous couplings has been
investigated by the Louvain group~\cite{louvain}. They focussed upon
$\gamma\gamma\rightarrow W^+W^-\rightarrow l^+l^-\nu\bar{\nu}$ and $\gamma\gamma\rightarrow ZZ\rightarrow l^+l^- j j$
using the signature of two leptons ($e$ or $\mu$) within the
acceptance cuts  $|\eta|<2.5$ and $p_T>10$~\GeVc.
With just a few fb$^{-1}$ of data, the limits are expected to
be orders of magnitude better than the best limits established at
LEP2~\cite{OPAL.limits} and significantly better than those that would
be obtained using the $W\gamma\gamma$ final state at the LHC
\cite{Eboli:2000ad,Bell:2009vh}. Similar
conclusions have been reached in the studies presented in
\cite{Chapon:2009hh}.

A study of the anomalous triple gauge couplings has also be performed. However, in this
case the expected sensitivities are not as impressive ~\cite{Kepka:2008yx}.


\subsubsection*{Supersymmetric pairs}
\label{sec:SusyBSM}

\begin{figure}[htb]
\centering
\includegraphics[width=0.8\textwidth]{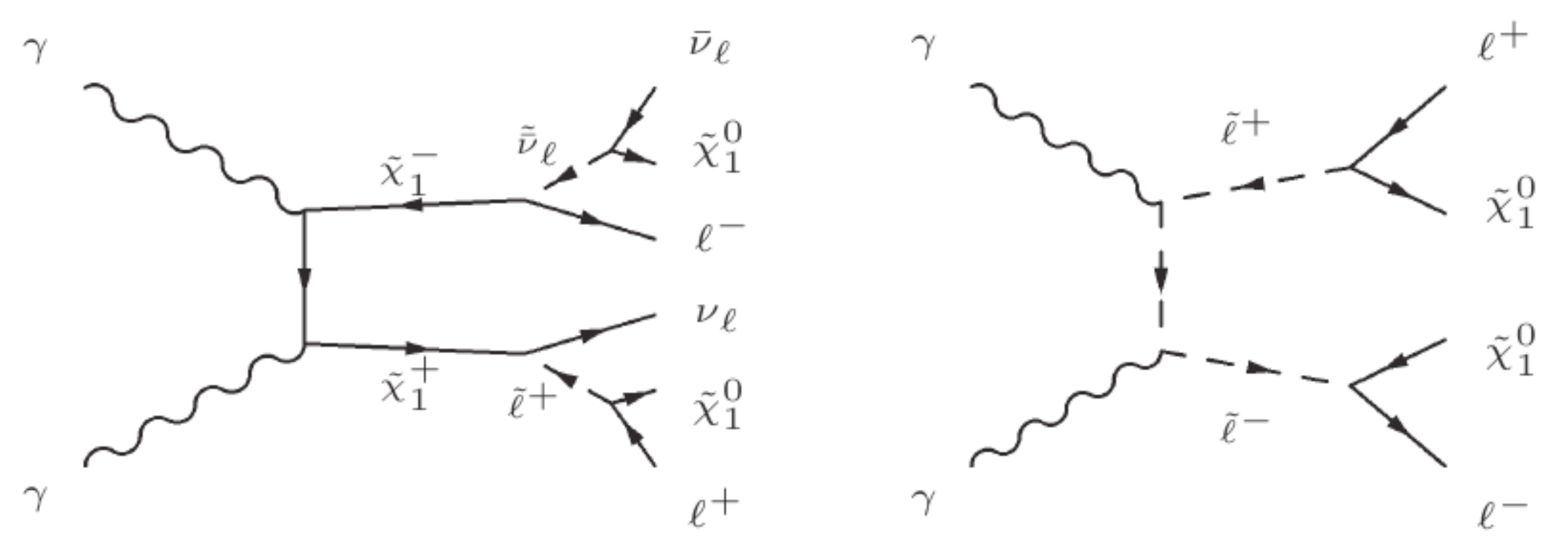}
\caption{Relevant Feynman diagrams for SUSY pair production with leptons in
 the final state: chargino pair production (left); slepton pair
 production (right). Figure from Ref.~\cite{louvain}.}
\label{fig:diag.susy}
\end{figure}

The LHC, of course, is geared up to discover and explore low-scale
supersymmetric particle production. However, SUSY pair production in
$\gamma\gamma$ collisions, with two tagged protons, offers the opportunity
to make direct measurements of the sparticle masses
\cite{louvain,Piotrzkowski:2009sa}. As we shall see, the cross
sections are not large and sufficient rate is only possible if the
sparticles are light. The two-photon
production of supersymmetric particles has been investigated
in~\cite{bib:Ohnemus,bib:Bhattacharya,bib:Drees}. 
For chargino or slepton pair production, the final states are simple,
see Fig.~\ref{fig:diag.susy}. Selecting a
final state composed of two charged leptons with large
missing energy and lepton acoplanarity means that the backgrounds should
be under control, and $\gamma \gamma \to W^+W^-$ generates the only
irreducible background. 

In Ref.~\cite{louvain}, three benchmark points in the parameter space of
the constrained minimal supersymmetric model (CMSSM) were
studied\footnote{The parameter space is constrained to be in
  agreement with the latest cosmological data~\cite{wmap}.}. Here we
will present the results for point LM1, which
predicts a 97~GeV$/c^2$ neutralino (the LSP), a $\tilde{l}^{\pm}_R$ of mass 118~GeV$/c^2$ 
($l=e,\mu$),  a $\tilde{l}^{\pm}_L$ of mass 184~GeV$/c^2$  and a lightest chargino, $\tilde{\chi}_1^+$, of mass 180~GeV$/c^2$.  Signal and background samples were produced using a modified
version of {\sc calchep}~\cite{Puk03} and the following acceptance cuts were applied: two leptons with $p_T >$ 3 \GeVc\ or 10 \GeVc\ and $\vert \eta \vert < 2.5$. 
Standard high level trigger efficiencies are high for all these
types of events.

\begin{figure}
\centering
\includegraphics[width=.6\textwidth]{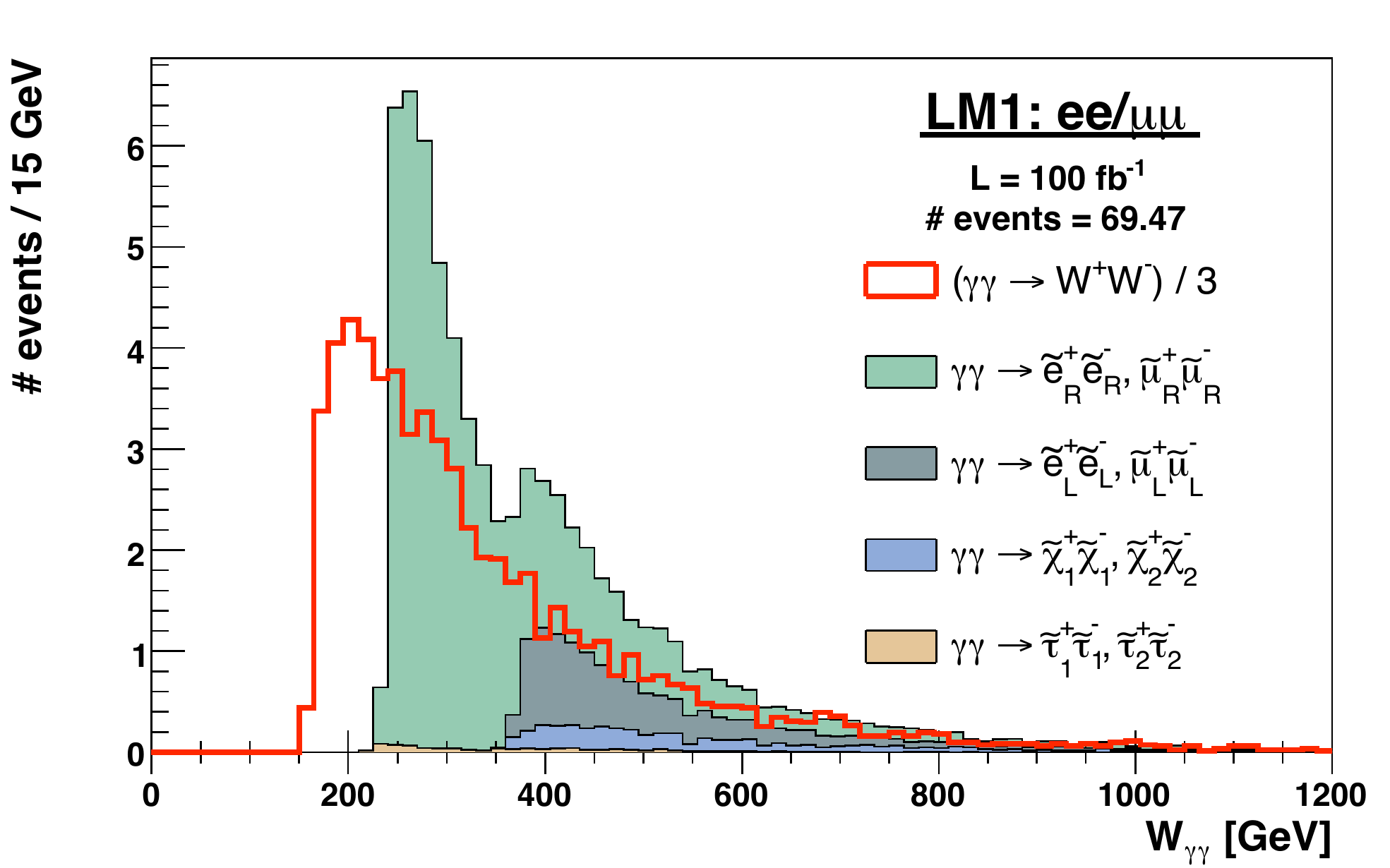}
\caption{The invariant mass spectrum for SUSY pair production in the
  LM1 scenario and assuming 100 fb$^{-1}$ of data. 
The $WW$ background  has been down-scaled by the quoted factor. See
text for event selection criteria. Figure from Ref.~\cite{louvain}.}
\label{fig:gamma-gamma-inv-mass}
\end{figure}

Measuring the outgoing protons allows one to measure the cross section
as a function of $M_{\gamma\gamma}$, and expected event
rates with 100~fb$^{-1}$ of data are presented in
Fig.~\ref{fig:gamma-gamma-inv-mass}. The two edges, at twice the mass
of the $\tilde{l}^{\pm}_R$ and at twice the mass of the
$\tilde{l}^{\pm}_L$ are visible. However, the event rates
are low and mass measurements with a precision of a few \GeVcc~will
require very high statistics.
{\sc hector}~\cite{hector} simulations of forward protons from slepton events consistent with the LM1 benchmark point indicate that
the 220~m detectors will have both protons tagged for only $30\%$ of
events.  Addition of detectors at 420~m increases that to $90\%$ of events. 

\begin{figure}
\centering
\includegraphics[width=.5\textwidth]{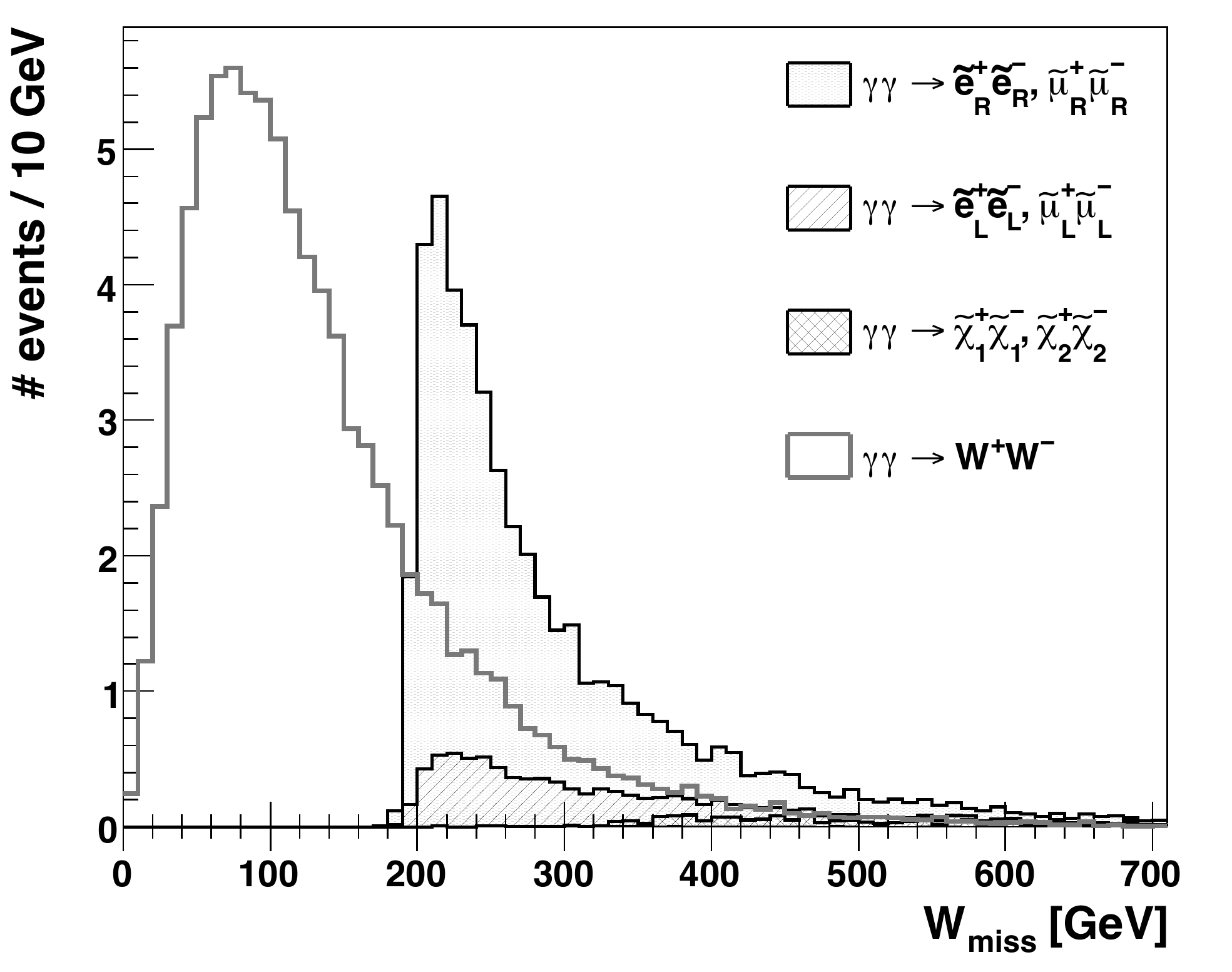}
\caption{The missing mass distribution for SUSY pair production in the
  LM1 scenario and assuming 100 fb$^{-1}$ of data. Event selection
  assumes two isolated opposite charge muons with $p_T > 7$~GeV$/c$ or two electrons with
  $p_T > 10$~GeV$/c$, $|\eta_l|<2.5$ and two tagged protons. 
Figure from Ref.~\cite{Piotrzkowski:2009sa}.}
\label{fig:wmiss}
\end{figure}

Another attractive feature of two-photon production of SUSY pairs with
tagged protons is that both the missing \emph{energy} $E\!\!\!\!/$ and $E\!\!\!\!/_T$ are measured,
as well as the missing mass (here called $W_{\mathrm{miss}}$). 
For the SUSY signal $W_{\mathrm{miss}}$
 starts at $2 \times m_{LSP}$\cite{Piotrzkowski:2009sa}, as
illustrated in Fig.~\ref{fig:wmiss}, while the background starts at $W_{\mathrm{miss}}=0$

\subsection{$\gamma p$ collisions}
\label{sec:GP}
In addition to $DI\!\!PE$ and two-photon collisions, it is also possible to study exclusive 
photoproduction ($\gamma I\!\!P$) at the LHC. A quasi-real photon is radiated off one of the 
protons and fluctuates into a $q\bar{q}$ pair, which scatters diffractively on the other proton. 
In such a reaction, only states with the same quantum numbers as the photon can be produced 
(the $\rho,\omega,\phi,J/\psi$ and $\Upsilon$ mesons, their excitations and the $Z$ boson). 
While vector meson photoproduction has been extensively studied in $ep$ collisions at HERA, 
it has only recently been observed (for $J/\psi$ and $\psi(2s)$ states) in hadron-hadron collisions 
at the Tevatron~\cite{Aaltonen:2009kg} (see section~\ref{tevcep}). It is not our aim here to present a complete survey of 
the physics which can be studied via exclusive photoproduction at the LHC. Instead, we focus on two 
examples, $\Upsilon$ production and $Z$ production. 

At the LHC, photoproduced $\Upsilon \rightarrow \mu^+\mu^-$ may be
measurable, with one proton detected (there is no acceptance to
measure both). The cross section is sensitive to QCD saturation effects, which arise as
a result of non-linear gluon dynamics. The large mass of the $\Upsilon$
offers hope that this effect could be quantified in QCD perturbation theory. 
Central exclusive $\Upsilon$ production at the LHC has been studied in
Refs.~\cite{motykaz,starlight,Rybarska:2008pk,Bzdak:2007cz,Goncalves:2007sa,Cox:2009ag}. The
study in Ref.~\cite{Cox:2009ag} presented results using a model for
diffractive scattering which fits both the existing HERA data, not
only for $\Upsilon$ production but also for the total inclusive deep inelastic
scattering cross section. The model includes a rudimentary modelling
of saturation effects and predicts a cross section of 5.1~pb for
14~TeV $pp$ collisions at the LHC assuming neither proton is detected. This
cross section is for $\Upsilon \rightarrow \mu^+\mu^-$ with $p_T^{\mu}>4$~GeV$/c$. 
The cross section falls to 3.0~pb if one proton is detected at 420~m from
the interaction point. These are large cross sections and can be
measured with early LHC data. Tagging one proton should help control
the pile-up background, and proton dissociation e.g. $p\rightarrow p\pi^+\pi^-$, thereby making the measurement possible also
at high luminosities. The large rates mean that it will also be
possible to measure the upsilon rapidity and this should
help distinguish between different models of high-energy
scattering. Measuring one proton makes possible a direct measurement
of the photoproduction cross section, $\sigma(\gamma p \to \Upsilon
p)$ since a cut on the transverse momentum (e.g. around $100-300$~MeV)
of the scattered proton
will permit one to divide the event sample into two halves: one in
which the detected proton predominantly radiated the photon and one in which it was
the undetected proton that radiated the photon. The low $p_T$ sample
corresponds to a measurement at higher values of the $\gamma p$
centre-of-mass energy, with sensitivity
up to 1~TeV in the $\gamma p$ centre-of-mass energy. The prediction of the 
photoproduction cross section is, however, complicated
by any dependence of gap survival on the scattered proton $p_T$
\cite{Khoze:2002dc}. On the other hand, the measurement gives information on $S^2$.
It should also not be forgotten that central
production of vector mesons can proceed also via odderon exchange, and it would be very interesting to observe that.

Also possible at the LHC is the diffractive photoproduction of
$Z$ bosons~\cite{goncalves,motykaz}. As discussed in section~\ref{tevcep}, the Standard Model cross section
for $Z$ production is negligibly small at the Tevatron:
$\d\sigma/\d y|_{y=0} = 0.077$~fb, compared with an upper limit from CDF of 0.25~pb at 95\% C.L.~\cite{Aaltonen:2009cj}. At the LHC however, the cross section is expected to be about 6 fb for $|y(Z)| < 2$, or 40 exclusive $Z \rightarrow e^+e^-$ or $\mu^+\mu^-$ events per 100 fb$^{-1}$~\cite{motykaz}.
Clearly one will have to work at high luminosities, which means that
pile-up backgrounds need bringing under control. This should be
possible by selecting events with only the $l^+l^-$ tracks on
the vertex, a dilepton invariant mass matching that of the $Z$, small
$p_T(l^+l^-)$ and $\pi -\Delta\phi(l^+l^-)$, in
conjunction with the requirement that the vertex be coincident with
that obtained using two measured protons (at 420~m).

\section{Concluding remarks}
\label{sec:conclude}

The subject of central exclusive production promises to deliver an
entirely new dimension to the physics that will be explored at the
LHC. With the installation of forward low-angle proton detectors, central
masses can be measured out to a few hundred GeV$/c^2$ and to a precision
at or below 1~GeV$/c^2$, even in the presence of
missing energy, using very few events. Moreover, tagging the protons
effectively filters the central system with a strong bias towards
$J=0$, even-parity systems. Such a filter will provide a unique
handle on the nature of any new physics that might be produced. It is
true to say that this programme constitutes a huge advance over
previous studies of CEP performed mainly at the ISR, S$p\bar{p}$S and
Tevatron. For the first time (with the exception of the glueball
searches), central masses can be produced in a
range where interesting new phenomena are expected. 

Measurements from CDF at the Tevatron are
very encouraging and support the validity of the theoretical
calculations based in perturbative QCD: The theory is probably not too far off the
mark. A second highlight of the recent past has
been the demonstration that high luminosity backgrounds can be brought
under control, which opens up the possibility to explore low
(i.e. below 1~fb) cross section processes at the LHC.

There are many concrete examples of new physics that may reveal itself in
CEP, and we have reviewed many of them here, but it should always be
understood that CEP is principally a direct source of glue-glue (or photon-photon)
interactions with a filter on the final state quantum numbers. It is
thus ideally suited to a study of any new
physics that couples to gluons (or photons).

\section*{Acknowledgements}
We should like to thank the very many of our colleagues who have made
our ventures into central exclusive production such a
pleasure. Particular thanks are due to Paul Bell, Mike Birse, Frank Close, Brian Cox, Dino Goulianos, Valery
Khoze, Krzysztof Piotrzkowski, Andy Pilkington and Misha Ryskin. We should
like also to thank the UK's STFC and the U.S. Dept. of Energy for financial support.